# Exact modes, hybridization and polarization rotation of electromagnetic fields propagating in topological insulating slab

Sebastián Filipini,[1,*] Mauro Cambiaso,[1,†] Alberto Martín-Ruiz,[2,‡] and Luis F. Urrutia[2,§]

[1]*Universidad Andres Bello, Departamento de Física y Astronomía, Facultad de Ciencias Exactas, Avenida República 220, Santiago, Chile*

[2]*Universidad Nacional Autónoma de México, Instituto de Ciencias Nucleares, Ciudad de México, 04510, México*

In this work, we study electromagnetic wave propagation in slab waveguides whose core is a topological insulator characterized by a topological magnetoelectric parameter. Topological insulators are materials that are electrically insulating in the bulk but support robust conducting states at their boundaries, protected by time-reversal symmetry. Their electromagnetic response is described by an axion-like term that modifies Maxwell's electrodynamics, leading to rich and unconventional phenomena, as the topological magnetoelectric effect. We find that all modes supported by these structures are all exact hybrid modes with nonvanishing longitudinal field components. This hybridization is a direct consequence of the modified boundary conditions produced by the topological term and is absent in topologically trivial, reciprocal and non-chiral slab waveguides. Modifications to the propagation condition and ensuing modes are shown for the case of an asymmetric slab, however, the detailed solution of the exact modes, coupling of modes and the dispersion relations is made for the symmetric slab. By solving the full $\Theta$-electrodynamics equations nonperturbatively, we derive the modal dispersion relations and explore polarization rotation and power transfer between modes. Our approach reveals qualitative and quantitative deviations from standard coupled-mode theory and captures new signatures of the topological magnetoelectric response. To compare with other approaches to the subject and owing to the smallness of the $\Theta$-effects, we perform a perturbative analysis of mode propagation, also based on writing a general solution as a superposition of exact modes of $\Theta$-ED but expanding to first non-vanishing order in $\Theta$. Also, we apply the methodology of coupled-mode theory. This last approach is predicated on building solutions as superpositions of modes of ordinary electrodynamics that fail to satisfy the boundary conditions imposed by the $\Theta$-term but compensate at the expense of modifying the field profiles. These findings provide a comprehensive framework for light control in topological photonics and potential routes to experimentally probe the magnetoelectric effect in guided settings.

## I. INTRODUCTION

Slab waveguides are a particular kind of waveguides for electromagnetic (EM) radiation. In their simplest form, slab waveguides consist of three planar media adjacent to each other, piled up along a *stacking* direction. The central medium is a high refractive index material, usually called *core*, and lies between lower refractive index layers, usually called the *cladding* layers. Along the core material, the electromagnetic fields are confined and propagate. The width of the core medium in the stacking direction will be called the opening of the slab waveguide, and is the characteristic length scale of the system. All media are of infinite extent in the directions parallel to their interfaces. The cladding layers may also be of infinite extent in the direction transverse to the interfaces and opposite to the core. Of infinite extent actually meaning that the dimensions in these directions are much larger than the width of the core medium. When both cladding layers have the same/different refractive indexes we speak of a symmetric/asymmetric slab waveguide. The boundary conditions (BCs) for the EM field at the core-cladding interfaces allow that the EM field propagating within the core undergoes total internal reflection at the core-cladding interfaces, provided the incident angle be larger than the critical angle and, as assumed, that the refractive index of the core be higher than those of the claddings. The same BCs also allow for evanescent waves in the claddings but only very close to the interfaces to decay exponentially away from the interface. The extent to which these evanescent waves carry energy allows to discriminate how good a given waveguide is in confining the transmitted energy. In fact, this determines an efficiency parameter of the guide. This confinement mechanism allows for efficient light propagation along the waveguide propagation direction.

Slab waveguides occupy a foundational role in integrated optics and their application pervades integrated photonic devices used in modern photonics and telecommunications [1, 2], in [3] to confine light in low-index materials for integrated photonics applications, or in [4] for guiding laser beams to generate THz signals in photomixer sources, and even for highly sensitive biochemical detection as in [5]. Ultimately, the interplay between the BCs at the core-cladding interfaces; the condition for total internal reflection; and the condition for having (or avoiding) evanescent waves in the claddings, lead to the characteristic features of slabs, namely: a) the EM field propagates along the waveguide with definite and restricted field distributions and propagation constants; b) precise dispersion relations depending on the core and cladding refractive indexes; c) a modal structure given the fact that the dispersion relation dictates that only certain frequencies (and hence certain propagation constants) can propagate along the waveguide in a confined manner; and d) the absence of "hybrid" modes, that is modes that are neither transverse electric (TE) nor transverse magnetic (TM). In standard electrodynamics, in the context of guided EM waves, TE and TM, are two sets of independent solutions of the field equations. That is to say, the solutions are either combinations of completely TE modes or combinations of completely TM modes or a combination of both TE and TM modes. The particular kinds of modes a given waveguide can host depend critically on the BCs that are determined by the geometry of the guide and the spatial dependence of the optical parameters. The modes are thus classified as TEM if both longitudinal components of the field vanish, $E_z = 0$ and $B_z = 0$; as TE if the longitudinal component of the electric field vanishes, $E_z = 0$ and $B_z \neq 0$; TM if the longitudinal component of the magnetic field vanishes, $B_z = 0$ and $E_z \neq 0$; or hybrid if neither longitudinal component vanishes, i.e., $E_z \neq 0$ and $B_z \neq 0$. With slab waveguides in conventional electrodynamics, however, the assumed infinite extension in the transverse direction that is not the stacking direction, mandates that the modes are in fact either TE or TM. These characteristic features are essential for light-manipulating purposes and are present in several other kinds of waveguides, optical fibers and integrated photonic circuits. Thus, slab waveguides offer a simplified model for understanding and analyzing more complex photonic structures and also potentially new kinds of photonic integrated circuits whenever new materials or new techniques are available.

* s.filipiniparra@uandresbello.edu
† mcambiaso@unab.cl
‡ alberto.martin@nucleares.unam.mx
§ urrutia@nucleares.unam.mx

In the recent years, there has been an increased interest on using Topological Insulators (TIs) and more generally metamaterials and their interaction with EM radiation in the context of waveguiding of electromagnetic fields. [6–9]. As far as the electromagnetic response of TIs is concerned and provided that the energy per photon associated with EM radiation of a given frequency be smaller [1], than the bandgap energy of the TI, $\hbar\omega \ll E_g$ [6, 10], the effect of the material being a TI is encoded in the so-called topological magnetoelectric parameter (TMEP) denoted by $\Theta$, which, at the macroscopic level plays a role similar to those played by the permittivities and permeabilities, that can lead to observable phenomena as signals of the topological magnetoelectric effect (TME) as if they were optical responses but of topological origin, [11–14]. For waveguiding purposes, the interest is two-fold, on the one hand TIs change the BCs that the electromagnetic field must satisfy [15–18], therefore, for slab waveguides this can result in a modification of: the dispersion relation, the allowed field modes and other of their subtle properties. The idea behind being that the core-cladding interface is now appended with the additional feature associated with the TMEP $\Theta$ characterizing the TI. This can be achieved by considering that the core of the slab waveguide be a TI itself. This provides an additional way in which slab waveguides can be functionalized to modify or improve the properties of light propagating along TI slabs. On the other hand, it is very interesting for understanding and further characterizing the topological magnetoelectric response of TIs to the EM radiation. Optics and photonics allow for highly sensitive measurements, hence this opens the possibility to probe or measure signals of the elusive topological magnetoelectric effect of TIs in ways that are different to what has been done so far.

The electromagnetic response of topological matter can be obtained from the standard macroscopic Maxwell's equations in terms of $\mathbf{D}$ and $\mathbf{H}$ via the following constitutive relations in the spacetime domain

$$\mathbf{D} = \epsilon \mathbf{E} + \Theta(\mathbf{x}, t)\mathbf{B}, \qquad \mathbf{H} = \frac{1}{\mu}\mathbf{B} - \Theta(\mathbf{x}, t)\mathbf{E}, \tag{1}$$

where $\Theta(\mathbf{x}, t)$ is a real non-dynamical field (called the axion field), i.e. a spacetime dependent parameter without governing equations that characterizes the medium in a similar way as the permittivity $\epsilon$ and permeability $\mu$. This is why we denote the resulting electrodynamics by $\Theta$-ED, instead of axion-ED, in the following. Substituting the relations in Eq. (1) in the macroscopic Gauss and Ampere laws produces Maxwell's equations that not depend on $\Theta$, but only on their gradients $\nabla\Theta$ and $\partial_t\Theta$. The choices of $\Theta(\mathbf{x},t)$ either as a time independent piecewise constant function, or as $\Theta(\mathbf{x}, t) = \mathbf{b}\cdot\mathbf{x} - b_0 t$ describe the electromagnetic response of TIs and Weyl semimetals, respectively. As an important conceptual issue we emphasize that the corresponding effective action yielding the constitutive relations in Eq. (1) is obtained from the microscopic structure of each material [10, 19]. Since the constitutive relations in Eq. (1) are reminiscent of those describing a bi-isotropic medium (for a review see for example Refs. [20, 21]), we find it convenient to briefly survey their main differences, together with some previous related work in this well established area of electromagnetism. We start from the bi-isotropic constitutive relations as presented in Refs. [21–23]

$$\mathbf{D} = \epsilon_{\rm EH}\mathbf{E} + (\chi_{\rm EH} - i\kappa_{\rm EH})\mathbf{H}, \qquad \mathbf{B} = \mu_{\rm EH}\mathbf{H} + (\chi_{\rm EH} + i\kappa_{\rm EH})\mathbf{E}, \tag{2}$$

that is written explicitly according to the conventions summarized in Ref. [23]. Taking $\mathbf{E}$ and $\mathbf{B}$ as the fundamental fields we can re-express them in the Post-Jaggard form [22]

$$\mathbf{D} = \Big(\epsilon_{\rm EH} - \frac{1}{\mu_{\rm EH}}(\chi_{\rm EH}^2 + \kappa_{\rm EH}^2)\Big)\mathbf{E} + \frac{1}{\mu_{\rm EH}}(\chi_{\rm EH} - i\kappa_{\rm EH})\mathbf{B}, \qquad \mathbf{H} = \frac{1}{\mu_{\rm EH}}\mathbf{B} - \frac{1}{\mu_{\rm EH}}(\chi_{\rm EH} + i\kappa_{\rm EH})\mathbf{E}. \tag{3}$$

[1] In fact, the bandgap energies are specific to each material. Furthermore, the bulk and surface bandgap energies of a solid differ, so for this argument we care for the smallest bandgap energies (between bulk or surface) of the TI in question. For several present day TIs these energies are close to $E_g \sim 0.3$ eV.

We observe that putting $\kappa_{\rm EH}=0$, and identifying

$$\epsilon=\epsilon_{\rm EH}-\frac{1}{\mu_{\rm EH}}\chi^2_{\rm EH}, \qquad \Theta(\mathbf{x},t)=\frac{\chi_{\rm EH}}{\mu_{\rm EH}}, \tag{4}$$

one can bring the representations of the constitutive relations of bi-isotropic media Eq. (3) and of topological matter Eq. (1) to an equivalence. Let us remark that in the notation of [23] (Eqs. 2.8 - 2.11), $\epsilon=\epsilon_{\rm EB}$ and $\Theta=\Theta_{\rm EB}$. However, despite the formal equivalence, in this work we will be concerned with topological matter as described by the constitutive relations in Eq. (1).

The bisotropic constitutive relations are written in the space-frequency domain, are valid only for harmonic time-dependent electromagnetic fields and are obtained by modeling the response of macroscopic objects randomly embedded in a dielectric. In general the parameters $\kappa_{\rm EH}$ and $\chi_{\rm EH}$ can depend on frequency with the reality conditions $\chi^*_{\rm EH}(\omega)=\chi_{\rm EH}(-\omega)$ and $\kappa^*_{\rm EH}(\omega)=-\kappa_{\rm EH}(-\omega)$. The parameter $\kappa_{\rm EH}$ describes a reciprocal chiral media (Pasteur media), modeled in Ref. [24], while $\chi_{\rm EH}$ corresponds to a non-reciprocal media dubbed also a Tellegen media [25] which can be engineered as Tellegen particles [26–28]. However, from the point of view of a Tellegen media an unexpected situation occurs: the Maxwell's equations arising from the constitutive relations of Eq. (2) are independent of the constant parameter $\chi_{\rm EH}$ and so are the boundary conditions that emerge from the differential equation. This has raised some doubts regarding the physical existence of such media, a much discussed topic centered on the validity of the Post constraint [29]. A similar situation occurs in the case of a topological insulator described by a constant axion field $\theta$ in a medium without boundaries. Nevertheless, this consequence applies perfectly in this case since TIs are characterized as being standard insulators in the bulk ($\theta=0$), with conducting properties arising solely from topological surface states at the boundaries.

In other words, the modification to Maxwell equation in TIs, as well as in Tellegen media, arises only at the interface between two such materials having different values of $\theta$ and $\chi_{\rm EH}$, respectively. For example, in the case of two semi infinite TI's having $\theta_1$ and $\theta_2$ respectively, and separated by a plane interface at $x=0$ the correct identification is

$$\Theta(\mathbf{x},t)=\theta_1\,H(-x)+\theta_2\,H(x), \tag{5}$$

where $H(x)$ is the Heaviside function, yielding $\nabla\Theta=(\theta_2-\theta_1)\delta(x)\,\hat{\mathbf{e}}_x$ which guarantees the appearance of a correction term in Maxwell equations. As it will be evident in section II, the extra terms appearing in the equations can be interpreted as an effective surface density and an effective surface current which turn out to be proportional to $\nabla\Theta\cdot\mathbf{B}$ and $\nabla\Theta\times\mathbf{E}$, respectively. These correspond to the electromagnetic response of the fermionic surface states in a TI. The need to parametrize the Tellegen parameter $\chi_{\rm EH}$ as indicated in Eq. (5) was early recognized in Ref. [30]. In summary, a generalization of the Tellegen media by formulating its electrodynamics in terms of the constitutive relations Eq. (1), together with the parametrization Eq. (5), presents at least two advantages: (i) it does not restrict applications to harmonic fields and (ii) it highlights the consistency between the equations of motion, with the parameter $\chi_{\rm EH}$ appearing through the contribution at the interface, and the boundary conditions that are calculated in the usual way from the equations via the well known application of the integral forms of the Gauss and Ampere laws. In addition, effective surface charges and currents can be identified as a direct consequence of $\chi_{\rm EH}$, rather than by additional contributions as considered in Refs. [31, 32]. It is important to emphasize a fundamental difference with standard bi-isotropic materials: the topological properties of TIs demand $\theta$ to be quantized in odd units of $\pi$, as a consequence of time reversal invariance in regions with no boundary [10, 19]. This produces additional observable contributions to Faraday and Kerr rotations and yields the correct form of the anomalous Hall effect, among other consequences.

Previous work on bi-isotropic waveguides has been mainly concentrated on chiral media ($\chi_{\rm EH}=0$), using alternative approaches to solve Maxwell's equations. The corresponding waveguides were introduced in Ref. [33] and christened chirowaveguides. In this reference, together with the more detailed version [34] the authors discussed cylindrical wave guides of arbitrary section with a perfect conductor as the boundary surface, together with the particular case of a parallel plate waveguide. They showed that only hybrid modes propagate and that there is a bifurcation of modes with a common cut-off frequency. A next step was taken in Ref. [35] using the Condon parametrization for the constitutive relations [36] and discussing in detail the case of a perfectly conducting circular chirowaveguide, finding similar general properties to the previous publications. Slab chirowaveguides were discussed in Refs. [37, 38] considering a central semi-infinite rectangular slab of bi-isotropic material surrounded by a standard dielectric. The previous slab configuration was modified to admit more general boundary conditions at the bi-isotropic central media including metallic and magnetic surfaces in Refs. [39, 40]. More recently, an additional modification of the original slab geometry was considered in the study of H chirowaveguides [41]. Inclusion of the Tellegen parameter in a general bi-isotropic media ($\chi_{\rm EH}\neq 0$, $\kappa_{\rm EH}\neq 0$) is considered in the study of cylindrical waveguides with arbitrary cross section and different boundary impedance conditions [42].

It is worth mentioning that in [43] a configuration of a the parallel-plate waveguide filled with a Tellegen (nonreciprocal) bi-isotropic medium is considered. In that work the authors derive the existence of TE modes and also hybrid modes, while showing that pure TM modes are excluded. In this case, however, the boundary conditions are imposed by perfect conductors playing the role of the core-cladding interface that we will exploit.

In this work we will be concerned with EM wave propagation along slab waveguides made of TIs and ordinary dielectrics. Specifically, the core material will be a TI and the claddings made of conventional dielectrics. In this context, we find: confined and radiation modes, evanescent waves at the core-cladding interfaces, modal dispersion relations, restriction on the allowed incidence angle of the EM field, rotation of the plane of polarization at total internal reflection at each core-cladding interface and hybridization of the confined modes. This is, in fact, a feature that is not present in conventional slab waveguides and, therefore, is a distinctive mark of the topological magnetoelectric effect. Some of these effects are exclusive of $\Theta$-ED and others occur naturally in the conventional electrodynamics of slab waveguides. In the latter case, the features will acquire slight $\Theta$ modifications which are subdominant compared to the usual optical responses, but in any case, the effects are, in principle, observable due to the high precision of optical, spectroscopic and other experimental techniques.

In the literature this has been addressed but, to the best of our knowledge, mostly in a "perturbative" sense. Namely to find solutions to the EM field equations that, compared to the case where the TMEP of the core material would be zero, receive only first order corrections in the TMEP of the core medium to the allowed confined modes, as done in [44]. Though sensible, owing to the smallness of the topological magnetoelectric effect of TIs (fundamentally because of its quantum origin and quantitatively because the topological magnetoelectric coupling comes along with the fine-structure constant), this approach misses some features. As far as the confined modes are concerned, a complete or exact solution not only is possible but also reveals features of the allowed confined modes that are lost in the approximation. For a slab waveguide characterized by a given geometry and set of optical parameters $\{\epsilon_i, \mu_i, \Theta_i\}$, where $i$ labels each medium, the "approximate" solution entails finding solutions to the $\Theta$-ED equations as linear combinations of the conventional confined modes that the same slab with $\Theta_i = 0$ would host (*aka* 0-ED modes) and imposing the $\Theta$ boundary conditions. This is basically the idea of coupled-mode theory.

Alternatively, we set out to obtain EM field solutions by finding the exact modes allowed by the $\Theta$-slab waveguide. This approach leads to confined modes that become $\Theta$-dependent and thus differ from those of 0-ED used in the coupled-mode theory. In [45] the author considers anisotropic magnetoelectric slabs in which the magnetoelectricity can be due to chirality and topological magnetoelectric polarizability and finds modified dispersion relations and a kind of hybridization among the individual transversal components of the TM modes, however a clear derivation and final expression of the exact hybrid (neither TE nor TM) modes are not provided. Similarly, a clear analysis of how external arbitrary EM fields couple to the waveguide, and hence which modes propagate is lacking. Interestingly, our exact approach leads to altogether new solutions that are not possible neither in 0-ED, nor when using coupled-mode theory to solve a $\Theta$-slab waveguide. For example, in our $\Theta$-slab all EM field modes are hybrid, a fact that does not occur for conventional slabs, as previously mentioned. The hybridization is triggered by the $\Theta$ modification of the BCs and, albeit small, implies an ineluctable mixing of the modes as the EM wave propagates along the slab. Even in conventional electrodynamics (meaning in topologically trivial dielectric waveguides), the description of general EM fields in terms of linear combinations of hybrid modes is highly non-trivial. The presence of degenerate modes implies there is no unique way to choose the basis set of modes. This justifies the use of the perturbative approach of coupled-mode theory, but therein lies a subtle but important contribution of this work. These different, and new signals of the topological magnetoelectric effect open the door to as yet unknown possibilities in the context of optics and photonics.

The paper is organized as follows. In Sec. II we introduce the general framework of $\Theta$-electrodynamics as a tool to describe the electromagnetic response of TI, the resulting modification to the boundary conditions for cases in which each media is characterized by constant $\Theta$ parameter. In the interest of plane wave solutions propagating in the slab waveguide along the $OZ$-direction, we write the general wave equations, the dispersion relations and obtain the transverse component of the fields in terms of their longitudinal components. We define the particular setup under consideration and for the dispersion relation in each media, we demand that in the TI core, the EM field be an oscillatory function of the longitudinal coordinate, while the fields decay exponentially in the transverse direction, and thus allow for the EM field to propagate along the waveguide in a confined manner. Section III, deals with the exact solutions once the boundary conditions of $\Theta$-ED for our particular setup are imposed. These boundary conditions are not only dependent on the parameters $\{\epsilon_i, \mu_i, \Theta_i\}$ of each media, but they also depend on the differences of $\Theta$ at the interfaces between media. In Sec. III A we show the derivation of EM fields in all regions of the waveguide in terms of the longitudinal components and also the propagation conditions for arbitrary $\epsilon, \mu, \Theta$. Different configurations are possible, therefore in Sec. III B we introduce the symmetric and asymmetric slab (in regards to the claddings) and in III C we focus on the so-called symmetric and antiparallel slab (in regards to the gradients of $\Theta$ at each core-cladding interface) . Applying the BCs leads to the fact that for a given frequency only certain "modal solutions" are possible, expressed as relations between the wavenumbers in each media written in terms of transcendental equations. Already here we find that the modes for a $\Theta$-slab differ from those of the equivalent topologically trivial one. In this section a thorough description of the field modes, their profiles, orthogonality relations and propagation is made. Section IV is devoted to the issue of coupling and external field to the slab waveguide. First, so as to connect with the modes of the 0-ED, in IV B we expand the $\Theta$-ED modes in the slab to first order in $\Theta$, evincing that the modes are coupled in a fashion that, captures the magnetoelectric effect (Eqs.(78) and (79)). Then in IV C, as a means of comparison with other approaches, we address the same issue of coupling external modes to the slab waveguide, but in IV A we answer the question non-perturbatively in $\theta$, i.e., in terms of the exact $\Theta$-ED modes.In Sec. V we analyze the dispersion relations and relate them with the conditions that the geometrical and optical parameters of the waveguide impose on the allowed operational frequencies in such a way that this is consistent with the constraint demanded by the validity of the topological field theory of $\Theta$-ED as a legitimate theory for the EM response of topological insulators. Given the dispersion relations are also predicated on the aforementioned transcendental equations, these have to be determined numerically. To this end we must choose some specific geometric and optical parameters, a frequency and, for illustrative purposes, some arbitrary

Θ-parameters. Section VI addresses Coupled-Mode theory as customarily done for ordinary waveguides, but here we do take into account the Θ-wave equations. Since the Θ-effects of TIs are usually suppressed by the fine-structure constant, it is sensible to assume that an arbitrary EM field in the Θ-slab can be expanded in the modes of the 0-ED. In so doing one finds that the expansion coefficients of a given mode, couple to those of a different mode as the wave propagates along the waveguide. Applying this again to a symmetric and antiparallel TI slab waveguide we show how the first two modes couple and hybridize. An analysis of the power carried by each mode, shows that energy is conserved despite the coupling. Finally in Sec. VII we summarize our results and conclude with some final remarks, emphasizing the difference of our approach and those followed by other authors. The experimental implications of this is commented and also we entertain the idea of Θ-slabs with large TMEP values or with ENZ (epsilon near zero values) materials, that are now feasible [8], as a means to improve the confinement of the EM waves in a Θ-slab.

## II. Θ-SLABS

### A. General methodology

Topological insulators are one of the paramount examples in condensed matter physics of states presenting topological order [46–55]. The origin of the emerging properties of TIs lies in their microscopic structure, where spin-orbit coupling plays a crucial role. Due to the latter, TIs present conducting edge/surface states [19, 56]. On the contrary, in the bulk, the states are gapped as in conventional insulators, which are then called topologically trivial insulators. These conducting edge/surface states, have been experimentally observed [57–61], are protected against disorder by time-reversal symmetry (TRS) and therefore have gathered great attention as possible candidates for low-dissipation electronic applications. In this work, we will be concerned with the fact that, insofar their interaction with the EM field, TIs are magnetoelectric. That is, the electric and magnetic fields, get intertwined by the presence of a TI. More precisely, when a TI interacts with the EM radiation, provided the energy carried by the EM field, say of frequency $\omega$ is much smaller than the band-gap energy of the surface states of the TI ($\hbar\omega \ll E_g$) [10], the EM response of TIs is described by the so-called Θ-electrodynamics (Θ-ED), defined by the following action:

$$S[\Phi, \mathbf{A}] = \int dt d^3\mathbf{x} \left[ \frac{1}{8\pi} \left( \epsilon \mathbf{E}^2 - \frac{1}{\mu} \mathbf{B}^2 \right) + \frac{1}{4\pi} \Theta(\mathbf{x}, t)\, \mathbf{E} \cdot \mathbf{B} - \rho \Phi + \frac{1}{c_0} \mathbf{J} \cdot \mathbf{A} \right], \tag{6}$$

where $\epsilon$ and $\mu$ are the permittivity and permeability of the medium and $\Theta$ is the TMEP, where it is understood that the fields are written in terms of the potentials:

$$\mathbf{E} = -\boldsymbol{\nabla}\Phi - \frac{1}{c_0}\partial_t \mathbf{A}, \tag{7}$$

$$\mathbf{B} = \boldsymbol{\nabla} \times \mathbf{A}. \tag{8}$$

The corresponding field equations are [2]:

$$\boldsymbol{\nabla} \cdot (\epsilon \mathbf{E}) = 4\pi\rho - \boldsymbol{\nabla}\Theta \cdot \mathbf{B}, \tag{9}$$

$$\boldsymbol{\nabla} \cdot \mathbf{B} = 0, \tag{10}$$

$$\boldsymbol{\nabla} \times \mathbf{E} + \frac{1}{c_0}\partial_t \mathbf{B} = 0, \tag{11}$$

$$c_0\, \boldsymbol{\nabla} \times (\mathbf{B}/\mu) - \partial_t(\epsilon \mathbf{E}) = 4\pi \mathbf{J} + c_0 \boldsymbol{\nabla}\Theta \times \mathbf{E} + \dot{\Theta}\, \mathbf{B}. \tag{12}$$

We consider $\Theta(\mathbf{x}, t)$ to be constant in time ($\dot{\Theta} = 0$) and throughout each medium $\mathcal{M}_i$, with constant and finite discontinuities at the interfaces between them, namely $\boldsymbol{\nabla}\Theta = \tilde{\theta}_i \delta(f_{\Sigma_i}(\mathbf{x}))\hat{\mathbf{n}}_i$, where,

$$\tilde{\theta}_i \equiv \Theta_{i+1} - \Theta_i, \tag{13}$$

and $\Theta_i$ being the value of the TMEP in medium $i$. The interface is defined by $f_{\Sigma_i}(\mathbf{x}) = 0$ and $\hat{\mathbf{n}}_i$ is perpendicular to $\Sigma_i$ going from medium $i$ to medium $i + 1$.

---

[2] Thus written Eqs. (9) and (12) are not specific to TIs. They describe a wider class of magnetoelectric media, sometimes referred as Tellegen media [62]. One might as well forget about the TI altogether and focus on the EM response of general magnetoelectrics. For this reason we write $\theta$ rather than $\theta_{\mathrm{TI}}$, but our calculations do consider the factor $\alpha/\pi$.

We observe that the spatial and temporal variations of $\Theta$ act as source for the EM field. These, do not modify the field equations in the bulk, but imply modification to the usual boundary conditions (BCs). In the absence of charge and current sources, the modified BCs read:

$$\Delta[\mathbf{E}_{\parallel}]|_{\Sigma} = 0, \qquad \Delta[\mathbf{B}_{\perp}]|_{\Sigma} = 0, \tag{14}$$

$$\Delta[\epsilon\mathbf{E}_{\perp}]|_{\Sigma} = -\tilde{\theta}\mathbf{B}_{\perp}|_{\Sigma}, \qquad \Delta[\frac{1}{\mu}\mathbf{B}_{\parallel}]|_{\Sigma} = \tilde{\theta}\mathbf{E}_{\parallel}|_{\Sigma}, \tag{15}$$

where $\Delta[A] \equiv A_{i+1} - A_i$, while $\mathbf{V}_{\perp}$ denotes the component of any vector $\mathbf{V}$ perpendicular to a given $\Sigma$ and $\mathbf{V}_{\parallel}$ denotes its component parallel or tangent to $\Sigma$. The expression $\Delta[\mathbf{V}]|_{\Sigma}$ denotes the discontinuity of a vector across an interface $\Sigma$ and $\mathbf{V}|_{\Sigma}$ is the continuous value of $\mathbf{V}$ evaluated at the interface. Eqs. (14, 15) lead to different solutions both at the interfaces and in the bulk.

We consider an EM waveguide system where we look for monochromatic solutions of Maxwell's equations that propagate in the direction of the guide ($\hat{\mathbf{z}}$-direction) and are confined in the transverse direction. Thus, the electric and magnetic fields are assumed to have the following form:

$$\mathbf{E}(\mathbf{r},t) = \mathbf{E}(\mathbf{r}_{\perp})e^{i(k_z z - wt)}, \qquad \mathbf{B}(\mathbf{r},t) = \mathbf{B}(\mathbf{r}_{\perp})e^{i(k_z z - wt)}, \tag{16}$$

where $k_z$ and $\omega$ are called the wave number and angular frequency. Other works have studied EM waves in contexts related to axion-electrodynamics [63, 64], but without focusing on exact solution for modes in confined scenarios. As is common [65], we decompose vectors in directions longitudinal and transverse to the direction of waveguide's axis: $\mathbf{V} = \mathbf{V}_{\perp} + V_z\hat{\mathbf{z}}$, with $\mathbf{V}_{\perp} \equiv (\hat{\mathbf{z}} \times \mathbf{V}) \times \hat{\mathbf{z}}$ and $V_z \equiv \hat{\mathbf{z}} \cdot \mathbf{V}$. Similar for the position vector, the EM fields, $\boldsymbol{\nabla} = \boldsymbol{\nabla}_{\perp} + \hat{\mathbf{z}}\partial_z$, and $\nabla^2 = \nabla^2_{\perp} + \partial^2_z$. Thus,

$$\mathbf{E}(\mathbf{r},t) = (\mathbf{E}_{\perp}(\mathbf{r}_{\perp}) + \hat{\mathbf{z}}E_z(\mathbf{r}_{\perp}))e^{i(k_z z - wt)}, \qquad \mathbf{B}(\mathbf{r},t) = (\mathbf{B}_{\perp}(\mathbf{r}_{\perp}) + \hat{\mathbf{z}}B_z(\mathbf{r}_{\perp}))e^{i(k_z z - wt)}. \tag{17}$$

Due to the confined propagation, the behavior of guided electromagnetic waves (EMW) is no longer transverse as in free space, because, within the guide, the EMWs have a longitudinal component and the wave number must have cross-sectional contributions, i.e., $\mathbf{k} = \pm\mathbf{k}_{\perp} + k_z\hat{\mathbf{z}}$. Interestingly, the interaction between leftward and rightward waves in the transverse direction leads to a characteristic general solution that manifests as traveling and standing waves across the longitudinal and transverse direction of the waveguide respectively.

Again, as done in ordinary electrodynamics, first by replacing the transverse-longitudinal decomposition in the field equations one can express the transverse components $\mathbf{E}_{\perp}, \mathbf{B}_{\perp}$ in terms of the longitudinal ones $E_z, B_z$ and finally by plugging these back in the $\Theta$-ED field equations, we arrive at:

$$\epsilon(\nabla^2_{\perp} + k^2_{\perp})E_z = \mu E_z(\boldsymbol{\nabla}_{\perp}\Theta)^2 - \boldsymbol{\nabla}_{\perp}\Theta \cdot \boldsymbol{\nabla}_{\perp}B_z \tag{18}$$

$$\frac{1}{\mu}(\nabla^2_{\perp} + k^2_{\perp})B_z = E_z\nabla^2_{\perp}\Theta + \boldsymbol{\nabla}_{\perp}\Theta \cdot \boldsymbol{\nabla}_{\perp}E_z \tag{19}$$

Here,

$$k^2_{\perp} = k^2_0\mu\epsilon - k^2_z, \tag{20}$$

is the dispersion relation, $k^2_{\perp} = k^2_x + k^2_y$ is the cut-off wavenumber, and $k_0 = \omega/c_0 \equiv 2\pi/\lambda_0$ determines the frequency of the EM wave along the waveguide, or the so-called operational frequency. We have taken $\partial_z\Theta = \dot{\Theta} = 0$, i.e. $\Theta$ is constant in time and does not vary in the longitudinal direction.

For stratified media; $\epsilon(\mathbf{r}), \mu(\mathbf{r})$ and $\Theta(\mathbf{r})$ are piecewise constant, i.e., the "topo-optical [3]" parameters are uniform in any media but change from one media to another. Then, the wave propagation can be studied using the known properties of EMW in homogeneous media, together with the appropriate BCs that dictate the laws of reflection and transmission. In this way, in the bulk of a homogeneous and uniform media we have that the transversal components correspond to those of Maxwell's usual electrodynamics,

$$\mathbf{E}_{\perp} = \frac{i}{k^2_{\perp}}(k_z\boldsymbol{\nabla}_{\perp}E_z - k_0\hat{\mathbf{z}} \times \boldsymbol{\nabla}_{\perp}B_z), \qquad \mathbf{B}_{\perp} = \frac{i}{k^2_{\perp}}(k_z\boldsymbol{\nabla}_{\perp}B_z + k_0\epsilon\mu\hat{\mathbf{z}} \times \boldsymbol{\nabla}_{\perp}E_z), \tag{21}$$

[3] We will use the term "topo-optical" parameters to refer collectively to the permittivity $\epsilon$, permeability $\mu$, and topological magnetoelectric polarizability $\Theta$ of a medium.

and the longitudinal components satisfy the Helmholtz equations,

$$(\nabla_\perp^2 + k_\perp^2)B_z = 0, \qquad\qquad (\nabla_\perp^2 + k_\perp^2)E_z = 0. \tag{22}$$

These last equations must be solved subject to the appropriate boundary conditions Eqs. (14,15) for each type of waveguide. Once the fields $E_z$ , $B_z$ are known, the transversal fields $\mathbf{E}_\perp$ , $\mathbf{B}_\perp$, are calculated from the Eq. (21), which results in a complete solution of $\theta$ Maxwell's equations for the waveguide structure.

It can be shown that all components satisfy the Helmholtz equation in a homogeneous media,

$$(\nabla_\perp^2 + k_\perp^2)\begin{Bmatrix}\mathbf{E}\\ \mathbf{B}\end{Bmatrix} = 0. \tag{23}$$

Therefore, as claimed before, the solutions correspond to standing waves in the transverse directions and a traveling wave in the $\hat{\mathbf{z}}$-direction.

### B. The particular setup

We consider a slab waveguide made up of three media stacked along the $\hat{\mathbf{x}}$ direction. All media are infinite in the $\hat{\mathbf{y}}$ and $\hat{\mathbf{z}}$ directions. The central medium ranges from $x = -L$ to $x = L$ and has topo-optical parameters $\epsilon_2, \mu_2, \Theta_2$ and we will refer to it as the TI core. The regions $L < |x|$ have topo-optical parameters $\epsilon_1, \mu_1, \Theta_1$ and $\epsilon_3, \mu_3, \Theta_3$ for $x < -L$ and $x > L$ respectively. We will refer to these regions as claddings. The whole setup is shown in Fig. 1. We will also consider that the propagation is effectively along the $\hat{\mathbf{z}}$ direction, which is equivalent to $\mathbf{k} = \mathbf{k}_\perp + k_z\hat{\mathbf{z}} = k_x\hat{\mathbf{x}} + k_z\hat{\mathbf{z}}$. The geometry of the propagation defines the incidence angle to be $\phi_0 = \arctan(k_z/k_x)$. In the sequel, we will indicate each medium by the index $m = 1, 2, 3$, where $m = 2$ corresponds to the TI core and $m = 1(3)$ to the cladding to its left(right). A ray of light, say in the $y = 0$ plane, making an angle $\phi_0$ with the normal to the walls (planes $x = \pm L$), reflects and bounces between them. Thus, the EMW is guided along the $\hat{\mathbf{z}}$-direction, provided that the incidence angle $\phi_0$ be greater than the greatest between the critical angles at the interfaces in $x = \pm L$, namely $\phi_{1,3}^c = \arcsin(n_{1,3}/n_2)$. We will assume this to be the case. In this system, the $y$-coordinate does not play a role in the equations, that is, $\partial_y = 0$ and the longitudinal components of the field satisfy the 1D-Helmholtz equation with the associated cutoff wavenumber $k_{xm}$ in each medium,

$$(\partial_x^2 + k_{xm}^2)B_z = 0, \qquad\qquad (\partial_x^2 + k_{xm}^2)E_z = 0. \tag{24}$$

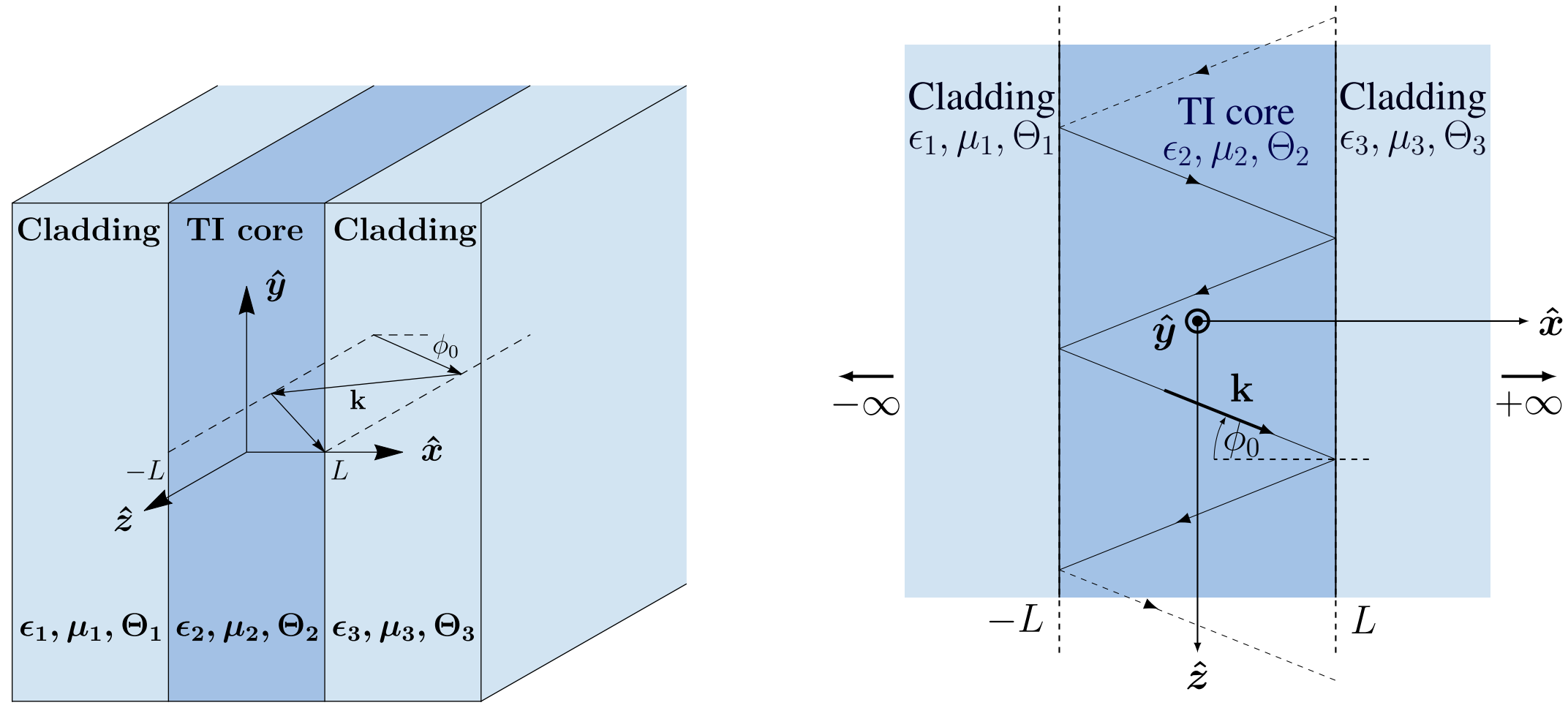

FIG. 1: Propagation of an EMW of wave vector $\mathbf{k} = \pm\mathbf{k}_\perp + k_z\hat{\mathbf{z}}$ in the slab waveguide. Only a portion of the media comprising the slab is shown. Due to the BCs at the core-cladding interfaces, the wave is confined in the core medium. Left: 3D-view of the wave propagating. The wave comes from $z = -\infty$ to $z = +\infty$.The angle of incidence $\phi_0$ of the EMW at each interface is measured with respect to the normal to the interface. Right: Top-view. All media extend uniformly in the $\pm z$ and $\pm y$ directions. The claddings may extend to infinity in the $\pm x$ directions.

The solutions correspond to sums of exponentials $\sim e^{\pm ik_{xm}x}$. To find solutions confined in the core of the slab we must impose that the EMW decays rapidly away from the core-cladding interface, this is achieved by considering solutions with an imaginary/real cutoff wavenumber at the cladding/core, i.e.,

$$k_{x1} = i\alpha_1, \qquad k_{x2} = \gamma, \qquad k_{x3} = i\alpha_3, \tag{25}$$

where $\{\alpha_1, \gamma, \alpha_3\} \in \mathbb{R}^+$. As a result, the dispersion relations in Eq. (20) in each media are related as follows,

$$\left.\begin{aligned} \alpha_1^2 &= k_z^2 - k_0^2 n_1^2, \\ \gamma^2 &= k_0^2 n_2^2 - k_z^2, \\ \alpha_3^2 &= k_z^2 - k_0^2 n_3^2, \end{aligned}\right\} \Rightarrow \begin{aligned} \gamma^2 + \alpha_1^2 &= k_0^2(n_2^2 - n_1^2), \\ \gamma^2 + \alpha_3^2 &= k_0^2(n_2^2 - n_3^2) = k_0^2(n_2^2 - n_1^2)(1-\delta), \\ \alpha_1^2 - \alpha_3^2 &= k_0^2(n_3^2 - n_1^2) = k_0^2(n_2^2 - n_1^2)\delta, \end{aligned} \tag{26}$$

where $n_m = \sqrt{\epsilon_m \mu_m}$ is the refractive index in each media and we defined the asymmetry parameter $\delta$,

$$\delta = \frac{n_3^2 - n_1^2}{n_2^2 - n_1^2}. \tag{27}$$

Note that $\delta \geq 0$ since we assumed $n_2 > n_3 \geq n_1$ to ensure total internal reflection of electromagnetic waves propagating through the core slab.

## III. EXACT SOLUTIONS

### A. General solutions for longitudinal components for the specific Θ boundary conditions

The interaction between leftward and rightward waves in the $x$-direction of core slab produces known patterns with defined parity which we will adopt here. In summary we have,

$$E_{z2}(x) = E_{2+}e^{i\gamma x} + E_{2-}e^{-i\gamma x} = E_e \sin\gamma x + E_o \cos\gamma x, \tag{28}$$

$$B_{z2}(x) = B_{2+}e^{i\gamma x} + B_{2-}e^{-i\gamma x} = B_e \sin\gamma x + B_o \cos\gamma x, \tag{29}$$

where, $C_e \equiv i(C_+ - C_-)$ and $C_o \equiv C_+ + C_-$ are the even and odd solutions of the EMW respectively. The even or odd refer to the parity of the solution of the transverse electric field, which depends on spatial derivatives of the longitudinal components. The longitudinal components of the field in each medium are as follows,

$$E_z(x) = \begin{cases} E_1 e^{\alpha_1 x} & \text{for} \quad x \leq -L, \\ E_e \sin\gamma x + E_o \cos\gamma x & \text{for} \quad -L < x < L, \\ E_3 e^{-\alpha_3 x} & \text{for} \quad x \geq L, \end{cases} \qquad B_z(x) = \begin{cases} B_1 e^{\alpha_1 x} & \text{for} \quad x \leq -L, \\ B_e \sin\gamma x + B_o \cos\gamma x & \text{for} \quad -L < x < L, \\ B_3 e^{-\alpha_3 x} & \text{for} \quad x \geq L, \end{cases} \tag{30}$$

where we have eliminated solutions that diverge in $x = \pm\infty$. We can see from Eq. (30) there are a total of 12 unknowns to determine, these are $E_1$, $E_e$, $E_o$, $E_3$, $B_1$, $B_e$, $B_o$, $B_3$, $\alpha_1$, $\gamma$, $\alpha_3$ and $k_z$. To determine the unknowns we apply all the BCs,

$$\Delta[E_z]|_{x=\pm L} = 0, \qquad \Delta[\frac{1}{\mu}B_z]|_{x=\pm L}, = (\tilde{\theta}_m E_z)|_{x=\pm L}, \tag{31}$$

$$\Delta[E_y]|_{x=\pm L} = 0, \qquad \Delta[\frac{1}{\mu}B_y]|_{x=\pm L} = (\tilde{\theta}_m E_y)|_{x=\pm L}, \tag{32}$$

$$\Delta[\epsilon E_x]|_{x=\pm L} = -(\tilde{\theta}_m B_x)|_{x=\pm L}, \qquad \Delta B_x|_{x=\pm L} = 0. \tag{33}$$

It can be shown that the BCs in Eqs. (32) and (33) are actually the same, this reduces to 8 BCs. We are still missing four equations to completely determine the system. The three equations in Eqs. (26) allow us to relate the unknowns $\alpha_1$, $\gamma$ and $\alpha_3$. In this way, we can leave the system as a function of a single amplitude of the wave that we can define by normalizing the modes allowed in the waveguide.

Let us first consider the 4 BCs of the longitudinal components of the fields, Eq. (31), thus finding the amplitude of the media 1 and 3,

$$E_1 = e^v[E_o \cos u - E_e \sin u], \qquad B_1 = e^v \frac{\mu_1}{\mu_2}[(B_o - E_o\mu_2\tilde{\theta}_1)\cos u - (B_e - E_e\mu_2\tilde{\theta}_1)\sin u], \tag{34}$$

$$E_3 = e^w[E_o \cos u + E_e \sin u], \qquad B_3 = e^w \frac{\mu_3}{\mu_2}[(B_o + E_o\mu_2\tilde{\theta}_2)\cos u + (B_e + E_e\mu_2\tilde{\theta}_2)\sin u], \tag{35}$$

where we are introducing the dimensionless wavenumbers:

$$v \equiv \alpha_1 L, \qquad u \equiv \gamma L, \qquad \text{and} \qquad w \equiv \alpha_3 L. \tag{36}$$

Applying the remaining 4 BCs, Eq. (32) or (33), we can write a linear system of equations in matrix form,

$$\mathbb{M} \cdot \mathbf{s} = \mathbf{0}, \qquad \text{where,} \qquad \mathbf{s} = \begin{pmatrix} E_e & E_o & B_e & B_o \end{pmatrix}^t, \tag{37}$$

and,

$$\mathbb{M} = \begin{pmatrix} \epsilon_3 u \sin u - \epsilon_2 w \cos u & \epsilon_3 u \cos u + \epsilon_2 w \sin u & \tilde{\theta}_2 w \cos u & -\tilde{\theta}_2 w \sin u \\ \epsilon_2 v \cos u - \epsilon_1 u \sin u & \epsilon_1 u \cos u + \epsilon_2 v \sin u & \tilde{\theta}_1 v \cos u & \tilde{\theta}_1 v \sin u \\ \mu_2 \mu_3 \tilde{\theta}_2 u \sin u & \mu_2 \mu_3 \tilde{\theta}_2 u \cos u & \mu_3 u \sin u - \mu_2 w \cos u & \mu_2 w \sin u + \mu_3 u \cos u \\ \mu_2 \mu_1 \tilde{\theta}_1 u \sin u & -\mu_2 \mu_1 \tilde{\theta}_1 u \cos u & \mu_2 v \cos u - \mu_1 u \sin u & \mu_2 v \sin u + \mu_1 u \cos u \end{pmatrix}. \tag{38}$$

To find a non-trivial solution for the vector of coefficients s, the matrix must be such that $\det(\mathbb{M}) = 0$. This condition leads to the characteristic equation or propagation condition, which establishes a relationship between $v$, $u$, and $w$ for certain specified topo-optical parameters. The characteristic equation, together with the three Eqs. (26), determines the complete set of propagation constants $v$, $u$, $w$, and $k_z$ necessary for the EM field to propagate properly along the waveguide, fulfilling all the BCs in each instance of total internal reflection.

### B. Choices of claddings: symmetric and asymmetric Θ-slab waveguide case

In ordinary electrodynamics, slab waveguides are usually distinguished in two categories. A symmetric slab corresponds to stratified media where the media 1 and 3 (the claddings) have the same optical properties, i.e., $\mu_3 = \mu_1$ and $\epsilon_3 = \epsilon_1$, which implies $\alpha_3 = \alpha_1 \equiv \alpha$ or $w = v$. On the other hand, an asymmetric slab is one in which the claddings differ. Solving the characteristic equation det$(\mathbb{M}) = 0$ analytically for arbitrary $\epsilon, \mu$ and $\Theta$ in each media to then find the exact modes, mode couplings, dispersion relations, etc. is quite complex.

#### *1. Asymmetric slab waveguide with topologically trivial claddings*

Without attempting a full discussion of all possible combinations of the topo-optical parameters, we can provide some quantitative and qualitative insights regarding the effects of considering an asymmetric waveguide, despite the aforementioned complexity. To illustrate the effects of the asymmetry of the slab, we go back to the effective dispersion relations in each media, Eqs. (26) and the definition of the asymmetry parameter $\delta$ of Eq. (27). We begin by representing the characteristic equation using the so-called normalized waveguide index $b$, and the normalized waveguide thickness $\mathcal{R}$:

$$b \equiv \frac{v^2}{\mathcal{R}^2} = \frac{k_z^2 - k_0^2 n_1^2}{k_0^2 (n_2^2 - n_1^2)}, \tag{39}$$

where $\mathcal{R} \equiv L k_0 \sqrt{n_2^2 - n_1^2}$ is dimensionless and is related to the operational frequency of the waveguide. Writing the dispersion relations of Eq. (26) in terms of these, we observe that $u = \mathcal{R}\sqrt{1-b}$, $v = \mathcal{R}\sqrt{b}$ and $w = \mathcal{R}\sqrt{b+\delta}$, *i.e.*, the characteristic equation can now be represented in terms of three parameters ($b, \mathcal{R}$ and $\delta$) instead of the previous five ($n_1$, $n_2$, $n_3$, $L$ and $k_0$).

In Fig. (2), we show the characteristic equation in terms of these normalized parameters for the case of the slab whose core is a TI, while the claddings are topologically trivial, *i.e.*, $\theta_1 = \theta_3 = 0$ y $\theta_2 \neq 0$; non magnetic, *i.e.*, $\mu_1 = \mu_2 = \mu_3 = 1$; medium 3 being empty space *i.e.*, $\epsilon_3 = 1$; medium 2 such that $\epsilon_2 = 16$, while the permittivity of medium 1 is allowed to vary. In so doing we are exploring the effects of the asymmetry $\delta$ on the propagation conditions. In particular, for the plots we vary $\epsilon_1$ in such a way to have $\delta = \{0, 1, 10, 100\}$ in (a), (b), (c) y (d) respectively. Furthermore, each plot also shows the propagation modes for different values of the topological parameter of the TI core: $\theta_2 = (0, 1, 2, 3)$. These values for $\theta_2$ are related to what we later call $\theta_{\text{TI}}$ via $\theta_{\text{TI}} = \theta_2\ \pi/\alpha$.

Note that by choosing a specific operational frequency and the optical parameters of the slab, the normalized thickness $\mathcal{R}$ is determined. In fig. (2) we see that for that $\mathcal{R}$, several values of $b$ satisfy the propagation condition, corresponding the different solutions of the EM field, which we define as the different propagation modes. In each figure the modes are labeled by $n$ and we see that the larger the frequency, the larger the amount of modes supported by in the slab.

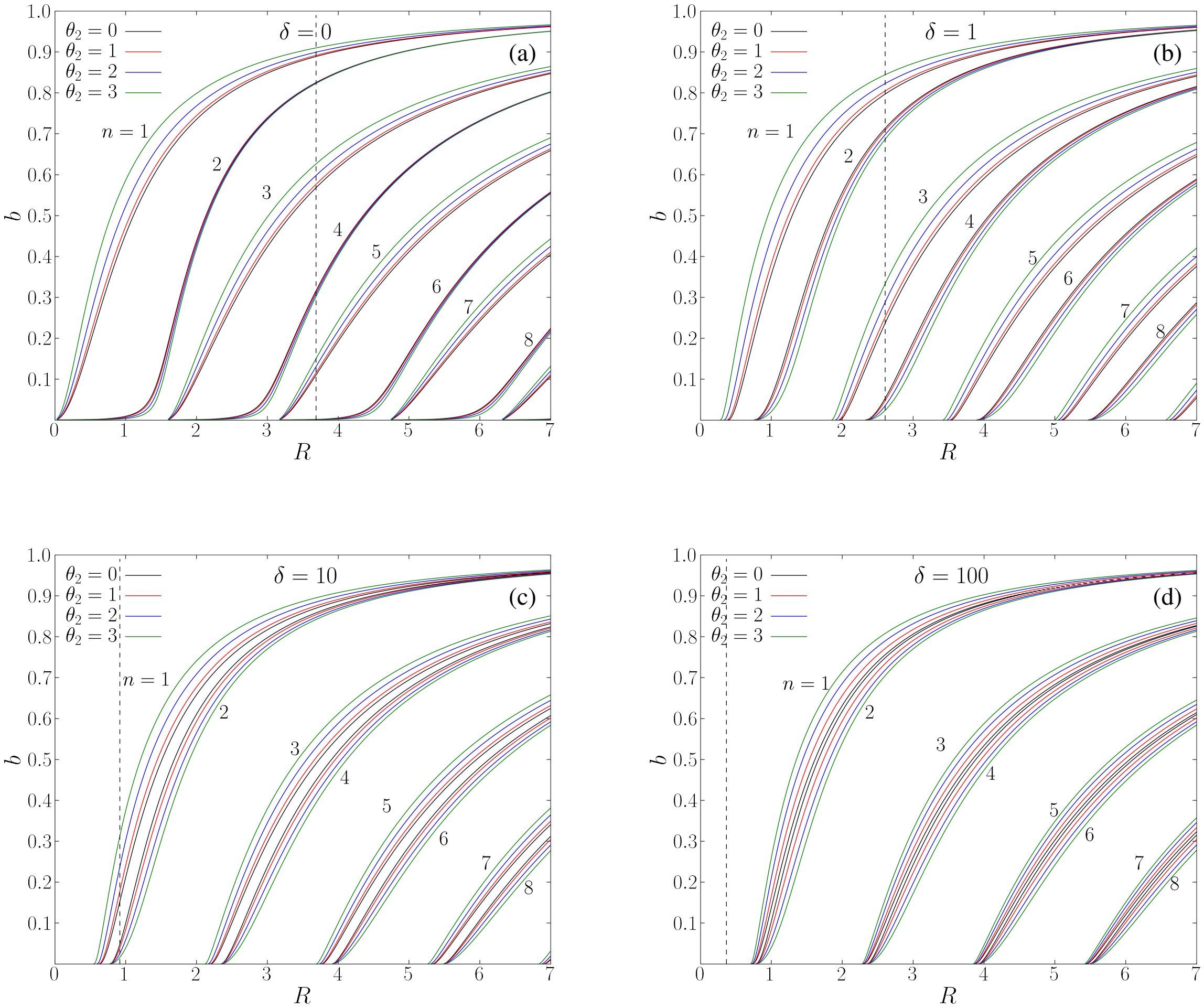

FIG. 2: The characteristic equation $\det(\mathbb{M}) = 0$ as a function of $\mathcal{R}$ and $b$ for a slab with topologically trivial claddings and a TI core. The effects are shown for different values of $\theta_2$ of the top-left panel. In all the plots we fix $\mu_1 = \mu_2 = \mu_3 = 1$ and $\epsilon_3 = 1$, while $\epsilon_2 = 16$. Mild and feasible permittivities of the right-most cladding, (a) $\epsilon_1 = 1$, (b) $\epsilon_1 = 8.5$, (c) $\epsilon_1 = 14.64$, y (d) $\epsilon_1 = 15.85$, allows to explore a wide range of values for the asymmetry parameter $\delta = 0, 1, 10, 100$ shown at the top-center of each plot. For all plots we choose an operational wavelength $\lambda_0 = 3.3(2L)$ that corresponds to the particular value of $\mathcal{R} = \pi\sqrt{16 - \epsilon_1}/3.3$, shown by the dashed-vertical line. This particular value of $\mathcal{R}$ define the allowed values of $b$ for the wave propagation along the slab. The labels $n$ indicate the propagation modes. Therefore, for a given $\theta$, the amount of allowed propagating modes is determined by the number of branches that are intersected by the dashed-vertical line. For example, for the given optical parameters and operational frequency, we observe that for a $\delta = 100$ asymmetric slab, no propagation modes are allowed in the waveguide.

In regards with the asymmetry of the waveguide, we observe that as $\delta \to 0$, only the odd modes show significant variation for the different values of $\theta$, or said in other way, for $\delta \to 0$ the even modes seem to be insensitive to the topological magnetoelectric response of the TI core. However, the more asymmetric the waveguide the more the even modes get modified by $\theta$. Interestingly, we note that for large $\delta$ the modes 1 and 2 (3 and 4, 5 and 6, ...) seem to coalesce smearing the clear distinction between them. Also, for all $\delta$ and $\theta_2$, we see that $b_{\text{n=odd}}(\theta_2) > b_{\text{n=odd}}(0)$ and $b_{\text{n=even}}(\theta_2) < b_{\text{n=even}}(0)$. It can be shown that this $b$ parameter is related to the inverse penetration length into the claddings, implying that the odd modes are more confined and uniform than the even modes when the slab's core is a TI. As can be seen, allowing more generality in the topo-optical parameter space of the slab waveguide gives rise for wider range of possibilities. This could be relevant for practical purposes at the moment of functionalizing a topological slab waveguides to achieve a specific response. One could also explore to possibility of considering the role of the permeabilitties $\mu$. In fact in [66], the inclusion of $\mu$-metals in the EM response of TIs has been considered, though not for propagation in slab waveguides. The analysis above show some qualitative and quantitative effects introduced by the asymmetry of the waveguide. For the goals of this work, keeping the asymmetry parameter in general till the very end is not conceptually challenging, but computationally costly.

### 2. *Symmetric slab waveguide of arbitrary* $\tilde{\theta}_i$

As mentioned previously, with the goal of obtaining exact mode description, mode coupling, dispersion relations, etc. we choose the optical parameters of the claddings in a restricted sense, such that the both claddings are identical and topologically trivial, but allowing for arbitrary values of $\Theta$. This choice introduces a further simplification that, though lacking full generality, still allows to explore in depth the effects of $\Theta$ on the propagation of wave along the slab.

First, we note that Eqs. (26) reduce to,

$$\left.\begin{aligned} \alpha^2 &= k_z^2 - k_0^2 n_1^2 \\ \gamma^2 &= k_0^2 n_2^2 - k_z^2 \end{aligned}\right\} \Rightarrow u^2 + v^2 = L^2 k_0^2 (n_2^2 - n_1^2) \equiv \mathcal{R}^2. \tag{40}$$

Observe that, given Eq. (30), the quantity $1/\alpha$ is a length, measured from the cladding-core interface and into the claddings, beyond which the longitudinal field amplitudes decay to $1/e$ its value at the interface. Hence, it is a penetration length into the claddings. For further utility, we introduce the concept of "dimensionless penetration length", $\ell_x = 1/\alpha L$. Replacing the simplifications of the symmetric case we can rewrite the matrix in Eq. (38) and the determinant is,

$$\det(\mathbb{M}) = v^4 + v^3 u \zeta_1 (\cot u - \tan u) + v^2 u^2 (\zeta_2 \cot^2 (2u) - \zeta_4) - v u^3 \zeta_1 \zeta_3 (\cot u - \tan u) + u^4 \zeta_3^2 = 0 \tag{41}$$

where,

$$\zeta_1 = \frac{\mu_1}{2\epsilon_2}\left(\tilde{\theta}_1^2 + \tilde{\theta}_2^2\right) + \frac{\mu_1}{\mu_2} + \frac{\epsilon_1}{\epsilon_2}, \qquad \zeta_2 = \frac{\mu_1}{\mu_2}\left(\frac{\mu_1}{\epsilon_2}\left(\tilde{\theta}_1 + \tilde{\theta}_2\right)^2 + 4\frac{\epsilon_1}{\epsilon_2}\right) \tag{42}$$

$$\zeta_3 = \frac{\mu_1 \epsilon_1}{\mu_2 \epsilon_2}, \qquad \zeta_4 = \frac{\mu_1^2}{\mu_2^2 \epsilon_2^2}\left(\epsilon_2 - \mu_2 \tilde{\theta}_1 \tilde{\theta}_2\right)^2 + \frac{\mu_1 \epsilon_1}{\epsilon_2^2}\left(\tilde{\theta}_1^2 + \tilde{\theta}_2^2\right) + \frac{\epsilon_1^2}{\epsilon_2^2}. \tag{43}$$

The explicit solutions of this polynomial of degree 4 are complicated. Although one can solve for $v$ as a function of $u$, it is laborious to do so. Observe that, so far, the difference in $\Theta$ at each interface is arbitrary, allowing for different values of $\Theta$ and different signs for their differences. To further simplify the system, we will consider the case in which the gradients of $\Theta$ at both interfaces have the same magnitude, but point in opposite directions. With this simplified choice of topo-optical parameters, in the next subsection we will derive several results that will be of importance for the rest od the manuscript, therefore we devote special attention to the following subsection.

## C. Symmetric and antiparallel Θ-slab waveguide case

Let us consider the antiparallel case where $\tilde{\theta}_1 = -\tilde{\theta}_2 \equiv \theta$, which also implies that we are considering the cladding media as topologically trivial ($\Theta_1 = \Theta_3 = 0$). Replacing in Eq. (38) the determinant is as follows,

$$\det(\mathbb{M}) = (v - \xi_+^\theta u \tan u)(v - \xi_-^\theta u \tan u)(v + \xi_+^\theta u \cot u)(v + \xi_-^\theta u \cot u) = 0, \tag{44}$$

with,

$$\xi_\pm^\theta = \frac{1}{2}\left(\xi_- + \xi_+ + \xi_0 \theta^2 \pm \sqrt{(\xi_- + \xi_+ + \xi_0 \theta^2)^2 - 4\xi_+ \xi_-}\right), \tag{45}$$

where,

$$\xi_+ = \frac{\mu_1}{\mu_2}, \qquad \xi_- = \frac{\epsilon_1}{\epsilon_2}, \qquad \text{and,} \qquad \xi_0 = \frac{\mu_1}{\epsilon_2}. \tag{46}$$

In Eq. (44), suffice for any term in parentheses to vanish in order to satisfy the propagation condition and, it turns out that, only one of them can vanish at the same time. For given topo-optical parameters $\epsilon_1, \epsilon_2, \mu_1, \mu_2, \Theta_2$, the vanishing of each individual term independently leads to different analytic equations for $v$, whose $\theta$-dependence we make explicit from now on: $v = v_p^\theta(u_p^\theta)$, for $p = 1, 2, 3, 4$ that satisfy the propagation condition. Each couple $(u_p^\theta, v_p^\theta)$ that satisfies the vanishing of any of the terms $v = \xi_\pm^\theta u \tan u$ or $v = -\xi_\pm^\theta u \cot u$ satisfies the propagation condition. Since the functions $\tan u$ and $\cot u$ have many branches, depending on the value of the frequency that is given by $\mathcal{R} = k_0 L\sqrt{n_2^2 - n_1^2}$, there may be several possible solution pairs $(u_p^\theta, v_p^\theta)$ for a given $\mathcal{R}$. Those specific $(u_p^\theta, v_p^\theta)$ such that $\mathcal{R}^2 = (u_p^\theta)^2 + (v_p^\theta)^2$ define the propagation modes that determine the behavior and polarization of the EMW in the waveguide. Further on we will elaborate on the property of the polarization of each solution. However, the modes on repeated branches (those related by the periodicity of the trigonometric functions) are not identical, as will be discussed below. The periodicity of the branches can be understood by looking at the solutions $u_0$ of Eq. (44) for $v = 0$. This produces two families of branches: one associated with $\tan u_0 = 0$, yielding $u_0 = (n_\mathrm{e})\pi/2$ and the other corresponding to $\cot u_0 = 0$ with $u_0 = (n_\mathrm{o})\pi/2$, with $n_\mathrm{e}(n_\mathrm{o})$ being even (odd) integers, respectively. We order the branches in increasing values of $n$ with $n_\mathrm{e} = 0, 2, 4, \ldots$ and $n_\mathrm{o} = 1, 3, 5, \ldots$. As each branch requires separate solutions of the Eq. (44) involving the two values $\xi_\pm^\theta$, one has a two-fold multiplicity for each tan or cot branch. This results in an alternating pattern between two tan even and two cot odd branches.

In Fig. 3, we identify the first four branches as $p = 1, 2, 3, 4$, corresponding to the pairs $(u_p^\theta, v_p^\theta)$ that cancel each term of the characteristic equation (44) and the intersection points with $\mathcal{R}^2 = (u_p^\theta)^2 + (v_p^\theta)^2$ show the first four modes.

Before describing each mode, we note that,

$$\xi_\pm^\theta \approx \xi_\pm \left(1 \pm \frac{\xi_0}{(\xi_+ - \xi_-)}\theta^2 + \mathcal{O}(\theta^4)\right), \qquad \text{i.e.,} \qquad \lim_{\theta\to 0} \xi_\pm^\theta = \xi_\pm, \tag{47}$$

which means that the topological magnetoelectric polarizability modifies the propagating modes of the usual electromagnetism and that for small $\theta$ and they are fully recovered in the $\theta \to 0$ limit.

Now, as mentioned above that $v = v(u)$ both have explicit $\theta$-dependence, so do the wavenumbers, $k_x$ and $k_z$. Recall that via Eq. (36) the dimensionless wavenumbers: $u, v = w$ (for the symmetric case under consideration), are the $\alpha$ and $\gamma$, which, modulo the $i$'s depending whether we are in the interior slab region or the outer cladding regions, are $k_x$. Together with Eq. (40) leads to the conclusion that the wavenumbers $k_x$ and $k_z$ too, have $\theta$-dependence. So as not the clutter the notation we have not introduced an explicit $\theta$ index for the wave numbers, but defer this for later on, where it will help clarify and important observation.

The fact that $(v_p^\theta)^2 + (u_p^\theta)^2 = \mathcal{R}^2$ , where $\mathcal{R} = k_0 L\sqrt{n_2^2 - n_1^2}$ requires considering a certain operating angular frequency $\omega = 2\pi\nu_0$ or vacuum wavelength $k_0 = 2\pi/\lambda_0$. In Fig.3, the branches' solutions and circle equation are shown for $\mu_1 = \mu_2 = 1$, $\epsilon_1 = 12$, $\epsilon_2 = 16$ and $\lambda_0 = 3.3 \times 2L$. At this frequency there are four guided modes - the lowest order TE and TM, odd and even modes (described below).

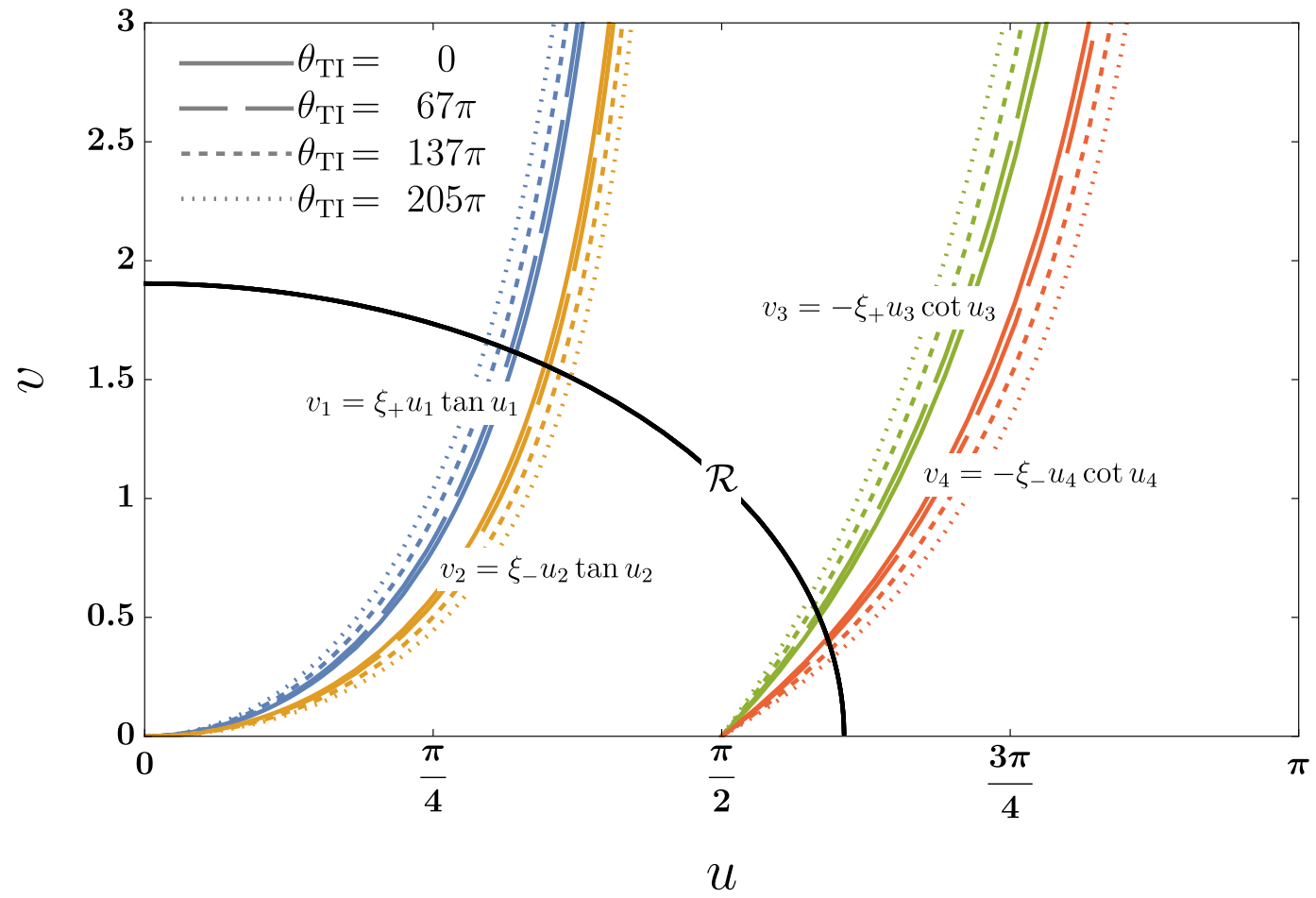

FIG. 3: Solutions to the characteristic equation (44) for the symmetric case ($\delta = 0$) and the antiparallel case with different values of $\Theta = \alpha\theta_{\mathrm{TI}}/\pi$. The values used are $\mu_1 = \mu_2 = 1$, $\epsilon_1 = \epsilon_3 = 12$, $\epsilon_2 = 16$, and $\lambda_0 = 3.3 \times 2L$. The first four propagation modes in the waveguide are shown for the operating frequency $k_0 = 2\pi/\lambda_0$. Each branch is represented by a different color: $p = 1$ in blue, $p = 2$ in yellow, $p = 3$ in green, and $p = 4$ in red. The black curve represents the dispersion relation $(v_p^\theta)^2 + (u_p^\theta)^2 = \mathcal{R}^2$, which allows us to find the pairs $(u_p^\theta, v_p^\theta)$ that satisfy both the characteristic equation and the dispersion relation simultaneously, hence the intersection points represent the first four propagating modes.

The intersection points $(u_p^\theta, v_p^\theta)$ in Fig. 3, are detailed in Table I, along with the allowed angles $\phi_0$ in degrees, the wavelength along the waveguide $\lambda_z = 2\pi/k_z$, the transverse wavelength $\lambda_x = 2\pi/\gamma$, and the dimensionless penetration length $\ell_x = 1/\alpha L$. For values of $Bi_2Se_3$, where $\epsilon_2 = 16$ and $\theta_{\mathrm{TI}} = \pi$ [44], the changes in the propagation constants are on the order of $0.01\%$.

| **Mode** $p$ | $\theta_{\text{TI}}$ | $(u_p, v_p)$ | $k_z L$ | $\phi_0$ [deg] | $\lambda_z/L$ | $\lambda_x/L$ | $\ell_x$ |
|---|---|---|---|---|---|---|---|
| 1 | $0$ | $(1.011, 1.613)$ | $3.671$ | $74.603$ | $1.711$ | $6.215$ | $0.62$ |
| | $11\pi$ | $(1.011, 1.614)$ | $3.671$ | $74.610$ | $1.711$ | $6.217$ | $0.62$ |
| | $137\pi$ | $(0.966, 1.641)$ | $3.683$ | $75.302$ | $1.706$ | $6.503$ | $0.61$ |
| 2 | $0$ | $(1.089, 1.562)$ | $3.649$ | $73.383$ | $1.722$ | $5.77$ | $0.64$ |
| | $11\pi$ | $(1.089, 1.562)$ | $3.649$ | $73.376$ | $1.722$ | $5.768$ | $0.64$ |
| | $137\pi$ | $(1.13, 1.532)$ | $3.636$ | $72.735$ | $1.728$ | $5.559$ | $0.653$ |
| 3 | $0$ | $(1.837, 0.501)$ | $3.336$ | $61.158$ | $1.884$ | $3.42$ | $1.997$ |
| | $11\pi$ | $(1.837, 0.501)$ | $3.336$ | $61.160$ | $1.884$ | $3.421$ | $1.995$ |
| | $137\pi$ | $(1.823, 0.55)$ | $3.343$ | $61.403$ | $1.879$ | $3.447$ | $1.817$ |
| 4 | $0$ | $(1.859, 0.413)$ | $3.324$ | $60.784$ | $1.891$ | $3.38$ | $2.422$ |
| | $11\pi$ | $(1.859, 0.412)$ | $3.323$ | $60.782$ | $1.891$ | $3.38$ | $2.425$ |
| | $137\pi$ | $(1.868, 0.366)$ | $3.318$ | $60, 616$ | $1.894$ | $3.363$ | $2.73$ |

TABLE I: The propagation constants of the first four modes are determined by the intersections of the branches of the characteristic equation and the circular equation of transverse wavenumbers, as shown in Fig. 3. We have included the numerical values of the propagation constants for $\theta_{\text{TI}} = 11\pi$, as these values will be used later to classify the propagation modes.

For each solution of $v_n^\theta$ of Eq. (44) we must replace in Eq. (37) and solve the system. If we do this in the 0-ED, i.e., $\Theta_i = 0$, we will notice that 3 of the 4 amplitudes of the vector $\mathbf{s} = (E_e \, E_o \, B_e \, B_o)^t$ vanish and, as can be seen from Eq. (30), the mode will be a purely TE or TM mode, even or odd. In summary, for the 0-ED we have,

| **Propagation Condition satisfied** | **Null Amplitudes** | **Non-Zero Amplitudes** | **Mode Classification** |
|---|---|---|---|
| $v_1 = \xi_+ u_1 \tan u_1$ | $E_{e1} = E_{o1} = B_{o1} = 0$ | $B_{e1} \neq 0$ | TE-even-1 |
| $v_2 = \xi_- u_2 \tan u_2$ | $E_{o2} = B_{e2} = B_{o2} = 0$ | $E_{e2} \neq 0$ | TM-even-1 |
| $v_3 = -\xi_+ u_2 \cot u_3$ | $E_{e3} = E_{o3} = B_{e3} = 0$ | $B_{o3} \neq 0$ | TE-odd-1 |
| $v_4 = -\xi_- u_4 \cot u_4$ | $E_{e4} = B_{e4} = B_{o4} = 0$ | $E_{o4} \neq 0$ | TM-odd-1 |

TABLE II: EM field amplitudes and TE/TM mode classification of 0-ED modes, according to the satisfied propagation condition. Recall that the parity of the modes is determined by the spatial derivatives of the longitudinal components. The index 1 in the mode classification denotes the first four modes labeled by $p = 1, 2, 3, 4$.

In this classification, each propagation condition defines a specific type of mode. For example, TE-even-1 corresponds to the first mode where the EM field is transverse electric (TE), and the transverse electric field is even. Although this classification repeats in cycles of four, it is important to note that each cycle represents different modes; that is, TE-even-1 $\neq$ TE-even-2. The classification depends on the non-zero amplitude resulting from applying the BCs. In summary, in the 0-ED, the EM wave propagates with a defined polarization, TE or TM, maintaining the profile of the transverse components of the EM field, given by Eq. (21), along the entire guide.

Applying the same analysis to the Θ-ED, we find that only two of the four amplitudes vanish. Observe that the vanishing amplitudes are of the same parity, either both odd or both even. This has two consequences. The resulting mode still has a definite parity and furthermore, both longitudinal components pick additional Θ contributions that render the modes neither totally TE nor TM. The results are summarized in Table III, highlighting the modes obtained for the Θ-ED:

| **Propagation Condition that satisfies** | **Null Amplitudes** | **Non-Zero Amplitudes** | **Hybrid Mode Classification** |
|---|---|---|---|
| $v_1^\theta = \xi_+^\theta u_1^\theta \tan u_1^\theta$ | $E_{o1}^\theta = B_{o1}^\theta = 0$ | $E_{e1}^\theta = \frac{1}{\mathcal{F}_+} B_{e1}^\theta, \quad B_{e1}^\theta \neq 0$ | QTE-even-1 |
| $v_2^\theta = \xi_-^\theta u_2^\theta \tan u_2^\theta$ | $E_{o2}^\theta = B_{o2}^\theta = 0$ | $B_{e2}^\theta = \mathcal{F}_- E_{e2}^\theta, \quad E_{e2}^\theta \neq 0$ | QTM-even-1 |
| $v_3^\theta = -\xi_+^\theta u_3^\theta \cot u_3^\theta$ | $E_{e3}^\theta = B_{e3}^\theta = 0$ | $E_{o3}^\theta = \frac{1}{\mathcal{F}_+} B_{o3}^\theta, \quad B_{o3}^\theta \neq 0$ | QTE-odd-1 |
| $v_4^\theta = -\xi_-^\theta u_4^\theta \cot u_4^\theta$ | $E_{e4}^\theta = B_{e4}^\theta = 0$ | $B_{o4}^\theta = \mathcal{F}_- E_{o4}^\theta, \quad E_{o4}^\theta \neq 0$ | QTM-odd-1 |

TABLE III: EM field amplitudes and hybrid mode classification of Θ-ED modes, according to the satisfied propagation condition.

where,

$$\mathcal{F}_\pm = \frac{\mu_1 \theta}{\xi_+ - \xi_\pm^\theta}. \tag{48}$$

In the Θ-ED, we identify four different propagation modes, all of which are hybrid in the sense that a longitudinal electric amplitude is always accompanied by a longitudinal magnetic amplitude, and vice versa. Since the value of $\theta$ is small compared to the usual optical parameters, these modes represent slight perturbations of the 0-ED modes, indicating an almost transverse polarization. Hence, for example, what was the TE-even-1 mode of 0-ED, now becomes hybridized and given it fails to be totally TE, it is now called a Quasi-TE-even (QTE) hybrid mode. The reason to call it Quasi-TE is because the $\theta$-induced coefficient governing the failure to have a totally transverse polarization is small, in fact,

$$\mathcal{F}_- \approx \frac{\mu_1 \theta}{(\xi_+ - \xi_-)} + \mathcal{O}(\theta^3), \qquad \lim_{\theta \to 0} \frac{1}{\mathcal{F}_+} = 0, \qquad \text{and} \qquad \lim_{\theta \to 0} \mathcal{F}_- = 0, \tag{49}$$

recovering the modes of usual ED. Therefore, hybridization is caused only by the presence of a non-zero Θ in the TIs.

Moreover, the field amplitudes with the same parity are coupled; for example, an even TE polarization induces an even TM polarization, and vice versa. The same coupling occurs for odd polarizations. We can express the induced electric and magnetic amplitudes solely in terms of $\mathcal{F}_+$ or $\mathcal{F}_-$, since

$$\mathcal{F}_+ \mathcal{F}_- = -\mu_2 \epsilon_2. \tag{50}$$

In summary, EM fields propagate in four distinct modes with different behaviors. Table IV presents the longitudinal fields for each of these four modes:

| Hybrid Mode | $-iE_z^\theta(x)$ | $-iB_z^\theta(x)$ | Region |
|---|---|---|---|
| QTE-even-1 | $-B_{e1}^\theta \frac{\sin u_1^\theta}{\mathcal{F}_+} e^{(x/L+1)v_1^\theta}$ | $-B_{e1}^\theta \xi_+^\theta \sin u_1^\theta e^{(x/L+1)v_1^\theta}$ | $x<-L$ |
| | $B_{e1}^\theta \frac{1}{\mathcal{F}_+} \sin(u_1^\theta x/L)$ | $B_{e1}^\theta \sin(u_1^\theta x/L)$ | $-L \leq x \leq L$ |
| | $B_{e1}^\theta \frac{\sin u_1^\theta}{\mathcal{F}_+} e^{(-x/L+1)v_1^\theta}$ | $B_{e1}^\theta \xi_+^\theta \sin u_1^\theta e^{(-x/L+1)v_1^\theta}$ | $L<x$ |
| QTM-even-1 | $-E_{e2}^\theta \sin u_2^\theta e^{(x/L+1)v_2^\theta}$ | $-E_{e2}^\theta \xi_-^\theta \mathcal{F}_- \sin u_2 e^{(x/L+1)v_2^\theta}$ | $x<-L$ |
| | $E_{e2}^\theta \sin(u_2^\theta x/L)$ | $E_{e2}^\theta \mathcal{F}_- \sin(u_2^\theta x/L)$ | $-L \leq x \leq L$ |
| | $E_{e2}^\theta \sin u_2^\theta e^{(-x/L+1)v_2^\theta}$ | $E_{e2}^\theta \xi_-^\theta \mathcal{F}_- \sin u_2^\theta e^{(-x/L+1)v_2^\theta}$ | $L<x$ |
| QTE-odd-1 | $B_{o3}^\theta \frac{\cos u_3^\theta}{\mathcal{F}_+} e^{(x/L+1)v_3^\theta}$ | $B_{o3}^\theta \xi_+^\theta \cos u_3^\theta e^{(x/L+1)v_3^\theta}$ | $x<-L$ |
| | $B_{o3}^\theta \frac{1}{\mathcal{F}_+} \cos(u_3^\theta x/L)$ | $B_{o3}^\theta \cos(u_3^\theta x/L)$ | $-L \leq x \leq L$ |
| | $B_{o3}^\theta \frac{\cos u_3^\theta}{\mathcal{F}_+} e^{(-x/L+1)v_3^\theta}$ | $B_{o3}^\theta \xi_+^\theta \cos u_3^\theta e^{(-x/L+1)v_3^\theta}$ | $L<x$ |
| QTM-odd-1 | $E_{o4}^\theta \cos u_4^\theta e^{(x/L+1)v_4^\theta}$ | $E_{o4}^\theta \xi_-^\theta \mathcal{F}_- \cos u_4^\theta e^{(x/L+1)v_4^\theta}$ | $x<-L$ |
| | $E_{o4}^\theta \cos(u_4^\theta x/L)$ | $E_{o4}^\theta \mathcal{F}_- \cos(u_4^\theta x/L)$ | $-L \leq x \leq L$ |
| | $E_{o4}^\theta \sin u_4^\theta e^{(-x/L+1)v_4^\theta}$ | $E_{o4}^\theta \xi_-^\theta \mathcal{F}_- \sin u_4^\theta e^{(-x/L+1)v_4^\theta}$ | $L<x$ |

TABLE IV: Longitudinal EM fields for each hybrid mode and in each media 1, 2 and 3 or regions $x<-L; -L<x<L; L<x$ respectively.

For higher order modes, the mathematical expressions of the fields remain the same, although the values of $u_n^\theta$, and consequently $v_n^\theta$, change. The transverse EM fields are described by Eqs. (21).

At this stage we can introduce a new way of labeling the modes. This is useful to better understand the structure of the modes (considering the many branches of the trigonometric functions, especially when one goes beyond the first four modes), and also for the normalization of the modes of Θ-ED with which we will deal in the next subsection. So far, we specified the modes by a single label $p=1,2,3,4,5,\dots$ each mode corresponding to the intersection of the dispersion relation Eq. (26) and the many branches of the propagation condition Eq. (44). This single label carries information about the parity (even/odd) and also of the transversality (TE/TM), which is confusing. Hence, we introduce two indices, $n,\eta$. The index $n=0,1,2,3,\dots$ labels the parity of the modes, all even/odd modes having even/odd $n$, and coincides with the index previously introduced to label the periodicity of the solutions $u_0$ of Eq. (44) for $v=0$. The index $\eta$ labels its (quasi)transversality, QTE/QTM being labeled by $\eta=+/-$, respectively. As shown in Table III, this index also labels the two entries $\xi_\pm^\theta$ in Eq. (52), as previously indicated when discussing the emergence of the branches in the waveguide. With this reinterpretation, the indices $(n,\eta)$ describe now two important physical properties of the modes: the parity and the polarization content. For example, the first five modes as labeled in 0-ED are: $p=1$ and it is QTE-even, $p=2$ is QTM-even, $p=3$ is QTE-odd, $p=4$ is QTM-odd and $p=5$ is QTE-even (and different to the $p=1$ mode), ... Now these same modes are labeled by the couples $(n,\eta)=(0,+),(0,-),(1,+),(1,-),(2,+),\dots$

With this change of notation, the $(u_p^\theta, v_p^\theta)$ relations of the propagation condition are written in a single expression that defines all the allowed modes of the Θ-ED:

$$v_{n\eta}^\theta = \xi_\eta^\theta u_{n\eta}^\theta \tan\left(u_{n\eta}^\theta + \frac{n\pi}{2}\right), \qquad n=0,1,2,\dots \qquad \text{and} \qquad \eta=\pm, \tag{51}$$

where

$$\xi_\eta^\theta = \frac{1}{2}(\xi_+ + \xi_- + \xi_0\theta^2) + \frac{\eta}{2}\sqrt{(\xi_+ + \xi_- + \xi_0\theta^2)^2 - 4\xi_+\xi_-}, \tag{52}$$

and,

$$\xi_+ = \frac{\mu_1}{\mu_2}, \qquad \xi_- = \frac{\epsilon_1}{\epsilon_2}, \qquad \xi_0 = \frac{\mu_1}{\epsilon_2}. \tag{53}$$

The expansion of $\xi_\eta^\theta$ that allows to compare with the modes of 0-ED now reads,

$$\xi_\eta^\theta \approx \xi_\eta\Big(1 + \eta\frac{\xi_0}{(\xi_+ - \xi_-)}\theta^2\Big). \tag{54}$$

The longitudinal components of the Θ-ED modes, up to the respective amplitudes $a^{\theta}_{n\eta}$, are written as:

$$\frac{E^{\theta}_{zn\eta}(x)}{a^{\theta}_{n\eta}} = i\Big[\frac{1}{\mathcal{F}_+}\delta^+_\eta + \delta^-_\eta\Big] \times \begin{cases} -\sin(u^{\theta}_{n\eta} - \frac{n\pi}{2})e^{(1+x/L)v^{\theta}_{n\eta}} & x < -L, \\ \sin(\frac{u^{\theta}_{n\eta}}{L}x + \frac{n\pi}{2}), & -L < x < L, \\ \sin(u^{\theta}_{n\eta} + \frac{n\pi}{2})e^{(1-x/L)v^{\theta}_{n\eta}} & L < x, \end{cases} \tag{55}$$

$$\frac{B^{\theta}_{zn\eta}}{a^{\theta}_{n\eta}}(x) = i\Big[\delta^+_\eta + \mathcal{F}_-\,\delta^-_\eta\Big] \times \begin{cases} -\xi^{\theta}_\eta \sin(u^{\theta}_{n\eta} - \frac{n\pi}{2})e^{(1+x/L)v^{\theta}_{n\eta}} & x < -L, \\ \sin(\frac{u^{\theta}_{n\eta}}{L}x + \frac{n\pi}{2}), & -L < x < L, \\ \xi^{\theta}_\eta \sin(u^{\theta}_{n\eta} + \frac{n\pi}{2})e^{(1-x/L)v^{\theta}_{n\eta}} & L < x, \end{cases} \tag{56}$$

where $a^{\theta}_{n\eta} = a^{\theta}_{0+} = B_{e1}$, $a^{\theta}_{0-} = E_{e2}$, $a^{\theta}_{1+} = B_{o3}$, $a^{\theta}_{1-} = E_{o4}$ and $\delta^{\pm}_\eta$ is the Kronecker delta.

The transversal components of the fields are given by Eqs. (21) and thus the complete EM fields are given by:

$$\mathbf{E}^{\theta}_{n\eta} = \begin{pmatrix} \frac{i}{k_x^2}k_z\partial_x E^{\theta}_{zn\eta} \\ -\frac{i}{k_x^2}k_0\partial_x B^{\theta}_{zn\eta} \\ E^{\theta}_{zn\eta} \end{pmatrix}, \tag{57}$$

$$\mathbf{B}^{\theta}_{n\eta} = \begin{pmatrix} \frac{i}{k_x^2}k_z\partial_x B^{\theta}_{zn\eta} \\ \frac{i}{k_x^2}k_0\epsilon\mu\partial_x E^{\theta}_{zn\eta} \\ B^{\theta}_{zn\eta} \end{pmatrix}, \tag{58}$$

At this point we recall that the wavenumbers $k_x$ and $k_z$ also have $\theta$ dependence, as explained after Eq. (47), but we still defer the explicit use of additional indexes for these. With the above simplified notation, the longitudinal components in Eqs. (57), (58) and Eqs. (55), (56), reproduce the fields in Table IV, with the appropriate changes in the mode-labeling notation.

The orthogonality relation of the $\theta$-ED modes proceeds in a similar fashion as in the usual case. This entails a $\theta$-ED version of the Lorentz reciprocity theorem resulting in:

$$\begin{aligned} \frac{c_0}{4\pi}\int_{\mathbb{R}^2}\frac{1}{4\mu}[\mathbf{E}^{\theta}_r \times (\mathbf{B}^{\theta}_s)^* + (\mathbf{E}^{\theta}_s)^* \times \mathbf{B}^{\theta}_r]\cdot\hat{\mathbf{z}}da_\perp &= \delta_{rs}p_r, \\ \frac{c_0}{4\pi}\int_{\mathbb{R}^2}\frac{1}{4\mu}[\mathbf{E}^{\theta}_{n\eta} \times (\mathbf{B}^{\theta}_{n'\eta'})^* + (\mathbf{E}^{\theta}_{n'\eta'})^* \times \mathbf{B}^{\theta}_{n\eta}]\cdot\hat{\mathbf{z}}da_\perp &= \delta_{nn'}\delta_{\eta\eta'}p_{n\eta}, \\ \frac{c_0}{4\pi}\int_{\mathbb{R}^2}\frac{1}{4\mu}[\mathbf{E}^{\theta}_{\perp,n\eta} \times (\mathbf{B}^{\theta}_{\perp,n'\eta'})^* + (\mathbf{E}^{\theta}_{\perp,n'\eta'})^* \times \mathbf{B}^{\theta}_{\perp,n\eta}]\cdot\hat{\mathbf{z}}da_\perp &= \delta_{nn'}\delta_{\eta\eta'}p_{n\eta}, \end{aligned} \tag{59}$$

where, $\mathbf{E}^{\theta}_{\perp,n\eta} = (\hat{\mathbf{z}}\times\mathbf{E}^{\theta}_{n\eta})\times\hat{\mathbf{z}}$ is the transverse component of the $\theta$-E field in Eq. (57). The second line simply expresses that the longitudinal components of the fields do not contribute given one integrates in the perpendicular cross-section. The coefficient $p_{n\eta}$ denotes the power of the EM field in the $n\eta$ mode. Using Eq. (59) for $n = n'$ and $\eta = \eta'$ the amplitudes can be found as:

$$a^{\theta}_{n\eta} = \sqrt{\frac{p_{n\eta}/L_y}{\mathcal{N}^{\theta}_{n\eta}}}, \tag{60}$$

where $p_{n\eta}/L_y$ is the power per unit length in the $y$ direction of the mode $n\eta$ and the normalization constants are:

$$\begin{aligned} \mathcal{N}^{\theta}_{n\eta} &= \frac{c_0}{4\pi}\int\frac{1}{4\mu}(\boldsymbol{\mathcal{E}}^{\theta}_{n\eta}\times(\boldsymbol{\mathcal{B}}^{\theta}_{n'\eta'})^* + (\boldsymbol{\mathcal{E}}^{\theta}_{n'\eta'})^*\times\boldsymbol{\mathcal{B}}^{\theta}_{n\eta})\cdot\hat{\mathbf{z}}dx\ \delta_{n\eta,n'\eta'} \\ &= \frac{c_0}{4\pi}\int\frac{1}{4\mu}(\boldsymbol{\mathcal{E}}^{\theta}_{\perp,n\eta}\times(\boldsymbol{\mathcal{B}}^{\theta}_{\perp,n\eta})^* + (\boldsymbol{\mathcal{E}}^{\theta}_{\perp,n\eta})^*\times\boldsymbol{\mathcal{B}}^{\theta}_{\perp,n\eta})\cdot\hat{\mathbf{z}}dx \\ &= \frac{c_0}{4\pi}\int\frac{1}{2\mu}\mathfrak{Re}\left(\boldsymbol{\mathcal{E}}^{\theta}_{\perp n\eta}\times(\boldsymbol{\mathcal{B}}^{\theta}_{\perp n\eta})^*\right)\cdot\hat{\mathbf{z}}dx \\ &= \frac{L^3\omega k^{\theta}_{zn\eta}}{16\pi}\Big(\frac{\delta^+_\eta}{\mathcal{F}^2_\eta} + \delta^-_\eta\Big)\Big[\Big(\frac{1-(-1)^n\cos(2u^{\theta}_{n\eta})}{(v^{\theta}_{n\eta})^3}\Big)\Big(\epsilon_1 + \frac{(\xi^{\theta}_\eta)^2\mathcal{F}^2_\eta}{\mu_1}\Big) + \Big(\frac{2u^{\theta}_{n\eta} + (-1)^n\sin(2u^{\theta}_{n\eta})}{(u^{\theta}_{n\eta})^3}\Big)\Big(\epsilon_2 + \frac{\mathcal{F}^2_\eta}{\mu_2}\Big)\Big] \end{aligned} \tag{61}$$

where we have defined:

$$\mathbf{E}^{\theta}_{n\eta} = a^{\theta}_{n\eta}\boldsymbol{\mathcal{E}}^{\theta}_{n\eta}, \qquad \mathbf{B}^{\theta}_{n\eta} = a^{\theta}_{n\eta}\boldsymbol{\mathcal{B}}^{\theta}_{n\eta}. \tag{62}$$

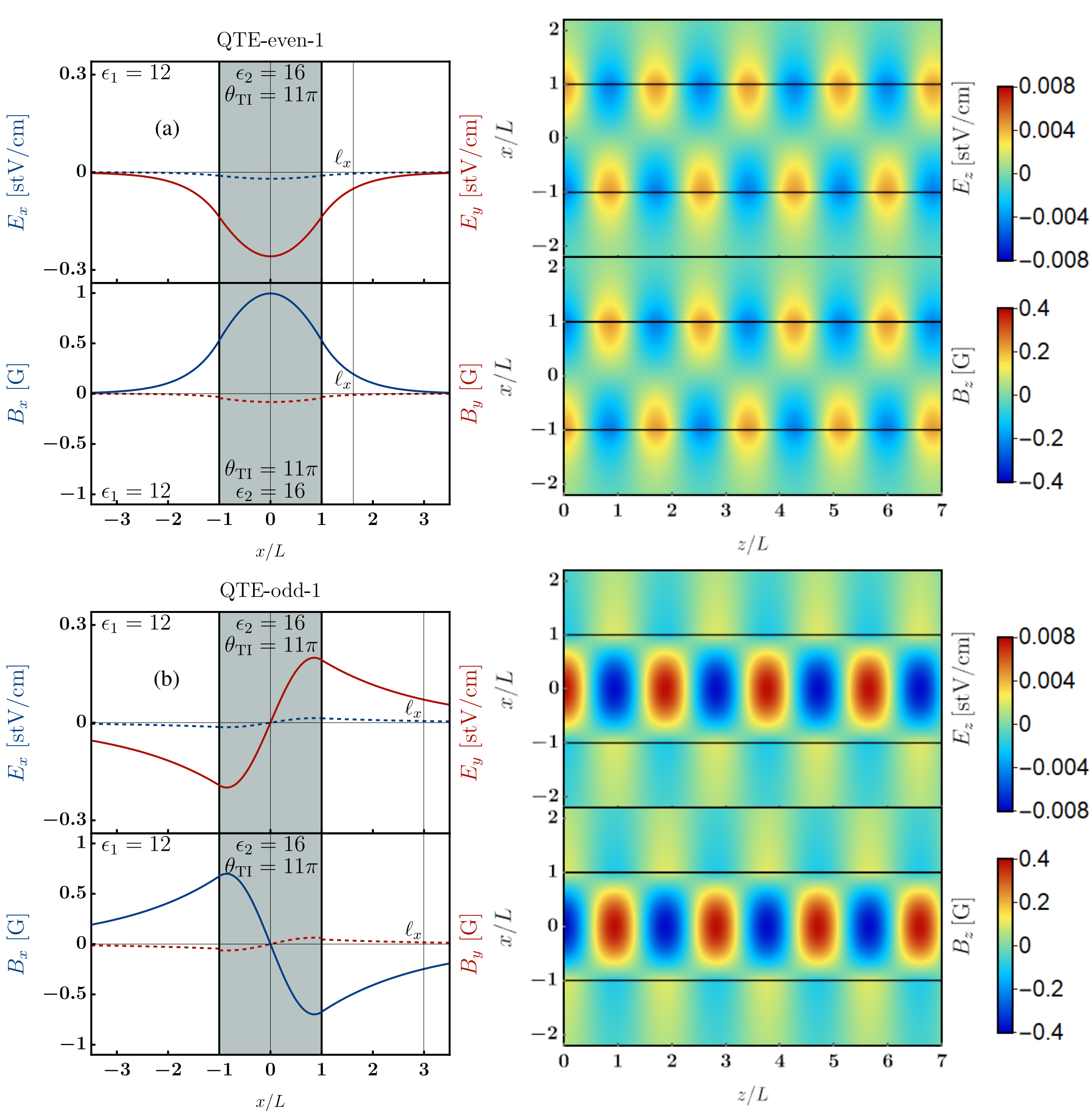

FIG. 4: EM fields of the first two quasi-transverse electric (QTE) {hybrid modes, even (a) and odd (b), are shown operating at a wavelength $\lambda_0 = 3.3 \times (2L)$ in a symmetric waveguide with a TI core width of $2L = 40\,\mu m$. The $x$ and $y$ components of the fields are depicted in blue and red, respectively, while the longitudinal components along the $z$ direction are illustrated using a density plot. The topo-optical parameters used are $\mu_1 = \mu_2 = 1$, $\epsilon_1 = 12$, $\epsilon_2 = 16$, and $\Theta_2 = \alpha\theta_{\mathrm{TI}}/\pi = 11\alpha$. The solid curves represent the dominant components of the QTE mode, while the dashed curves indicate the components induced solely by the TMEP $\Theta_2$ of the TI. The dimensionless penetration length $\ell_x$ is marked with a black vertical line. We observe that the even mode is more confined than the odd mode.

The normalization will be chosen such that all modes have the same power per unit length, i.e., $p^{\theta}_{n\eta}/L_y = p^{\theta}/L_y$ for all modes. In Figures 4 and 5, we present the EM fields of the first four allowed propagation modes, operating at a wavelength $\lambda_0 = 3.3 \times (2L)$ in a symmetrical waveguide with a TI-core of width $2L = 40\,\mu$m. The topo-optical parameters considered are $\mu_1 = \mu_2 = 1$, $\epsilon_1 = 12$, $\epsilon_2 = 16$, and $\Theta_2 = \alpha\theta_{\mathrm{TI}}/\pi = 11\alpha$. The blue and red curves illustrate the components of the fields in the $x$ and $y$ directions, respectively. The longitudinal components along the $z$ direction are shown with density plots in the right column of the figures. Each of these components of the EM fields is presented in their respective Gaussian units. The allowed modes are normalized according to Eq. (60) to a power of $p^{\theta}/L_y = 0.1$ W/cm. This means that the EM field amplitudes shown in Figs. 4 and 5 correspond to those required for the power transmitted by a specific mode in a 40 $\mu$m wide waveguide to be $0.1$ W per centimeter of transverse length. Observe that the electric and magnetic fields have discontinuities at the boundaries. Given we take all media as non-magnetic and that there are no free charge or current densities, the $y$-component of the magnetic field is only discontinuous due to the product of $\tilde{\theta}$ (Eq. (32)) times the $y$-component of the $E$ field at the interface. Since $\tilde{\theta}$ is small, this discontinuity is hardly perceptible in Fig. 4. The $x$-component of the $E$ field, however, is discontinuous also. This discontinuity has two contributions. One due to the different permittivities, leading to bound charges at the interfaces, and the other due to $\theta$ times the $x$-component of the $B$ field at the interface, as shown in Eq. (33). These two contributions together are more apparent in Fig. 5.

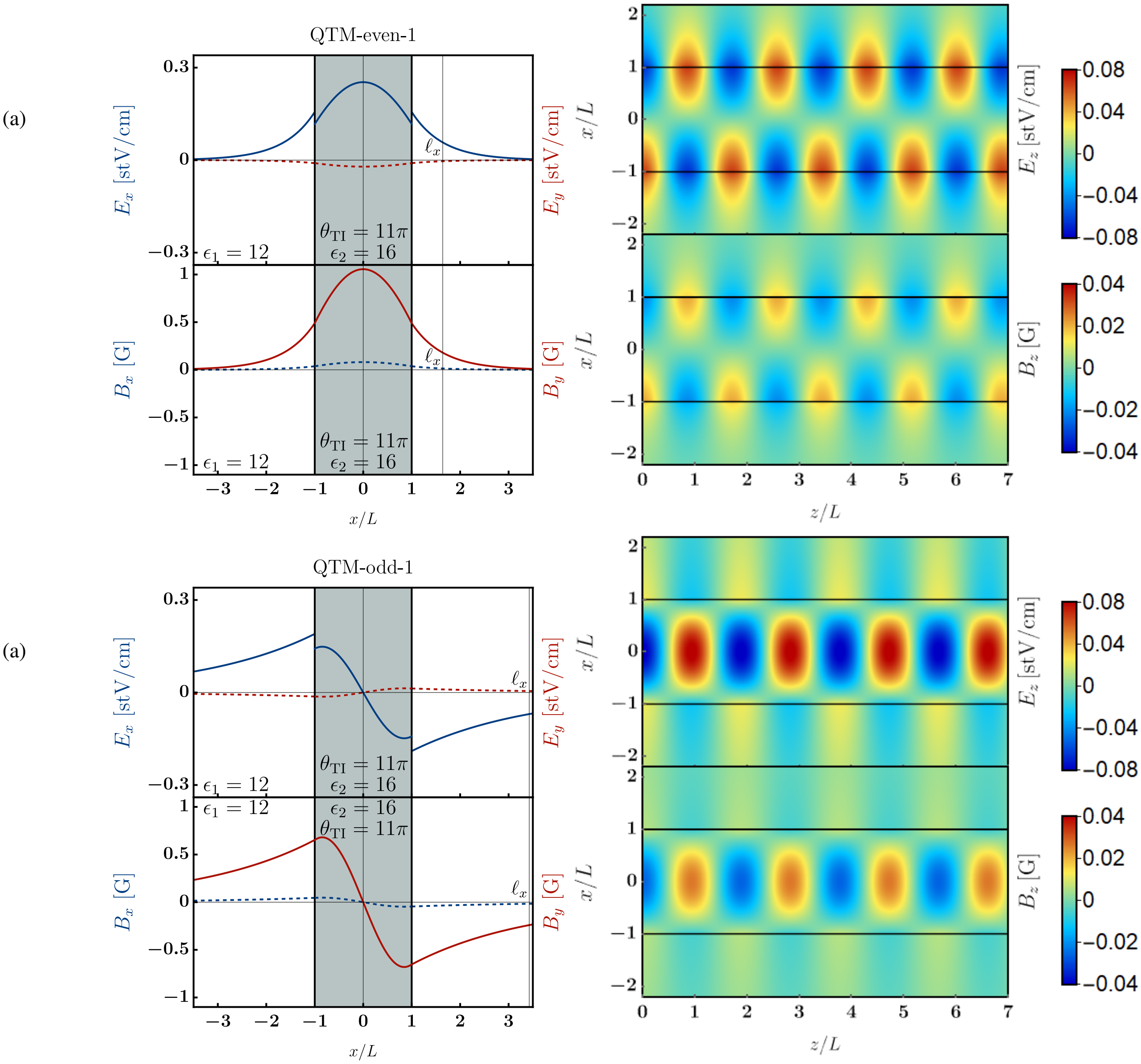

FIG. 5: EM fields of the first two quasi-transverse magnetic (QTM) hybrid modes, even (a) and odd (b), are shown operating at a wavelength $\lambda_0 = 3.3 \times (2L)$ in a symmetric waveguide with a TI core width of $2L = 40\,\mu m$. The $x$ and $y$ components of the fields are depicted in blue and red, respectively, while the longitudinal components along the $z$ direction are illustrated using a density plot. The topo-optical parameters used are $\mu_1 = \mu_2 = 1$, $\epsilon_1 = 12$, $\epsilon_2 = 16$, and $\Theta_2 = \alpha\theta_{TI}/\pi = 11\alpha$. The solid curves represent the dominant components of the TM mode, while the dashed curves indicate the components induced solely by the TMEP $\Theta_2$ of the TI. The dimensionless penetration length $\ell_x$ is marked with a black vertical line. We observe that the even mode is more confined than the odd mode.

### D. High $\theta$ limit and quasi-TEM modes

We end this section by observing an interesting feature of the exact solution. We entertain the idea of considering the high $\theta$ limit of our solutions. Such a limit seems untenable a priori, because typically, the $\theta$-effects are small compared to other optical responses, due to its underlying quantum nature. However, tuning the optical parameters $(\epsilon, \mu)$ to enhance the topological magnetoelectric response of TIs is feasible. Mu-metals were used in [66] to this end. Also in [67] the possible effects of ENZ (epsilon-near-zero) materials [8] have also been commented.

We begin by noting that Eq. (54) can be written as:

$$\xi_\eta^\theta \approx \xi_\eta \Big(1 + \eta\ Z_2^2\ \theta^2\Big), \qquad \text{if} \qquad Z_2 \ll Z_1, \tag{63}$$

where $Z_i \equiv \sqrt{\mu_i/\epsilon_i}$ are the impedances in the claddings ($i = 1$) and in the TI ($i = 2$) respectively. So in the limit of a low impedance TI slab (relative to the claddings) the effective parameter the $\xi_\eta^\theta$ is in fact the product $Z\theta$, i.e., from a practical point of view, we consider as equivalent the high-$\theta$ limit to the low-TI impedance limit, relative to that of the cladding. Hence increasing the impedance of the TI slab, while keeping it lower than that of the claddings has the effect of increasing $\theta$. With the latter as a physical argument to endorse the theoretical limit of taking high $\theta$, we analyze the effects on the EM fields. To fix the idea, we consider a frequency that allows only the first two modes, namely $(0+)$ and $(0-)$. For these modes, from Fig. 3 we see that the larger the $\theta$ the larger(smaller) the $v^\theta$ for the $0+$ $(0-)$ respectively. And larger $v$'s means higher damping for the EM fields into the cladding regions, see Eq. (30), i.e., for large $\theta$ the $(0+)$ mode gets more confined in the TI slab, while the $(0-)$ mode gets less confined, or leaks more into the claddings. Next, we focus on the longitudinal components of the EM fields, Eqs. (55) and (56) and we numerically evaluate them in this high $\theta$-limit, we see that in fact, for both modes the longitudinal components are nearly zero, i.e.:

$$E^\theta_{z,0+} \approx 0, \quad B^\theta_{z,0+} \approx 0, \qquad \text{and} \qquad E^\theta_{z,0-} \approx 0, \quad B^\theta_{z,0-} \approx 0, \qquad\qquad \text{in the slab, in the } Z_2 \ll Z_1 \text{ limit.} \tag{64}$$

But these longitudinal components are not exactly zero. Using these, to again numerically evaluate the transverse components inside and outside the TI slab via Eqs. (57) and (58), we find that:

1. The transverse components of the $\mathbf{E}$ and $\mathbf{B}$ fields are negligible for the $(0-)$ mode inside the TI slab, while non-vanishing and actually leaking to infinity into the claddings (i.e., the $(0-)$ mode ceases to be confined).

2. The transverse components of the $\mathbf{E}$ and $\mathbf{B}$ fields enhanced for the $(0+)$ mode inside the TI slab, while decaying rapidly into the claddings (i.e., the $(0+)$ mode gets highly confined).

The above results, together with the vanishing of the longitudinal components, allow us to state that the EM field propagation in the $Z_2 \ll Z_1$ limit of our $\Theta$-slab results in quasi-TEM (transverse electromagnetic) modes. In [67], for the first time exact TEM modes in the context of $\Theta$-ED were found, as a particular case of what had been envisoned in [16] as a possibility to escape Earnshaw's theorem. Here we have a 2D and high-$\theta$ version of the latter. These TEM waves are highly appreciated for waveguiding purposes in optics and photonics due to their low energy loss rates and high speed of information transmission [68, 69].

## IV. COUPLING EXTERNAL MODES TO THE $\Theta$-SLAB

Having found the exact solutions we now aim to answer how external field modes couple to the $\Theta$-slab waveguide with its $\Theta$ boundary conditions. The external field modes are arbitrary, they could be superpositions of 0-ED modes, superpositions of $\Theta$-modes, etc. In ordinary ED or in $\Theta$-ED, to tackle this issue, it is customary to follow the approach of coupled-mode theory. In it, one takes as modes (basis) a set of EM field profiles that do not necessarily match the boundary conditions of the specific problem, usually the modes of 0-ED. This happens, for example, for a waveguide with varying longitudinal geometry, or having geometric irregularities along the waveguide, having slight alterations to the assumed axial symmetry, or, due to modified boundary conditions as occurs with magnetoelectric media. And in order to make a superposition of the 0-ED modes to satisfy the altered boundary conditions, the expansion coefficients acquire additional dependence on the longitudinal coordinate $z$.

In this work, however, we have found the exact $\Theta$-ED modes that, by construction, satisfy the $\Theta$-boundary conditions. We thus have two approaches from where to address the issue of how external fields couple to the $\Theta$-slab waveguide, and therefore we would like to provide a quantitative comparison between the two approaches. To this end, in Sec. IV A we couple a 0-ED mode to the $\Theta$-slab, expressed in terms of the exact $\Theta$-ED modes obtained in Eqs. (55) through (58). To compare with other works that have approached this issue but instead of using exact modes have used a perturbative expansion and/or coupled-mode theory, as in e.g., [44], in Sec. IV B, we then compute an expansion of the $\Theta$-ED modes as a series in powers of $\theta$, in terms of the 0-ED modes. Finally, in Sec. IV C we couple a 0-ED mode to the $\Theta$-slab, expressed in terms of the $\Theta$-ED modes, but performing the expansion of the latter in terms of the 0-ED modes, i.e., the expansions obtained in IV C are replaced in the exact calculations of Sec. IV A. In this sense is that we call this approach perturbative in $\theta$ and resembles coupled-mode theory (Coupled-mode theory will in fact be dealt with in Sec. VI). The exact approach differs from the perturbative or coupled-mode theory approaches. Interestingly, as we will comment below, this difference has consequences on the power transfer between propagating modes.

### A. Coupling 0-ED external modes to the slab in terms of exact Θ-ED modes

Having found the complete EM field confined modes for the slab in the Θ-ED, we can use these to expand an arbitrary field propagating in the slab as linear combinations of these modes, and pose the following question. Suppose we want to propagate a EMW along the slab, given that a prescribed EM field is imposed at $z = 0$, say, $\{\mathbf{E}(\mathbf{r}_{|z=0}, t), \mathbf{B}(\mathbf{r}_{|z=0}, t)\} = \{\mathbf{E}_{\text{ext}}, \mathbf{B}_{\text{ext}}\}$ with the usual harmonic time dependence $\sim e^{-i\omega t}$, the question then is, which modes of the Θ-ED and with what amplitudes will propagate down the slab? Naturally, inside the slab, the total field will be a superposition of Θ-modes, with the harmonic time dependence and propagating in the $z$ direction, i.e., $\propto e^{i(k_z^\theta - \omega t)}$, with unknown expansion coefficients yet to be determined, but dependent of the specific external fields. To do so, we will use the orthogonality of the Θ-modes (see Appendix IX B).

Omitting the harmonic time dependence and taking the external fields as the values of the total field at $z = 0$, we can write:

$$\mathbf{E}_{\text{ext}} \equiv \mathbf{E}^\theta_{\text{Tot}}|_{z=0} = \sum_r A^\theta_r \mathbf{E}^\theta_r, \qquad \text{and} \qquad \mathbf{B}_{\text{ext}} \equiv \mathbf{B}^\theta_{\text{Tot}}|_{z=0} = \sum_r A^\theta_r \mathbf{B}^\theta_r, \tag{65}$$

where we introduced $r \equiv n\eta$ to simplify the notation, and $\mathbf{E}^\theta_{\text{Tot}}$, $\mathbf{B}^\theta_{\text{Tot}}$ are the total EM field resulting from the superposition of Θ-modes, i.e., at $z = 0$, the external field is expressed as a superposition of Θ-ED modes. For given external EM fields $\mathbf{E}_{\text{ext}}$ and $\mathbf{B}_{\text{ext}}$ at $z = 0$, the expansion coefficients are found by applying the orthogonality of the Θ-modes, we get:

$$A^\theta_r(\mathbf{E}_{\text{ext}}, \mathbf{B}_{\text{ext}}) = \frac{c_0}{4\pi} \int \frac{1}{4\mu} [\mathbf{E}_{\text{ext}} \times (\mathbf{B}^\theta_r)^* + (\mathbf{E}^\theta_r)^* \times \mathbf{B}_{\text{ext}}] \cdot \hat{\mathbf{z}} dx. \tag{66}$$

The external fields are arbitrary at this point and one is free to take them at will. So as to compare which modes are excited in the slab with the $\theta = 0$ case, we take as external fields a particular normalized mode of the 0-ED, namely: $\mathbf{E}_{\text{ext}} = \mathbf{E}^{\theta=0}_l \equiv \mathbf{E}_l$ y $\mathbf{B}_{\text{ext}} = \mathbf{B}^{\theta=0}_l \equiv \mathbf{B}_l$.

$$A^\theta_r(\mathbf{E}_l, \mathbf{B}_l) \equiv A^\theta_{rl} = \frac{c_0}{4\pi} \int \frac{1}{4\mu} [\mathbf{E}_l \times (\mathbf{B}^\theta_r)^* + (\mathbf{E}^\theta_r)^* \times \mathbf{B}_l] \cdot \hat{\mathbf{z}} dx \tag{67}$$

The power transmitted by the EM field when excited by an external $l$-th mode of 0-ED is:

$$\begin{aligned} P^\theta_l &= \frac{c_0}{4\pi} \int \frac{1}{2\mu} \text{Re}\, [\mathbf{E}_l \times (\mathbf{B}_l)^*] \cdot \hat{\mathbf{z}} dx, \\ &= \sum_{s,r} \frac{c_0}{4\pi} \int \frac{1}{2\mu} \text{Re}\, [A^\theta_{sl} (A^\theta_{rl})^* \mathbf{E}^\theta_s \times (\mathbf{B}^\theta_r)^* e^{i(k^\theta_s - k^\theta_r)z}] \cdot \hat{\mathbf{z}} dx \\ &= \sum_s P^\theta_{sl}, \qquad \text{where} \qquad P^\theta_{sl} \equiv \sum_r \frac{c_0}{4\pi} \int \frac{1}{2\mu} \text{Re}\, [A^\theta_{sl} (A^\theta_{rl})^* \mathbf{E}^\theta_s \times (\mathbf{B}^\theta_r)^* e^{i(k^\theta_s - k^\theta_r)z}] \cdot \hat{\mathbf{z}} dx. \end{aligned} \tag{68}$$

If the external field is a 0-ED mode, and in particular the TE even mode, then we set out to compute the power $P^\theta_{\text{TEe}}$, that is the power transmitted along the Θ-slab given it was excited by the TE-even mode of the 0-ED. Two important observations at this point are in order. The first one is that the $l$ index corresponds to a mode of 0-ED. However, the indexes $r, s$ label the modes of Θ-ED. Second, the sums above run over all possible modes supported by the waveguide, but those possible modes are restricted by the operational frequency. Now, as is seen in Fig. 3, for a given frequency the amount of modes of 0-ED and Θ-ED are always the same, and in one-to-one correspondence. To simplify the computation we assume that the frequency is such that only the first two modes of Θ-ED are allowed, namely $s, r = 0+, 0-$ or simply $r, s = +, -$. Thus $P^\theta_{\text{TEe}} = P^\theta_{+;\text{TEe}} + P^\theta_{-;\text{TEe}}$, where each $P^\theta_{\pm;\text{TEe}}$ are to be computed using Eq. (68) with $l = \text{TEe}$ and $r, s = +, -$. To compute each $P^\theta_{\pm;\text{TEe}}$, we use the exact expressions for the Θ-ED modes of Eqs. (55) through (58), and also recall that the $k^\theta_r$ in the exponentials are in fact the $k^\theta_{n\eta,z}$ that are given by the dispersion relation in Eq. (20) and the $(u^\theta_{n\eta}, v^\theta_{n\eta})$ of Eq. (51) through (53). Unfortunately, performing the integrations is computationally involved. As the details are not very illuminating and we simply to explaining the results.

1. The power transmitted by each mode $P^{\theta}_{\pm;\text{TEe}}$ are dependent of $z$, i.e., as the EM field propagates along the slab, power from one mode is transferred to the other.

2. The expression for the latter are are exact, i.e., valid to all orders in $\theta$.

3. The total power transmitted by the EM field as it propagates along the slab is constant, and not dependent on $z$.

The last point reveals two important conclusions. Firstly, we are verifying energy conservation and, secondly, the fact that we build our solutions with the exact $\Theta$-ED modes means that we are satisfying the boundary conditions at every instant of time and at every $z$ along the waveguide. This is a subtle but radically different approach from what is mostly done in the literature, as will be shown below.

### B. Expansion to first order in $\theta$ to connect with 0-ED

As shown before, we have written the EM field satisfying all of Maxwell's equations and the boundary conditions of $\Theta$-ED in the case of a symmetrical antiparallel slab waveguide. This EM field is exact, i.e., valid to all orders in $\theta$ and thus, is expressed in modes that differ from those of 0-ED. However, it is sensible to consider the case when the $\theta$-effects are small, for which it might be useful to expand the $\Theta$-ED modes in powers of $\theta$. To do this, we must remember that our solution is semi-analytical: we have analytical expressions for all the unknowns, but they depend on a parameter, $u^{\theta}_{n\eta}$, that must be defined by a transcendental equation. Specifically, this equation is the intersection of the branches with the dispersion relation $(v^{\theta}_{n\eta})^2 + (u^{\theta}_{n\eta})^2 = k_0^2 L^2 (n_2^2 - n_1^2)$, equation that defines the modes. Now, the $\theta$-effects modify where the branches intersect the dispersion relation, and hence the numerical values of the modes $(u^{\theta}_{n\eta}, v^{\theta}_{n\eta})$. So for a given branch, say the one of the first mode $(n = 0)$, and given the dispersion relation we can thus determine the value of $u^{\theta}_{0\eta}$ by:

$$v^{\theta}_{0\eta} = \xi^{\theta}_{\eta} u^{\theta}_{0\eta} \tan(u^{\theta}_{0\eta}), \qquad \text{and} \qquad (u^{\theta}_{0\eta})^2 + (v^{\theta}_{0\eta})^2 = \mathcal{R}^2 = k_0^2 L^2 (n_2^2 - n_1^2), \tag{69}$$

from which we obtain an equation for $u^{\theta}_{0\eta}$ as:

$$\mathcal{R}^2 = (u^{\theta}_{0\eta})^2 + (\xi^{\theta}_{\eta})^2 (u^{\theta}_{0\eta})^2 \tan^2(u^{\theta}_{0\eta}). \tag{70}$$

Here we use the expression for $\xi^{\theta}_{\eta}$ in Eq. (52) expanded in powers of $\theta$. So in order to obtain how the $\theta$ modes for this branch differ from the usual ones we propose the ansatz:

$$u^{\theta}_{0\eta} = u_{0\eta} \left(1 + a\,\theta + b\,\theta^2 + \mathcal{O}(\theta^3)\right). \tag{71}$$

Replacing the latter equation in (70) and also using that the 0-ED modes also satisfy the same equation (69) (because they also lie in the intersection of the propagation condition with the dispersion relation circle), the zeroth order terms cancel with $\mathcal{R}^2$ and the linear and quadratic terms in $\theta$ yield two equations for the two unknowns $a, b$, because the left-hand-side in Eq. (70) does not have $\theta$ contributions. The result of this reveals that:

$$a = 0, \tag{72}$$

$$b = \frac{\eta\,\xi_\eta^2\,\xi_0 \tan^2(u)}{(\xi_- - \xi_+)\left(\xi_\eta^2 u \tan^3(u) + \xi_\eta^2 \tan^2(u) + \xi_\eta^2 u \tan(u) + 1\right)}. \tag{73}$$

Then, replacing this in the definition of the $v^{\theta}_{0\eta}$ mode and again expanding $\xi^{\theta}_{\eta}$ we can read the modification to the $v^{\theta}_{0\eta}$ mode, for this branch, resulting in:

$$v^{\theta}_{0\eta} = v_{0\eta} \left(1 - \frac{\eta\,\xi_0}{(\xi_- - \xi_+)\left(\xi_\eta^2 u \tan^3(u) + \xi_\eta^2 \tan^2(u) + \xi_\eta^2 u \tan(u) + 1\right)} \theta^2\right). \tag{74}$$

Therefore, we found that both dimensionless wavenumbers acquire only second order corrections due to $\theta$, i.e.,

$$u^{\theta}_{0\eta} \approx u_{0\eta} + \mathcal{O}(\theta^2), \qquad \text{and} \qquad v^{\theta}_{0\eta} \approx v_{0\eta} + \mathcal{O}(\theta^2). \tag{75}$$

For higher modes a similar situation occurs. Now it is time to emphasize that, as mentioned after Eq. (47), the wavenumbers $k_x$ and $k_z$ of each mode do have a $\theta$ dependence inherited from the dimensionless wavenumbers $(u^{\theta}_{n\eta}, v^{\theta}_{n\eta})$. The previous analysis implies that in all media $k_{x,n\eta}$ depend on $\theta$, which we write as $k^{\theta}_{x,n\eta}$ and, in fact, they also acquire only $\theta^2$ corrections, and therefore (by Eq. (20)), so do $k^{\theta}_{z,n\eta}$. With this, the $\theta$-ED modes are written with all $\theta$ dependence explicit as:

$$\boldsymbol{\mathcal{E}}^{\theta}_{n\eta} \equiv \hat{\mathbf{x}}\frac{i}{(k^{\theta}_{xn\eta})^2}k^{\theta}_{zn\eta}\partial_x E^{\theta}_{zn\eta} - \hat{\mathbf{y}}\frac{i}{(k^{\theta}_{xn\eta})^2}k_0\partial_x B^{\theta}_{zn\eta} + \hat{\mathbf{z}}E^{\theta}_{zn\eta}, \tag{76}$$

$$\boldsymbol{\mathcal{B}}^{\theta}_{n\eta} \equiv \hat{\mathbf{x}}\frac{i}{(k^{\theta}_{xn\eta})^2}k^{\theta}_{zn\eta}\partial_x B^{\theta}_{zn\eta} + \hat{\mathbf{y}}\frac{i}{(k^{\theta}_{xn\eta})^2}k_0\mu\epsilon\partial_x E^{\theta}_{zn\eta} + \hat{\mathbf{z}}B^{\theta}_{zn\eta}, \tag{77}$$

Thus, replacing the latter and expanding the trigonometric functions in Eqs. (55), (56) together with the overall linear and zeroth order terms (in $\theta$) in the square brackets, once replaced in Eqs. (76), (77) yields only first order corrections to the $\theta$-ED modes that result in:

$$\boldsymbol{\mathcal{E}}^{\theta}_{n\eta} = \boldsymbol{\mathcal{E}}_{n\eta} - \theta\,\frac{1}{\mu}\,\frac{\xi_0}{\xi_+ - \xi_-}\,\boldsymbol{\mathcal{B}}_{n\eta} + \mathcal{O}(\theta^2), \tag{78}$$

$$\boldsymbol{\mathcal{B}}^{\theta}_{n\eta} = \boldsymbol{\mathcal{B}}_{n\eta} + \theta\,\epsilon\,\frac{\xi_0}{\xi_+ - \xi_-}\,\boldsymbol{\mathcal{E}}_{n\eta} + \mathcal{O}(\theta^2), \tag{79}$$

where the EM field mode vectors without the $\theta$ superindex are the EM modes of 0-ED.
The explicit form of the modified modes of the Θ-ED for a TI slab, and its corresponding first order expansion, comprise one of the important results of this work. In the literature, usually when dealing with field propagation in TI waveguides (slabs included) it has been assumed that all possible solutions for the fields can be built from the modes of the 0-ED with amplitudes that vary and acquire a $\theta$-correction, and also a dependence on the longitudinal coordinate $z$, as in [44] . What we have found here is that not only are the amplitudes modified, but also the modes, i.e., the base of solutions is different. Of course, the expectation is that the $\theta$-effects are minute, however, the field structure differs at a fundamental level and in the $\theta \to 0$ limit we recover the results of 0-ED.

### C. Coupling 0-ED external mode to the slab in terms of exact Θ-ED modes, expanded as 0-ED modes in powers of $\theta$

Now we can expand the Θ-ED modes in terms of the 0-ED modes using Eqs. (78) and (79). But also one has to take into account the normalization of each mode, demanding an expansion of the normalization factors also. Close inspection of Eq. (61) shows the first correction to $\mathcal{N}^{\theta}_r$ is of order $\theta^2$, because the $\mathcal{O}(\theta)$ contributions indeed cancel each other. Therefore, the first-order in $\theta$ contribution to $A^{\theta}_{rl}$ come from Eqs. (78) and (79) yielding:

$$A^{\theta}_{rl} \approx \delta_{lr} + \theta\sigma_{lr}, \qquad \text{where,} \qquad \sigma_{lr} \equiv \frac{\xi_0}{\xi_+ - \xi_-}\frac{c_0}{4\pi}\int \frac{1}{4\mu}[\epsilon \mathbf{E}_l \times \mathbf{E}^*_r - \frac{1}{\mu}\mathbf{B}^*_r \times \mathbf{B}_l]\cdot\hat{\mathbf{z}}dx = -\sigma^*_{rl}. \tag{80}$$

Thus, an arbitrary EM field, with the harmonic time dependence omitted, that propagates in the slab having "excited" the waveguide at $z = 0$ with the $l$-th mode of 0-ED is expressed as:

$$\mathbf{E}_l(\mathbf{r}) = \sum_r A^{\theta}_{rl}\mathbf{E}^{\theta}_r(\mathbf{r}_\perp)e^{ik^{\theta}_r z} \approx \sum_r [(\delta_{lr} + \theta\sigma_{lr})\mathbf{E}_r - \theta\delta_{lr}\frac{1}{\mu}\frac{\xi_0}{\xi_+ - \xi_-}\mathbf{B}_r]e^{ik_r z} \tag{81}$$

And similarly,

$$\mathbf{B}_l(\mathbf{r}) = \sum_r A^{\theta}_{rl}\mathbf{B}^{\theta}_r(\mathbf{r}_\perp)e^{ik^{\theta}_r z} \approx \sum_r [(\delta_{lr} + \theta\sigma_{lr})\mathbf{B}_r + \theta\delta_{lr}\epsilon\frac{\xi_0}{\xi_+ - \xi_-}\mathbf{E}_r]e^{ik_r z} \tag{82}$$

where in the expressions of the rhs we have omitted the dependence on the transverse coordinates $\mathbf{r}_\perp$.

$$\begin{aligned} P^{\theta}_{+;\mathrm{TEe}} &= \frac{1}{\sqrt{\mathcal{N}_+}}\frac{c_0}{4\pi}\int \frac{1}{2\mu}\mathrm{Re}\,[(\frac{1}{\sqrt{\mathcal{N}_+}}[\boldsymbol{\mathcal{E}}_+ \times \boldsymbol{\mathcal{B}}^*_+ + \theta\frac{\xi_0}{\xi_+ - \xi_-}(\epsilon\boldsymbol{\mathcal{E}}_+ \times \boldsymbol{\mathcal{E}}^*_+ - \frac{1}{\mu}\boldsymbol{\mathcal{B}}_+ \times \boldsymbol{\mathcal{B}}^*_+)] + \frac{1}{\sqrt{\mathcal{N}_-}}\theta\sigma^*_{+-}\boldsymbol{\mathcal{E}}_+ \times \boldsymbol{\mathcal{B}}^*_- e^{i(k_+ - k_-)z})]\cdot\hat{\mathbf{z}}dx \\ &= \frac{1}{\sqrt{\mathcal{N}_+}}\Big\{\sqrt{\mathcal{N}_+} + \theta\frac{c_0}{4\pi}\int \frac{1}{2\mu}\mathrm{Re}\,[\frac{1}{\sqrt{\mathcal{N}_-}}\sigma^*_{+-}\boldsymbol{\mathcal{E}}_+ \times \boldsymbol{\mathcal{B}}^*_- e^{i(k_+ - k_-)z}]\cdot\hat{\mathbf{z}}dx\Big\} \\ &= 1 + \theta\frac{1}{\sqrt{\mathcal{N}_+\mathcal{N}_-}}\frac{c_0}{4\pi}\int \frac{1}{2\mu}\mathrm{Re}\,[\sigma^*_{+-}\boldsymbol{\mathcal{E}}_+ \times \boldsymbol{\mathcal{B}}^*_- e^{i(k_+ - k_-)z}]\cdot\hat{\mathbf{z}}dx\Big\} \end{aligned} \tag{83}$$

And similarly, $P^{\theta}_{-;\text{TEe}}$

$$P^{\theta}_{-;\text{TEe}} = \theta\frac{1}{\sqrt{\mathcal{N}_-}}\frac{c_0}{4\pi}\int\frac{1}{2\mu}\text{Re}\,[\sum_r\frac{1}{\sqrt{\mathcal{N}_r}}\sigma_{+-}\delta_{+r}\boldsymbol{\mathcal{E}}_-\times\boldsymbol{\mathcal{B}}^*_r e^{i(k_--k_+)z}]\cdot\hat{\mathbf{z}}dx$$
$$= \theta\frac{1}{\sqrt{\mathcal{N}_+\mathcal{N}_-}}\frac{c_0}{4\pi}\int\frac{1}{2\mu}\text{Re}\,[\sigma_{+-}\boldsymbol{\mathcal{E}}_-\times\boldsymbol{\mathcal{B}}^*_+ e^{i(k_--k_+)z}]\cdot\hat{\mathbf{z}}dx \tag{84}$$

We observe that the power transmitted by the EM field in the $\Theta$-slab, given the external field was the TE-even mode of 0-ED, and having assumed a frequency such that only the first two modes are available, is $1+\theta\delta^{\theta}P_{\text{TEe}}+\mathcal{O}(\theta^2)$, where $\delta^{\theta}P_{\text{TEe}}$ expresses a correction to the power due to the $\Theta$-slab, terms that can be read off from Eqs. (83) and (84). Seemingly, these contributions to the power carried by each $\Theta$-ED mode (the $0+$ and $0-$) are dependent on the coordinate $z$, however, the integrals in these equations vanish identically. The computation entails the evaluation of the $\sigma_{\pm,\mp}$ coefficients above, the vectorial expressions of the 0-ED modes and their corresponding cross products. It is clearly cumbersome and thus have been done in the computer (Mathematica). Reassuring as this may seem, it is only a reflection of the fact that the first order correction to the powers come from $\mathcal{O}(\theta^2)$ contributions. Hence, to make a consistent expansion one must therefore expand the field themselves to $\mathcal{O}(\theta^2)$ too. This is even more elaborated. It amounts to doing an $\mathcal{O}(\theta^2)$ expansion in Eqs. (78) and (79), to then substitute in Eqs. (81) and (82) and carry along the computation of the powers. Doing the computation numerically shows that the $\mathcal{O}(\theta^2)$ contributions to the powers $P^{\theta}_{+;\text{TEe}}$ and $P^{\theta}_{-;\text{TEe}}$ do not vanish, and they do depend on $z$. Hence we can summarize by saying that, having excited the $\Theta$-slab with the TEe external mode, the power transmitted by the $\Theta$-ED fields, expanded to $\mathcal{O}(\theta^2)$ is:

$$P^{\theta}_{\text{TEe}} = P^{\theta}_{+;\text{TEe}} + P^{\theta}_{-;\text{TEe}} = 1 + 0\cdot\theta + \theta^2\left(P^{\theta}_{+} + P^{\theta}_{-}\right) + \mathcal{O}(\theta^3), \tag{85}$$

where the sum in brackets is independent of $z$. In retrospect, the fact that the first order corrections to the power come from the $\sim\theta^2$ terms traces back to symmetry arguments. We know that a power must transform under time-reversal (TR) as the Poynting vector, which is odd. Now $\theta$ is odd under TR and so is $\mathbf{B}$, while $\mathbf{E}$ is even, therefore terms like $\theta(\boldsymbol{\mathcal{E}}\times\boldsymbol{\mathcal{B}}^*)$ like the ones in Eqs. (83) and (84) are TR even and hence forbidden. Interestingly, there are other terms contributing linearly in $\theta$, the ones shown in the middle parenthesis in the first line of Eq. (83), that, although not forbidden by symmetry arguments, do vanish due to their explicit vectorial nature.

Finally, to put the two approaches in perspective, we summarize the results in this section. We started writing an arbitrary EM field as a superposition of exact $\Theta$-ED modes. Then we ask how the amplitudes evolve for each of the allowed modes, given that the input on the slab was one 0-ED mode (in particular the TE-even). To fix the idea, we consider an operational frequency allowing only the first two modes[4]. We argued that the exact results are consistent with energy conservation and reflect the hybridization of modes, transferring power from one mode to the other as the wave propagates, but keeping the total power independent of $z$. In the perturbative approach, which is similar to coupled-mode theory, however, we expanded the exact $\Theta$-ED modes in powers of $\theta$ in terms of the 0-ED modes. Here the total power is only consistent with energy conservation up to $\mathcal{O}(\theta)$, but fails to do so up to higher orders. For example, in [44], a coupled-mode theory is used. Nevertheless, the comparison is in order. In that work, the author finds a power transfer between modes that is linear in $\theta$ but depending on $z$. However, in there the $\mathcal{O}(\theta^2)$ contributions to the wave equations of $\Theta$-ED (see Eqs. (18) and (19)) are discarded from the very beginning, which is different to what we do here.

## V. DISPERSION RELATIONS

The dispersion relation is a fundamental physical relation in any propagating wave solution. In fact, phase velocity, group velocity and other properties are determined from it.

The allowed $k_{n\eta}$ values, for a given frequency, cannot be determined analytically, because they are defined via transcendental equations. Therefore, we have to solve them numerically. Let us first consider finding the values of $u_{n\eta}$ and $v_{n\eta}$ allowed. For this we need to find the intersection values of the determinant in Eq. (44) with that of the wavenumbers, Eq. (40), by rearranging we have that $u_{n\eta}$ and $v_{n\eta}$ satisfy the following equation as a function of frequency

$$u^2 + v^2 = \frac{\omega^2}{c_0^2}L^2(n_2^2 - n_1^2) \equiv \mathcal{R}(\omega)^2, \tag{86}$$

which corresponds, in the $(u, v)$ coordinates, to a circle of radius $\mathcal{R}(\omega)$, and the $n\eta$ label for the polarizations have been omitted. For all numerical calculations we will use the speed of light in vacuum as $c_0 = 299792458$ m/s. Let us begin by considering the

---

[4] Recall, that for a given frequency the amounts of modes are the same, regardless if we label them as 0-ED or $\Theta$-ED modes. Specifically here, there are: TE-e and TM-e in the language of the 0-ED or the $0+$ and $0-$ in the language of the $\Theta$-ED modes

first two modes, which correspond to a maximum radius of $\mathcal{R}^* = \pi/2$. This means that the operating frequency, $f = \omega/2\pi$, is only allowed in the range of $0 < f < f_0$, with $f_0 = c_0/\left(4L\sqrt{n_2^2 - n_1^2}\right)$. For the other modes with $\mathcal{R}^* = \pi, 3\pi/2, 2\pi, \ldots$ the situation can be represented in Table V:

| $\mathcal{R}^*$ | Operational frequency $f$ [Hz] | n° of even modes | n° of odd modes | n° of total modes |
|---|---|---|---|---|
| $\pi/2$ | $0 < f < f_0$ | 2 | 0 | 2 |
| $\pi$ | $f_0 < f < 2f_0$ | 2 | 2 | 4 |
| $3\pi/2$ | $2f_0 < f < 3f_0$ | 4 | 2 | 6 |
| $2\pi$ | $3f_0 < f < 4f_0$ | 4 | 4 | 8 |

TABLE V: Even and odd modes and allowed operational frequency range, for given maximum $\mathcal{R}^*$.

In general, the frequency $f_0$ depends on the waveguide width $d = 2L$, however, we must be careful with the operational frequency because for Θ-ED to be a valid description of TI's response to the EM field we must ensure that the energy carried by the EM field, $E_\gamma$, does not exceed the (typical) bandgaps of TIs, which are in the order of $E_g \sim 0.3$ eV [10]. This corresponds to a frequency of $\sim 75$ THz. Plausible slab thicknesses and operational frequencies are shown in Table VI.

| Slab thickness | $0 < f < f_0$ | $E_\gamma = hf$ |
|---|---|---|
| $d = 1$ cm | $0 < f < 5$ GHz | $0 < E_\gamma < 20.7\ \mu$eV |
| $d = 1$ mm | $0 < f < 50$ GHz | $0 < E_\gamma < 207\ \mu$eV |
| $d = 100\ \mu$m | $0 < f < 500$ GHz | $0 < E_\gamma < 2.07$ meV |
| $d = 10\ \mu$m | $0 < f < 5$ THz | $0 < E_\gamma < 20.7$ meV |
| $d = 1\ \mu$m | $0 < f < 50$ THz | $0 < E_\gamma < 207$ meV |
| $d = 10$ nm | $0 < f < 5$ PHz | $0 < E_\gamma < 20.7$ eV |

TABLE VI: Examples of slab thicknesses and operational frequencies consistent with EM response of Θ-ED, for $n_2^2 - n_1^2 = 4$.

In many practical applications, only a few modes are preferred, e.g., in order to reduce interference and noise. Therefore, the above information allows us to determine the slab thickness and/or the optical properties of the TI and cladding, according to the number of modes desired and complying with the bound for the operational frequency imposed by the validity of the Θ-ED.

### A. Numerical solution for $u$ and $v$.

The following steps lead to the numerical equation to be solved in the case of the first two modes $(0, \pm)$. To simplify the notation we delete the subindices $n\eta$. The starting point is Eq. (86) in which, on the left-hand-side, we use Eq. (69) to express $v$ and on the right-hand-side we use Eq. (20) with $\mu\epsilon = n_2^2$ together with $u = k_x L$ and $\kappa = k_z L$, to arrive at:

$$u^2 + \left(\xi_\pm^\theta u \tan u\right)^2 = \frac{(n_2^2 - n_1^2)}{n_2^2}(u^2 + \kappa^2), \tag{87}$$

which is the numerical equuation to be solved for $u = u(\kappa)$, with the remaining coefficients being function of the topo-optical parameters. Analogously we can solve numerically for $v = v(\kappa)$.

TIs with widths of $d = 10$ nm has been possible, i.e., $L = 5$ nm. As the $f_0$ depends on the optical properties, we will make a set of plots with different values of $\epsilon_1$, $\epsilon_2$, $\mu_1$ and $\mu_2$. For all the following figures we will consider the following labels: The colors blue, yellow, green and red correspond to the modes: $(0, +)$, $(0, -)$, $(1, +)$ and $(1, -)$ in increasing order respectively.

Continuous, semi-dashed, dashed, and dotted lines correspond to $\tilde{\theta}_{AP} = 0$, $\tilde{\theta}_{AP} = 1$, $\tilde{\theta}_{AP} = 2$ and $\tilde{\theta}_{AP} = 3$ respectively. Figure 6 shows $u$ as a function of $\kappa$. As $\lambda_x = 2\pi L/u$ and $\lambda_z = 2\pi L/\kappa$, we observe that the upper right sectors correspond to small wavelengths in the $x$ and $z$ directions. The plots have been made up to a given fixed frequency, and for that frequency the $u$'s of the different modes have support in decreasing $\kappa$ domains for increasing modes. This is why the plots seem to cut-off at ever increasing $\kappa$ the higher the mode.

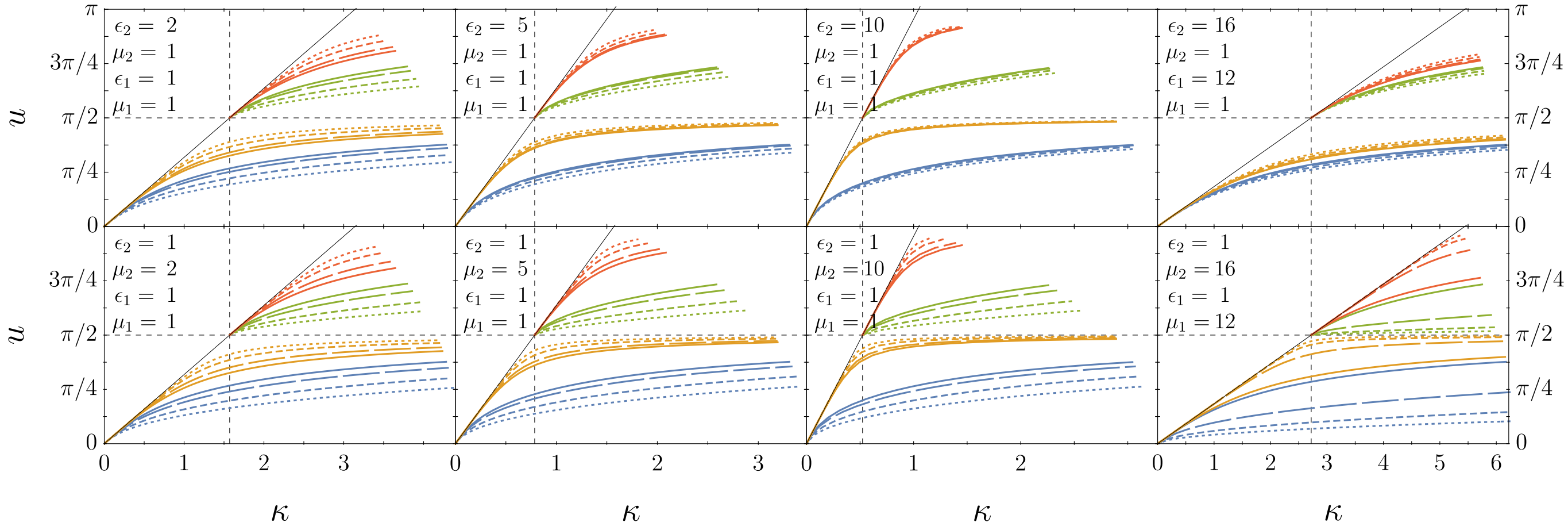

FIG. 6: $u(\kappa)$ for different optical values of the system. The colors blue, yellow, green and red correspond to the modes: $(0,+)$, $(0,-)$, $(1,+)$ and $(1,-)$ in increasing order respectively. Continuous, semi-dashed, dashed, and dotted lines correspond to $\theta_{\text{TI}} = 0$, $\theta_{\text{TI}} = 67\pi$, $\theta_{\text{TI}} = 137\pi$ and $\theta_{\text{TI}} = 205\pi$ respectively. The horizontal line corresponds to the value $u = \pi/2$. For $\kappa \geq \tilde{\kappa} = \frac{\pi}{2}\sqrt{\frac{\mu_1\epsilon_1}{\mu_2\epsilon_2 - \mu_1\epsilon_1}}$, shown with a vertical line, the next two modes $(1+)$ and $(1-)$ appear. The horizontal and vertical lines through the point $u = \pi/2, \kappa = \tilde{\kappa}$ divide the figure into four sectors.

Figure 7 shows $v$ as a function of $\kappa$. As $v$ is related to the penetration length $v = \frac{1}{\ell_x}$, the upper right sectors correspond to modes with small penetration lengths in $x$, i,e., into the claddigns.

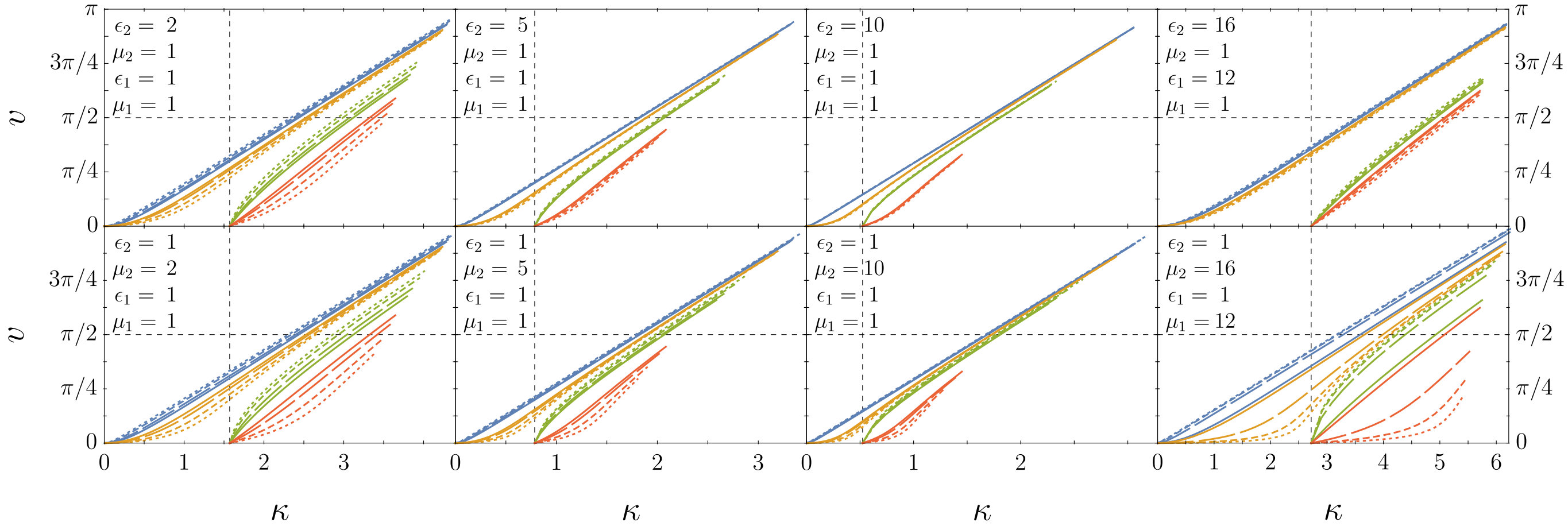

FIG. 7: $v(\kappa)$ for different optical values of the system. The colors blue, yellow, green and red correspond to the modes: $(0,+)$, $(0,-)$, $(1,+)$ and $(1,-)$ in increasing order respectively. Continuous, semi-dashed, dashed, and dotted lines correspond to $\theta_{\text{TI}} = 0$, $\theta_{\text{TI}} = 67\pi$, $\theta_{\text{TI}} = 137\pi$ and $\theta_{\text{TI}} = 205\pi$ respectively. The vertical line at $\kappa = \tilde{\kappa}$ is the same as in Fig. 6, to the right of which the next two modes appear.

### B. Numerical dispersion relations

Having obtained $u$ and $v$ numerically, we use Eq. (86) to determine $\omega$. In Fig. 8, we show the dimensionless dispersion relations by plotting $L\omega(\kappa)/c_0$ as a function of $\kappa$.

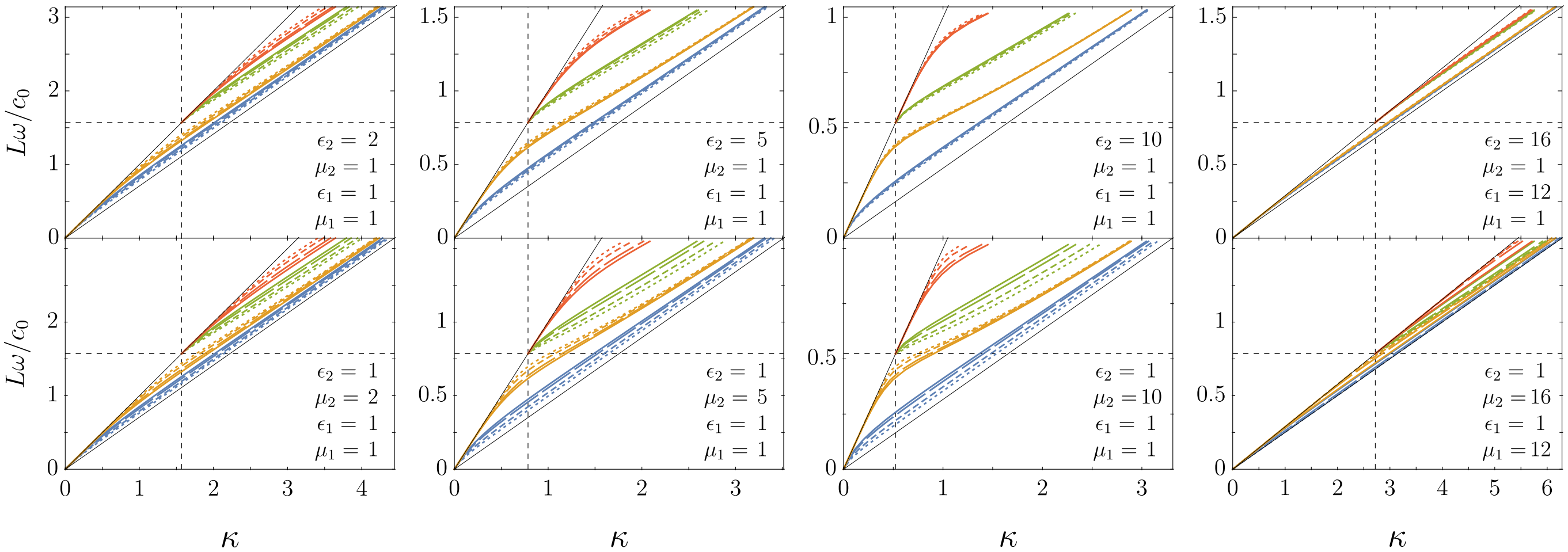

FIG. 8: $L\omega(\kappa)/c_0$ for different optical values of the system. The colors blue, yellow, green and red correspond to the modes: $(0,+)$, $(0,-)$, $(1,+)$ and $(1,-)$ in increasing order respectively. Continuous, semi-dashed, dashed, and dotted lines correspond to $\theta_{\rm TI}=0$, $\theta_{\rm TI}=67\pi$, $\theta_{\rm TI}=137\pi$ and $\theta_{\rm TI}=205\pi$ respectively. The horizontal line corresponds to $L\omega_0/c_0=2\pi L f_0/c_0=\pi/2\sqrt{n_2^2-n_1^2}$. The linear dispersion relations of electromagnetic waves in homogeneous medium 1 and 2, i.e., $\omega=c_0k_z/n_1$ and $\omega=c_0k_z/n_2$ can be seen by the plots of constant minimum and maximum slopes, shown in thin black lines. The vertical line at $\kappa=\tilde{\kappa}$ is the same as in Fig. 6, to the right of which the next two modes appear.

### C. Numerical phase and group velocities

In Fig. 9, we show the dimensionless phase velocity $v_p/c_0$ in the TI slab. This is defined as follows,

$$\frac{v_p}{c_0}=\frac{\omega}{c_0 k}=\frac{1}{n_2}\sqrt{1+\frac{u^2}{\kappa^2}}. \tag{88}$$

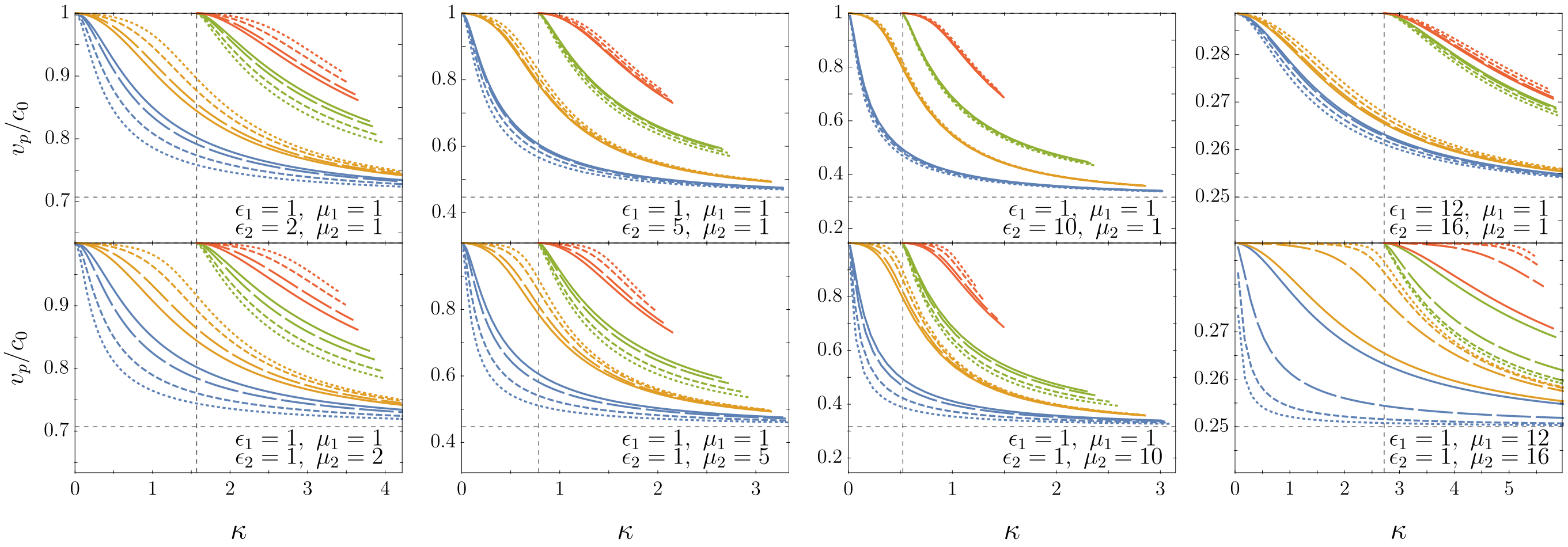

FIG. 9: $v_p/c_0$ for different optical values of the system. The colors blue, yellow, green and red correspond to the modes: $(0,+)$, $(0,-)$, $(1,+)$ and $(1,-)$ in increasing order respectively. Continuous, semi-dashed, dashed, and dotted lines correspond to $\theta_{\rm TI}=0$, $\theta_{\rm TI}=67\pi$, $\theta_{\rm TI}=137\pi$ and $\theta_{\rm TI}=205\pi$ respectively. The horizontal line represents the asymptotic value of the phase velocity in the slab $v_{p2}/c_0=1/n_2$. The vertical line at $\kappa=\tilde{\kappa}$ is the same as in Fig. 6, to the right of which the next two modes appear.

From the definition of the group velocity, and some algebra we find:

$$\frac{v_g}{c_0} = \frac{1}{c_0}\frac{\partial \omega}{\partial k_z} = \frac{c_0}{v_p}\left(\frac{u + v\partial_u v}{n_1^2 u + n_2^2 v \partial_u v}\right). \tag{89}$$

In Fig. 10, we plot the group velocities $v_g/c_0$ as a function of $\kappa$ and for different optical parameters $n_1, n_2$.

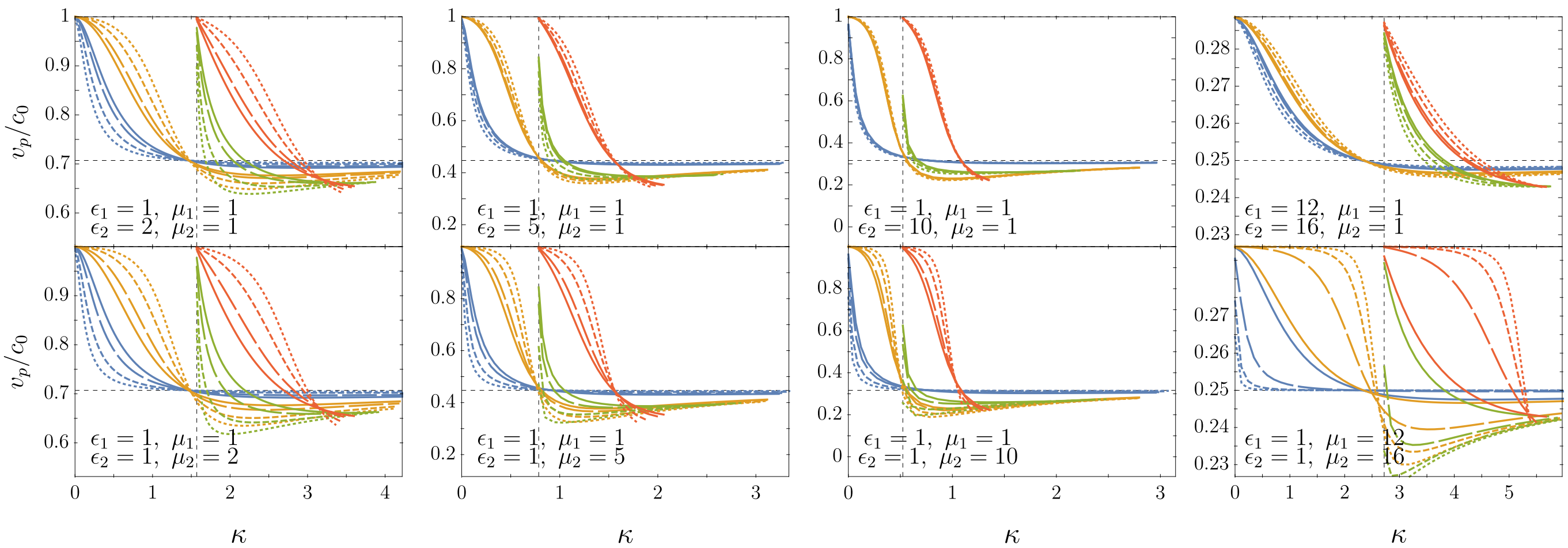

FIG. 10: $v_g/c_0$ for different optical values of the system. The colors blue, yellow, green and red correspond to the modes: $(0,+)$, $(0,-)$, $(1,+)$ and $(1,-)$ in increasing order respectively. Continuous, semi-dashed, dashed, and dotted lines correspond to $\theta_{\text{TI}} = 0$, $\theta_{\text{TI}} = 67\pi$, $\theta_{\text{TI}} = 137\pi$ and $\theta_{\text{TI}} = 205\pi$ respectively. The horizontal lines are the group velocities as if medium 2 were non dispersive, i.e., $v_g/c_0 = 1/n_2$. The vertical line at $\kappa = \tilde{\kappa}$ is the same as in Fig. 6, to the right of which the next two modes appear.

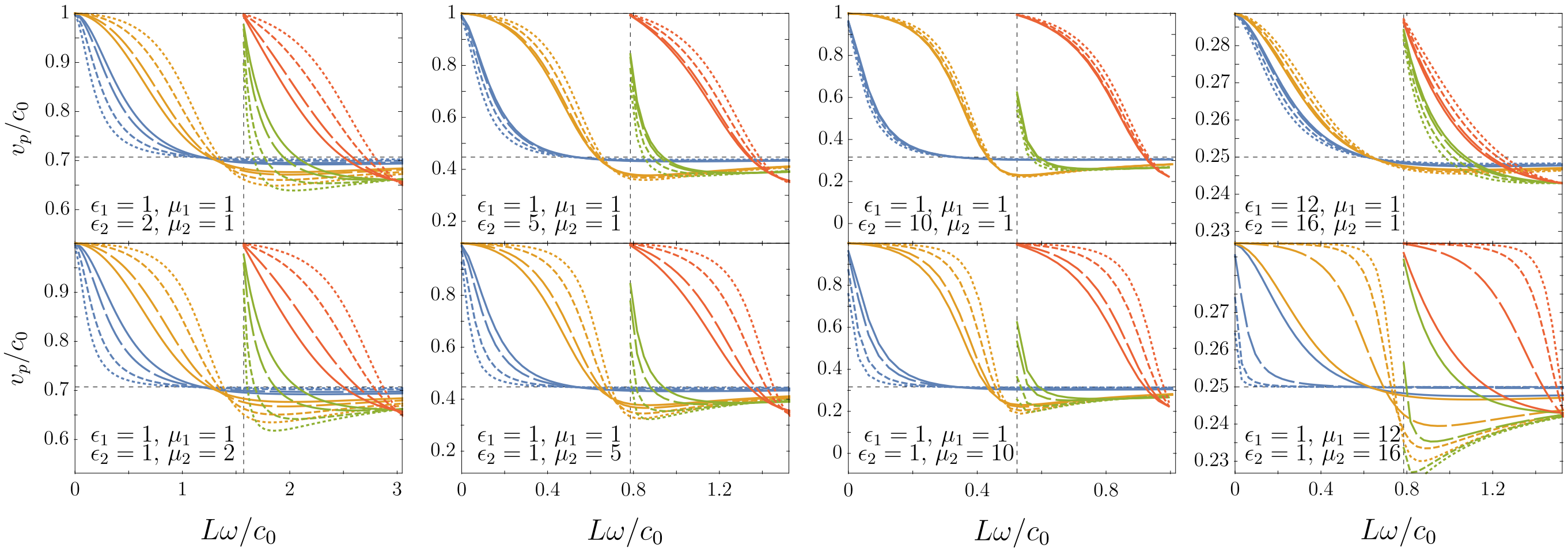

FIG. 11: Dimensionless group velocities as a function of the dimensionless EM wave frequency for different optical values of the system. The colors blue, yellow, green and red correspond to the modes: $(0,+)$, $(0,-)$, $(1,+)$ and $(1,-)$ in increasing order respectively. Continuous, semi-dashed, dashed, and dotted lines correspond to $\theta_{\text{TI}} = 0$, $\theta_{\text{TI}} = 67\pi$, $\theta_{\text{TI}} = 137\pi$ and $\theta_{\text{TI}} = 205\pi$ respectively. The horizontal lines are the group velocities as if medium 2 were non dispersive, i.e., $v_g/c_0 = 1/n_2$. The vertical line at $\kappa = \tilde{\kappa}$ is the same as in Fig. 6, to the right of which the next two modes appear.

## VI. SOLUTION VIA COUPLED-MODE THEORY

We will consider electromagnetic waves propagating in the $\hat{\mathbf{z}}$-direction within a dielectric waveguide with an arbitrary transverse geometry composed of topological materials (TIs). We will assume that the $\Theta$-term induces a small perturbation such that the EM fields are expanded in the "base" modes of the 0-ED with amplitudes that acquire a $\Theta$ modification and also a dependence on the longitudinal coordinate $z$.

### A. Helmholtz Equations

Taking the curl of the vectorial Maxwell equations (11) and (12), we find the following differential equations:

$$(\nabla^2 + k_0^2\mu\epsilon)\mathbf{E} + \frac{1}{\epsilon}\boldsymbol{\nabla}(\boldsymbol{\nabla}\Theta\cdot\mathbf{B}) + ik_0\mu(\boldsymbol{\nabla}\Theta\times\mathbf{E}) = 0, \tag{90}$$

$$(\nabla^2 + k_0^2\mu\epsilon)\mathbf{B} + \mu\boldsymbol{\nabla}\times(\boldsymbol{\nabla}\Theta\times\mathbf{E}) = 0, \tag{91}$$

where $k_0 = \omega/c_0$ is the dispersion relation in vacuum. These equations correspond to Helmholtz equations with modifications proportional to $\boldsymbol{\nabla}\Theta$, where we have assumed $\dot{\Theta} = 0$. As we are considering a perturbation in the EM fields of the usual electrodynamics, we expand the fields in the allowed modes of the usual ($\Theta = 0$) dielectric waveguide,

$$\mathbf{E}_{\text{tot}} = \sum_r A_r(z)\mathbf{E}_r(\mathbf{r}_\perp)e^{ik_r z}, \qquad \mathbf{B}_{\text{tot}} = \sum_r A_r(z)\mathbf{B}_r(\mathbf{r}_\perp)e^{ik_r z}, \tag{92}$$

where $\mathbf{E}_r(\mathbf{r}_\perp) = \mathbf{E}_r^{\theta=0} \equiv \mathbf{E}_r$ denotes the normalized transverse profile of mode $r = n\eta$ defined previously in the Eq. (62). The summation runs over modes $r = \{0+\}, \{0-\}, \{1+\}, \ldots, \{n\eta\}, \ldots, M(\omega)$ (where $M(\omega)$ is the number of modes supported at the operating angular frequency $\omega$), and $A_r(z)$ are the corresponding dimensionless amplitudes varying along the longitudinal coordinate $z$. Substituting Eq. (92) into Eqs. (90) and (91), and assuming $\Theta(\mathbf{r}) = \Theta(\mathbf{r}_\perp)$, we obtain:

$$\sum_r \left\{(\partial_z^2\tilde{A}_r + k_r^2\tilde{A}_r)\mathbf{E}_r + \tilde{A}_r\left(\frac{1}{\epsilon}\nabla_\perp(\nabla_\perp\Theta\cdot\mathbf{B}_r) + ik_0\mu\nabla_\perp\Theta\times\mathbf{E}_r\right) + \partial_z\tilde{A}_r\frac{1}{\epsilon}\hat{\mathbf{z}}(\nabla_\perp\Theta\cdot\mathbf{B}_r)\right\} = 0, \tag{93}$$

$$\sum_r \left\{(\partial_z^2\tilde{A}_r + k_r^2\tilde{A}_r)\mathbf{B}_r + \tilde{A}_r\mu\nabla_\perp\times(\nabla_\perp\Theta\times\mathbf{E}_r) + \partial_z\tilde{A}_r\mu\hat{\mathbf{z}}\times(\nabla_\perp\Theta\times\mathbf{E}_r)\right\} = 0, \tag{94}$$

where $\tilde{A}_r \equiv A_r(z)e^{ik_r z}$. We have assumed that the profiles of the EM fields satisfy the Helmholtz equation of the $\Theta = 0$ waveguide:

$$(\boldsymbol{\nabla}_\perp^2 + k_0^2\mu\epsilon - k_r^2)\mathbf{E}_r = 0, \qquad (\boldsymbol{\nabla}_\perp^2 + k_0^2\mu\epsilon - k_r^2)\mathbf{B}_r = 0, \tag{95}$$

These electromagnetic profiles satisfy the BCs of the conventional dielectric waveguide, along with their corresponding transcendental equations that define the modes. Therefore, we have all the information regarding the electromagnetic profiles. To find a particular $\tilde{A}_s$ we apply the normalized Lorentz reciprocity theorem,

$$\frac{c_0}{4\pi}\int_{\mathbb{R}^2}\frac{1}{4\mu}\left[\mathbf{E}_r\times\mathbf{B}_s^* + \mathbf{E}_s^*\times\mathbf{B}_r\right]\cdot\hat{\mathbf{z}}da_\perp = \delta_{rs}. \tag{96}$$

To this end we compute: $\frac{c_0}{4\pi}\int\frac{1}{4\mu}[(93)\times\mathbf{B}_s^* + \mathbf{E}_s^*\times(94)]\cdot\hat{\mathbf{z}}da_\perp$, leading to:

$$\partial_z^2\tilde{A}_s + k_s^2\tilde{A}_s + \sum_r(\kappa_{rs}\tilde{A}_r + \sigma_{rs}\partial_z\tilde{A}_r) = 0, \tag{97}$$

where,

$$\kappa_{rs} = \frac{c_0}{16\pi}\int\left[ik_0E_{zr}\nabla_\perp\Theta\cdot\mathbf{B}_s^* + \frac{1}{\mu\epsilon}\mathbf{B}_s^*\cdot(\hat{\mathbf{z}}\times\nabla_\perp(\nabla_\perp\Theta\cdot\mathbf{B}_r)) + \mathbf{E}_s^*\cdot\nabla_\perp(\nabla_\perp\Theta\cdot(\hat{\mathbf{z}}\times\mathbf{E}_r))\right]da_\perp, \tag{98}$$

$$\sigma_{rs} = \frac{c_0}{16\pi}\int\left[E_{zr}\nabla_\perp\Theta\cdot(\hat{\mathbf{z}}\times\mathbf{E}_s^*)\right]da_\perp. \tag{99}$$

Equation (97), together with the initial conditions, enables us to determine the behavior of the amplitude of a specific mode along the $\hat{\mathbf{z}}$ direction due to perturbations induced by the term $\boldsymbol{\nabla}_\perp\Theta$. By setting $\boldsymbol{\nabla}_\perp\Theta = 0$, the amplitudes satisfy $\partial_z^2\tilde{A}_s + k_s^2\tilde{A}_s = 0$, whose solution is given by $\tilde{A}_s = A_{s1}e^{ik_s z} + A_{s2}e^{-ik_s z}$. Since we are only interested in solutions corresponding to waves propagating in the $+\hat{\mathbf{z}}$ direction, we set $A_{s2} = 0$. Given that $\tilde{A}_s = A_s(z)e^{ik_s z}$, the amplitude $A_s(z) = A_{s1}$ remains constant along the entire $\hat{\mathbf{z}}$ direction for each mode of the conventional electromagnetic field. It is worth noting that Eq. (97) holds for a waveguide with arbitrary geometry. Subsequently, we will consider a slab waveguide stacked in the $x$-direction.

### B. Θ-Slab

For an EM wave propagating in a slab waveguide stacked in the $x$-direction with a geometry as shown in Fig. (1), and discontinuities in Θ in the $x$-direction $\Theta(\mathbf{r}_\perp) = \Theta(x)$, the profiles of the EM field modes, defined as solutions of the system in the absence of Θ discontinuities, are given by:

$$\mathbf{E}_{n+} = E_{yn+}\hat{\mathbf{y}}, \qquad \mathbf{B}_{n+} = B_{xn+}\hat{\mathbf{x}} + B_{zn+}\hat{\mathbf{z}}, \tag{100}$$

$$\mathbf{E}_{n-} = E_{xn-}\hat{\mathbf{x}} + E_{zn-}\hat{\mathbf{z}}, \qquad \mathbf{B}_{n-} = B_{yn-}\hat{\mathbf{y}}, \tag{101}$$

where the subscripts $\{n+\}$ and $\{n-\}$ correspond to the TE and TM modes reported in the literature [1]. These vectorial structures are important to determine the coupling between the modes. Replacing Eqs. (100) and (101) in Eqs. (98) and (99) we observe that some combinations of modes do not contribute to Eq. (97). In particular,

$$\kappa_{n'+,n+} = \kappa_{n'-,n-} = \sigma_{n'+,n-} = \sigma_{n'+,n+} = \sigma_{n'-,n-} = 0. \tag{102}$$

For example, in the case of TE-TE coupling, we have that $E_{zn+} = \mathbf{E}_{n+} \cdot \hat{\mathbf{z}} = 0$ and similarly $B_{yn+} = \mathbf{B}_{n+} \cdot \hat{\mathbf{y}} = 0$. Thus, in Eq. (98) the second and third terms vanish due to: the geometry of the TE fields above, the cross products, and that $\boldsymbol{\nabla}_\perp \Theta \parallel \hat{\mathbf{x}}$. For the other couplings, similar analyses apply. With this in consideration, in Eq. (97) when summing in the TE, TM modes, we have:

$$(\partial_z^2 + k_{n+}^2)\tilde{A}_{n+} + \sum_{n'} (\kappa_{n'-,n+}\tilde{A}_{n'-} + \sigma_{n'-,n+}\partial_z \tilde{A}_{n'-}) = 0, \tag{103}$$

$$(\partial_z^2 + k_{n-}^2)\tilde{A}_{n-} + \sum_{n'} \kappa_{n'+,n-}\tilde{A}_{n'+} = 0, \tag{104}$$

where,

$$\kappa_{n'+,n-} \equiv \frac{c_0}{16\pi} \int_{-\infty}^{\infty} (E_{yn'+}\partial_x E_{xn-}^* - \frac{1}{\epsilon\mu} B_{xn'+}\partial_x B_{yn-}^*)\, \partial_x \Theta \, dx, \tag{105}$$

$$\kappa_{n'-,n+} \equiv \frac{c_0}{16\pi} i k_0 \int_{-\infty}^{\infty} E_{zn'-} B_{xn+}^* \, \partial_x \Theta \, dx, \tag{106}$$

$$\sigma_{n'-,n+} \equiv -\frac{c_0}{16\pi} \int_{-\infty}^{\infty} E_{zn'-} E_{yn+}^* \, \partial_x \Theta \, dx. \tag{107}$$

To simplify, we can express the transversal components in terms of the longitudinal ones,

$$\kappa_{n'+,n-} = \frac{c_0}{16\pi} \frac{k_0}{\gamma_{n'+}^2} (k_{n'+} + k_{n-}) \int_{-\infty}^{\infty} \partial_x B_{zn'+} E_{zn-}^* \, \partial_x \Theta \, dx, \tag{108}$$

$$\kappa_{n'-,n+} = \frac{c_0}{16\pi} \frac{k_0 k_{n+}}{\gamma_{n+}^2} \int_{-\infty}^{\infty} E_{zn'-} \partial_x B_{zn+}^* \, \partial_x \Theta \, dx, \tag{109}$$

$$\sigma_{n'-,n+} = -\frac{c_0}{16\pi} \frac{i k_0}{\gamma_{n+}^2} \int_{-\infty}^{\infty} E_{zn'-} \partial_x B_{zn+}^* \, \partial_x \Theta \, dx. \tag{110}$$

Integrating we get,

$$\kappa_{n'-,n+} = \frac{k_{n+}}{\gamma_{n+}} \tilde{\kappa}_{n'-,n+}, \qquad \sigma_{n'-,n+} = -\frac{i}{\gamma_{n+}} \tilde{\kappa}_{n'-,n+}, \qquad \kappa_{n'+,n-} = \frac{(k_{n'+} + k_{n-})}{\gamma_{n'+}} \tilde{\kappa}_{n-,n'+}^* \tag{111}$$

with,

$$\tilde{\kappa}_{n'-,n+} = \frac{c_0 k_0}{16\pi} \left[ \tilde{\theta}_1 E_{zn'-}(-L) \partial_x B_{zn+}^*(-L) + \tilde{\theta}_2 E_{zn'-}(L) \partial_x B_{zn+}^*(L) \right], \qquad \text{and} \qquad \tilde{\theta}_i = \Theta_{i+1} - \Theta_i. \tag{112}$$

### C. Parallel and antiparallel couplings

At each Θ-interface we have several options. We call $\tilde{\theta}_1 = -\tilde{\theta}_2 = \tilde{\theta}_{AP}$ the antiparallel configuration and $\tilde{\theta}_1 = \tilde{\theta}_2 = \tilde{\theta}_P$ the parallel configuration. With this we can write a $\tilde{\kappa}$ for each configuration:

$$\tilde{\kappa}_{n'-,n+}^{(AP)} = \frac{c_0 k_0}{16\pi} \tilde{\theta}_{AP} \left[ E_{zn'-}(-L) \partial_x B_{zn+}^*(-L) - E_{zn'-}(L) \partial_x B_{zn+}^*(L) \right], \tag{113}$$

$$\tilde{\kappa}_{n'-,n+}^{(P)} = \frac{c_0 k_0}{16\pi} \tilde{\theta}_P \left[ E_{zn'-}(-L) \partial_x B_{zn+}^*(-L) + E_{zn'-}(L) \partial_x B_{zn+}^*(L) \right]. \tag{114}$$

Besides, each propagation mode, TE $\{n+\}$ and TM $\{n-\}$, have definite parity $(-1)^n$. Again, it is necessary to recall that the parity of the modes is determined by that of the transversal components, i.e., the spatial derivatives of the longitudinal components. In this way, the $n+$ mode will be even/odd depending on the parity of $\partial_x B_{zn+}$, thus, for the $n+$ to be even, we must have $\partial_x B_{zn+}(x) = \partial_x B_{zn+}(-x)$ and, in particular, $\partial_x B_{zn+}(L) = \partial_x B_{zn+}(-L)$. Similarly, for the $n-$ mode to be odd, we must have $\partial_x E_{zn-}(x) = -\partial_x E_{zn-}(-x)$ and therefore $E_{zn-}(x) = E_{zn-}(-x)$ and, in particular, $E_{zn-}(L) = E_{zn-}(-L)$. Of course, we must bear in mind that the previous argument is predicated on the fact that the longitudinal components are sines or cosines, hence the change in parity upon derivation. Thus, we can label the modes of the 0-ED as shown in Table VII.

| **Mode** $s$ | **Type** | $\{n\pm\}$ equivalence | **Symmetry** |
|---|---|---|---|
| 0 | TE-even-1 | 0+ | $\partial_x B_{z0+}(-L) = \partial_x B_{z0+}(L)$ |
| 1 | TM-even-1 | 0− | $E_{z0-}(-L) = -E_{z0-}(L)$ |
| 2 | TE-odd-1 | 1+ | $\partial_x B_{z1+}(-L) = -\partial_x B_{z1+}(L)$ |
| 3 | TM-odd-1 | 1− | $E_{z1-}(-L) = E_{z1-}(L)$ |

TABLE VII: Symmetries of the modes according to the TE/TM and even/odd type. The "1" in the second column alludes to the first set of four modes

Therefore, according to the symmetries of the modes shown in Table VII, and replacing in Eqs. (113) and (114), we observe that in the antiparallel configuration, $AP$, the TE-even modes couple only to the TM-even modes and the TE-odd couple only to the TM-odd modes. On the contrary, for the parallel configuration $P$, the TE-even modes couple to the TM-odd and the TE-odd couple only to the TM-even modes.

### D. Mode coupling between the lowest four modes

For simplicity of notation, in this subsection we will use the mode classification label of the first column in Table VII. Going back to Eqs. (103) and (104) we recall that these are coupled differential equations for the coefficients of the complete EM fields expanded in the 0-ED modes. Hence, in a particular configuration with given slab waveguide parameters, if we solve these equations for the expansion coefficients along the waveguide, then we can see how a given mode couples to other modes as the wave propagates. In the antiparallel configuration, the lowest TE and TM modes of the same parity couple with each other. The labeling of the modes is as defined in Table VII. The longitudinal components of the modes' profiles are:

$$B_{z0} = ia_0 \sin\gamma_0 x, \qquad E_{z1} = ia_1 \sin\gamma_1 x, \qquad B_{z2} = ia_2 \cos\gamma_2 x, \qquad E_{z3} = ia_3 \cos\gamma_3 x, \tag{115}$$

where $a_i$ are normalization coefficients. Choosing them such that the density of power per unit length transmitted along the waveguide be $1\,[W/cm]$ for each mode and $L_y = 1\,cm$, we find:

$$\begin{aligned}
a_0 &= \frac{2\mu_1\sqrt{\gamma_0^3\mu_2}}{\sqrt{k_0 k_{z0}[2(\mu_1^2\gamma_0 L + \mu_2^2\cot\gamma_0 L) - (\mu_2^2 - \mu_1^2)\sin(2\gamma_0 L)]}},\\
a_1 &= \frac{2\epsilon_1\sqrt{\gamma_1^3\epsilon_2}}{\sqrt{\epsilon_2 k_0 k_{z1}[2(\epsilon_1^2\gamma_1 L + \epsilon_2^2\cot\gamma_1 L) - (\epsilon_2^2 - \epsilon_1^2)\sin(2\gamma_1 L)]}},\\
a_2 &= \frac{2\mu_1\sqrt{\gamma_2^3\mu_2}}{\sqrt{k_0 k_{z2}[2(\mu_1^2\gamma_2 L - \mu_2^2\tan\gamma_2 L) + (\mu_2^2 - \mu_1^2)\sin(2\gamma_2 L)]}},\\
a_3 &= \frac{2\epsilon_1\sqrt{\gamma_3^3\mu_2}}{\sqrt{\epsilon_2 k_0 k_{z3}[2(\epsilon_1^2\gamma_3 L - \epsilon_2^2\tan\gamma_3 L) + (\epsilon_2^2 - \epsilon_1^2)\sin(2\gamma_3 L)]}}.
\end{aligned}$$

Replacing in Eq. (112), we get:

$$\tilde{\kappa}_{01}^{(AP)} = -a_0^* a_1 \frac{c_0 k_0}{8\pi\gamma_0}\tilde{\theta}_{AP}\cos\gamma_0 L\sin\gamma_1 L, \qquad \tilde{\kappa}_{23}^{(AP)} = a_2^* a_3 \frac{c_0 k_0}{8\pi\gamma_2}\tilde{\theta}_{AP}\cos\gamma_3 L\sin\gamma_2 L. \tag{116}$$

To visualize the mode coupling we take numerical values of some waveguide parameters while others are derived from previous equations. Setting the operational wavelength to be 3.3 times the width of the guide ($2L = 40\,\mu m$), i.e., $\lambda_0 = 132\,\mu m$, therefore, $k_0 = 0.0476\,\mu m^{-1}$. Also, we take $\epsilon_1 = 12$, $\epsilon_2 = 16$ y $\mu_1 = \mu_2 = 1$. At this wavelength the waveguide can indeed host the 4 lowest modes shown in Table VII. These numerical values define the $\gamma_i$ of the transcendental equation for each mode. Table VIII summarizes the constants that enter in the Eqs. (103) and (104), where the $k_{zs}$ are $k_{TE}$ or $k_{TM}$ respectively.

| **Mode $s$ in Eq. (97)** | $\gamma_s\ \left[\frac{1}{cm}\right]$ | $k_{zs}\ \left[\frac{1}{cm}\right]$ | $a_s\ \left[\frac{1}{s}\sqrt{\frac{g}{cm}}\right]$ |
|---|---|---|---|
| 0 | 505.509 | 1835.66 | 0.000275101 |
| 1 | 544.487 | 1824.48 | 0.0000757061 |
| 2 | 918.477 | 1667.81 | 0.000385529 |
| 3 | 929.352 | 1661.78 | 0.0000833669 |

TABLE VIII: Values of the parameters for the amplitudes in Eqs. (103) and (104).

Having given numerical values for all the parameters in the differential equations for the coupled amplitudes in Eqs. (103) and (104) we can now solve for these. As in the antiparallel case AP, the TE-even (0) mode couples to the TM-even (1) mode, we can plot the evolution along $z$ of the $A_{TE/TM}$ coefficient given an initial condition at $z = 0$, and as a function of $\theta_{AP}$. In Figs. 12 and 13, we show the coupling between the TE-even - TM-even modes (01-coupling). All plots in Figs. 12 represent solutions such that at $z = 0$ the fields start with only the TE-even component, i.e., at $z = 0$ the TM-even mode is turned off. As the equations between the modes are coupled, in the evolution the initially turned off components emerge. Each column represents the effect for increasing $\theta_{AP}$ values. For low $\theta_{AP}$ we see that if the fields starts at $z = 0$ as a TE-even mode it will mostly remain as TE-even modes as it propagates along the slab waveguide. However, as $\theta_{AP}$ increases, we see that up to a certain length, the TE-even amplitude decays to zero, while the TM-even amplitude locally is different from zero. We also observe that the larger the $\theta_{AP}$ the shorter the wave has to propagate along $z$ for the TM-even to dominate over the TE-even mode. Finally we observe that both amplitudes have an in-phase enveloping oscillations. Figs. 13 show the analogous situation for the case when the fields starts at $z = 0$ as a TM-even mode, and similar situations occur, however, in this case the enveloping oscillations of each amplitude evolve out-of-phase.

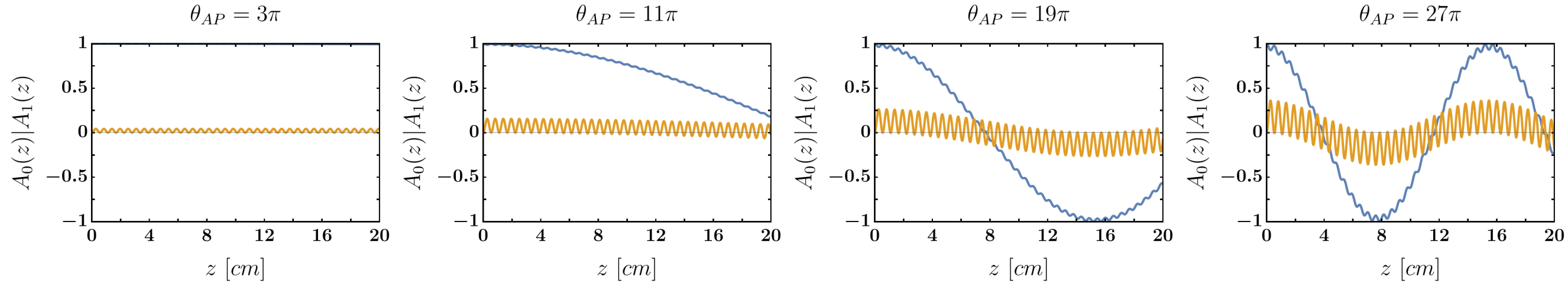

FIG. 12: TE-even – TM-even coupling. The field starts at $z = 0$ with $A_0(z = 0) = 1$ and $A_1(z = 0) = 0$. From left to right the effect of the mode coupling for increasing values of $\theta$.

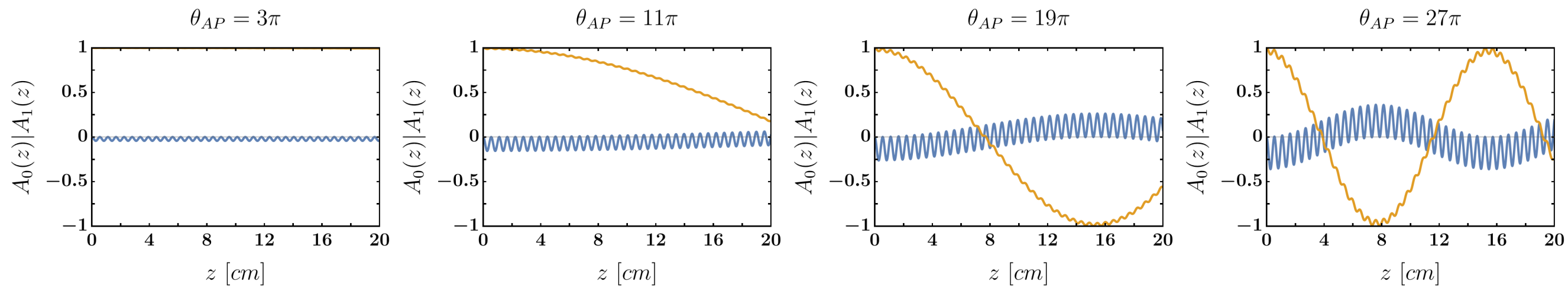

FIG. 13: TE-even – TM-even coupling. The field starts at $z = 0$ with $A_0(z = 0) = 0$ and $A_1(z = 0) = 1$. From left to right the effect of the mode coupling for increasing values of $\theta$.

Figures 14 and 15 shows the coupling between the TE-odd - TM-odd modes (23-coupling). The behaviour of the evolution of the amplitudes given a TE-odd or TM-odd initial condition is similar to what happened for the 01-coupling.

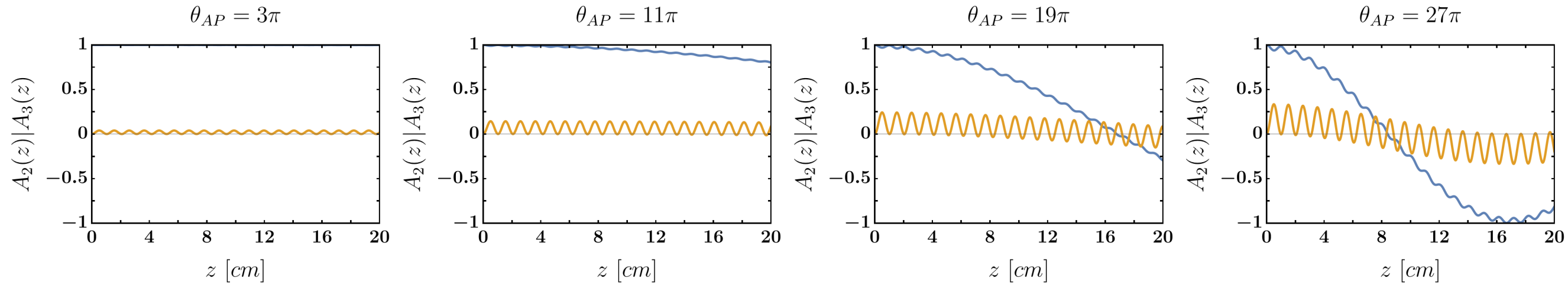

FIG. 14: TE-odd – TM-odd coupling. The field starts at $z = 0$ with $A_2(z = 0) = 1$ and $A_3(z = 0) = 0$. From left to right the effect of the mode coupling for increasing values of $\theta$.

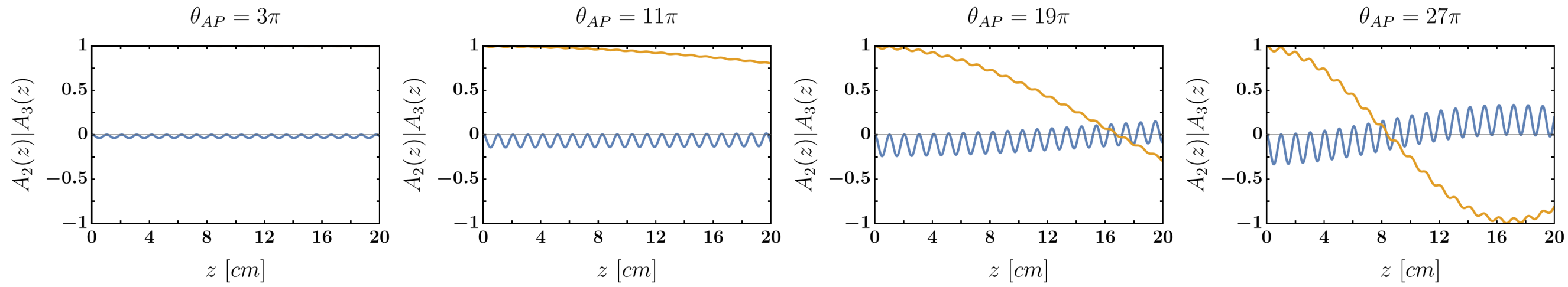

FIG. 15: TE-odd – TM-odd coupling. The field starts at $z = 0$ with $A_2(z = 0) = 0$ and $A_3(z = 0) = 1$. From left to right the effect of the mode coupling for increasing values of $\theta$.

One figure of merit to illustrate the coupling is to ask how far down the slab along the propagation does the initial mode vanishes. In figures 12 and 13, we observe that for $\theta_{AP} = 19\pi$ the first time the starting mode decays to zero occurs at $\approx 8\ cm$. These decay lengths grow linearly with the width of the slab.

### E. Power transmitted by the modes and by the full EM field

Having normalized the modes such that the density of power per unit length along the waveguide of 1 [W/cm], the dimensionless power density per unit length is given by:

$$p_z = \sum_n |A_n|^2. \tag{117}$$

This is useful to understand how the power is being "transferred" from one mode to the other and also serves as a means to check energy conservation. Figures 16 illustrate this for the TE-even – TM-even. For the power transferred between the TE-odd – TM-odd couplings a similar thing happens (not shown). The straight lines plot the sum of the powers of both even modes. Consistent with energy conservation, one sees that it corresponds to the total (unit) power density. The first row shows solutions for which at $z = 0$ the field starts with TE-even component only, while the TM-even component is turned off. Correspondingly, the second row shows the converse situation, i.e., the cases when at $z = 0$ the field starts with the TM-even component only, and the TE-even turned off.

## VII. CONCLUSIONS

In summary, we have studied the propagation of electromagnetic waves in a planar slab waveguide whose core medium is a generic reciprocal and non-chiral topological insulator characterized by the topological magnetoelectric parameter, $\Theta$. Working in the framework of $\Theta$-electrodynamics ($\Theta$-ED) to describe the electromagnetic response of Topological Insulators, we introduced the asymmetric and symmetric topological slab waveguide. For the asymmetric case we briefly show the effects of the asymmetry parameter $\delta$ on the mode distribution and their $\Theta$ modification. For given optical parameters and operational frequency, varying the asymmetry allows to control the number of propagating modes and also "lifts the degeneracy in $\Theta$" of the even modes. However, keeping the asymmetry parameter general to compute exact modes, mode couplings, dispersion relations is computationally costly. Therefore, we then focused on a symmetric slab geometry where both claddings share identical optical parameters, and an antiparallel configuration for which the gradient of $\Theta$ points in opposite directions at each interface. Under these conditions, we derived the exact field solutions subject to the corresponding $\Theta$-boundary conditions. From this analysis, we found that the presence of the $\Theta$-term modifies the electromagnetic modes in comparison to those of ordinary electrodynamics (denoted as 0-ED). In particular, for small $\Theta$ the modes revert smoothly to their 0-ED counterparts with albeit small magneto-electric behaviour (Eqs. (78) and (79), confirming that topological corrections vanish in the limit $\Theta \to 0$. However, away from this regime, both the wavevectors and the mode profiles acquire nontrivial $\Theta$-dependence, thus offering a distinctive signature of the topological magnetoelectric response. Another important implication of our results is the high-$\theta$ or, equivalently, the low TI-impedance relative to the claddings ($Z_2 \ll Z_1$) limit. In this case, we observe that the EM field propagates in quasi-TEM (transverse electromagnetic) manner. These are not exact TEM modes as those found for the first time in [67], however, from practical point of view quasi-TEM modes are still considered relevant as done in [68, 69].

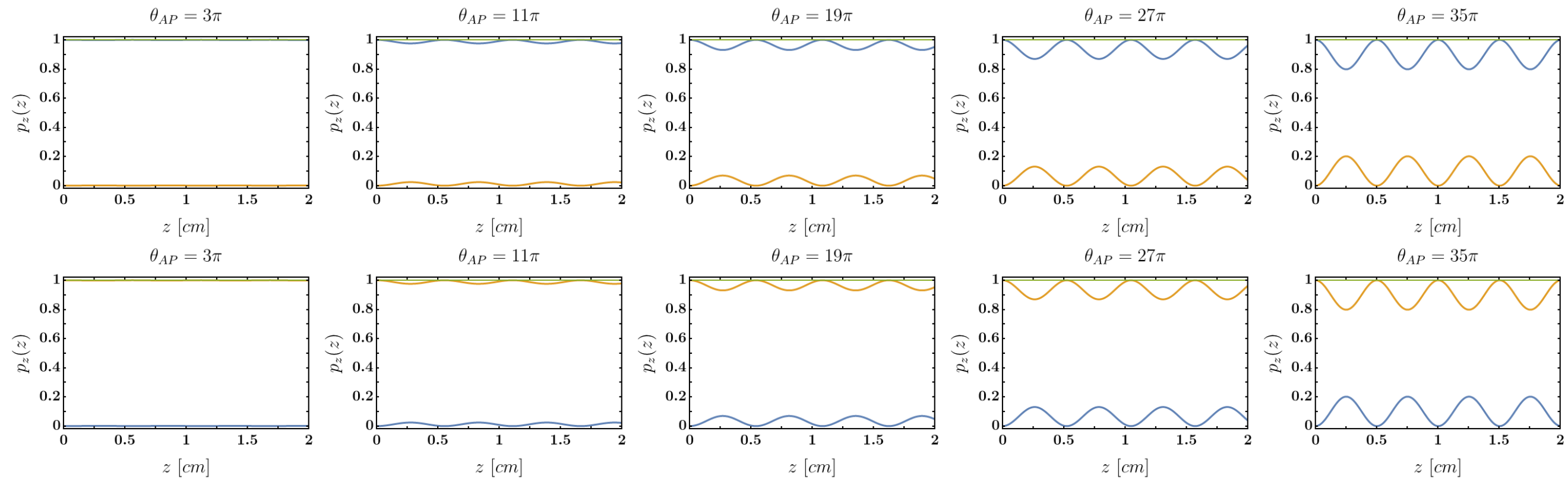

FIG. 16: Power transferred between TE and TM- even modes due to coupling triggered by $\theta$. Though the field starts with one particular mode, the $\theta$-coupling turns the other mode on, both with oscillatory behaviour along the guide. To total power transported by the EM wave, shown by the straight line, is constant, consistent with energy conservation.

Furthermore, by employing the exact solutions as a complete basis, we demonstrated how an initial field prescribed at the waveguide entrance ($z = 0$) can be expanded in the $\Theta$-ED modes, yielding coefficients that vary along the propagation direction and depend on $\Theta$. This is one of the important results and contribution of this work. Even in the case of 0-ED when there are "alterations" like, say, geometric inhomogeneities or anisotropies, time-varying perturbations, or even the use of general magnetoelectric media [45], to the date, most research on electromagnetic wave propagation in waveguides do not seek exact mode expansion for the optical confined modes and is mostly predicated in coupled-mode theory. Few examples that address the role of topological insulators [44] also employ either coupled-mode theory or, from the outset an approach that is perturbative in $\Theta$ i.e., capturing only $\mathcal{O}(\Theta)$ effects in the modified EM propagation. Our approach, however, for the first time provides exact solutions to all orders in $\Theta$. Though the modifications might not differ considerably as far as the observable magnitude of the effects, our solutions are exact and qualitatively different, presenting features that are not possible otherwise, namely, the strict hybridization of modes and a polarization rotation that arises as a direct consequence of the exact solutions that take into account the boundary conditions of $\Theta$-ED (and not as an artifact of coupled-mode theory). As we mentioned in the Introduction, the hybridization of modes is a prevalent phenomenon in waveguides and present in different situations. A particular case that is similar to ours is analyzed in [43]. In that work, the authors consider a parallel-plate waveguide filled with a Tellegen (nonreciprocal) bi-isotropic medium and in fact, they derive the existence of TE modes and a distinct family of hybrid modes, while showing that pure TM modes are excluded. Apart from the seeming similarities, some differences are in order, namely, in our case all guided modes supported by the $\Theta$-waveguide are hybrid modes with nonvanishing longitudinal components. In our case the hybridization is triggered by a surface $\Theta$-induced boundary-condition at the two TI interfaces. In the case of [43], the boundary conditions are imposed by perfect conductors. Therefore, these are two complementary scenarios in which hybridization arises from different surface effects. To gain deeper insights into the dispersion relations, we numerically solved the transcendental equations that govern them, revealing how the geometry, optical properties (permittivity and permeability), and topological character of the waveguide constrain the operational frequency ranges. In other words, for a given number of modes to exist in the $\Theta$-slab, the allowed frequencies must simultaneously satisfy the waveguide's geometric-optical requirements and remain in the valid regime of $\Theta$-ED for topological insulators.

We then determined how both phase and group velocities of the guided modes depend on $\Theta$, thus illustrating the interplay between topological and conventional waveguide parameters. Specifically, the dispersion relations in slab waveguides, either in 0-ED or in $\Theta$-ED, cannot be found analytically, as these are defined via transcendental equations. In our case, we also find the dispersion relations numerically for given waveguide parameters. The geometry and optical parameters of the waveguide restrict the operational frequencies for a given number of modes that are allowed inside the waveguide.

Lastly, inspired by the smallness of $\Theta$ for known three-dimensional topological insulators, we adopted a perturbative (coupled-mode) approach to treat the $\Theta$-effects as corrections to the 0-ED waveguide modes. In this framework, the amplitudes of the ordinary modes become coupled through a set of differential equations whose coupling coefficients scale with $\Theta$. Consequently, even a weak topological contribution can redistribute power among modes as the wave propagates, thus providing yet another pathway for detecting or exploiting topological signatures in guided-wave photonic applications.

Our findings highlight that even modest $\Theta$-values can produce measurable deviations from 0-ED wave propagation in slab geometries. These theoretical results are promising for designing next-generation waveguide devices where subtle magnetoelectric effects could be harnessed for improved control of light, and open new opportunities for probing the topological nature of real materials under electromagnetic fields.

## VIII. ACKNOWLEDGEMENTS

S. F thanks support from the Scholarship Program/BECAS DOCTORADO UNAB during earlier stages of this work. M.C. had been funded by Universidad Andrés Bello UNAB DI-16-20/REG internal project at earlier stages of this project and thanks Instituto de Ciencias Nucleares at UNAM, for the warm hospitality at several stages. All authors acknowledge partial support from DGAPA-UNAM Project No. IG100224 and by Project CONAHCyT (México) No. 428214. A.M.-R. also acknowledges financial support by SECIHTI project No. CBF-2025-I-1862 and by the Marcos Moshinsky Foundation.

## IX. APPENDIX

### A. Notation and definitions

$$\Theta = \frac{\alpha}{\pi}\theta_{\text{TI}}, \tag{118}$$

$$\tilde{\theta}_i \equiv \Theta_{i+1} - \Theta_i, \tag{119}$$

$$\tilde{\theta}_1 = \Theta_2 - \Theta_1, \qquad \tilde{\theta}_2 = \Theta_3 - \Theta_2, \tag{120}$$

$$\text{Antiparallel:} \quad \tilde{\theta}_2 = -\tilde{\theta}_1, \qquad \Rightarrow \qquad \Theta_1 = \Theta_3, \tag{121}$$

$$\text{Parallel:} \quad \tilde{\theta}_2 = \tilde{\theta}_1, \qquad \Rightarrow \qquad \Theta_1 + \Theta_3 = 2\Theta_2, \tag{122}$$

$$\tilde{\theta}_{AP} \equiv \tilde{\theta}_1 = -\tilde{\theta}_2, \tag{123}$$

$$\tilde{\theta}_P \equiv \tilde{\theta}_1 = \tilde{\theta}_2, \tag{124}$$

$$\text{For} \quad \Theta_3 = \Theta_1 = 0 \qquad \Rightarrow \qquad \tilde{\theta}_1 = \Theta_2 \quad \text{and} \quad \tilde{\theta}_2 = -\Theta_2, \tag{125}$$

$$k_{x1} = i\alpha_1, \qquad \alpha_1 \equiv v/L, \tag{126}$$

$$k_{x2} = \gamma, \qquad \gamma \equiv u/L, \tag{127}$$

$$k_{x_3} = i\alpha_3, \qquad \alpha_3 \equiv w/L, \tag{128}$$

$$\delta = \frac{n_3^2 - n_1^2}{n_2^2 - n_1^2}, \qquad \text{Asymmetry parameter of slab}, \tag{129}$$

$$\delta_x = \frac{1}{\alpha}, \qquad \text{penetration length}, \tag{130}$$

$$\ell_x = \frac{\delta_x}{L} = \frac{1}{v} = \frac{1}{\alpha L}, \qquad \text{penetration parameter or ``dimensionless penetration length''}, \tag{131}$$

$$u^2 + v^2 = L^2 k_0^2 (n_2^2 - n_1^2) \equiv \mathcal{R}^2, \tag{132}$$

$$\mathbf{k}^2 \equiv k_x^2 + k_y^2 + k_z^2 = k_\perp^2 + k_z^2, \tag{133}$$

$$k_0 \equiv \frac{\omega}{c_0} = \frac{2\pi}{\lambda_0}, \qquad c_0 = 1/\sqrt{\epsilon_0 \mu_0}, \tag{134}$$

$$k^2 = k_\perp^2 + k_z^2 = k_0^2 \mu\epsilon. \tag{135}$$

**B. Lorentz Reciprocity Theorem in Θ-ED**

Let us consider the curl equations of the fields for a specific mode,

$$\boldsymbol{\nabla}\times\mathbf{E}_n = ik_0\mathbf{B}_n, \qquad \boldsymbol{\nabla}\times(\frac{1}{\mu}\mathbf{B}_n) = -ik_0\epsilon\mathbf{E}_n + \boldsymbol{\nabla}\Theta\times\mathbf{E}_n. \tag{136}$$

We apply convenient dot products (with other mode $s$) and conjugate the equation on the right. Then we subtract the equations as follows,

$$\begin{aligned}\frac{1}{\mu}\mathbf{B}_s^*\cdot\boldsymbol{\nabla}\times\mathbf{E}_n - \mathbf{E}_n\cdot\boldsymbol{\nabla}\times\frac{1}{\mu}\mathbf{B}_s^* &= ik_0\left(\frac{1}{\mu}\mathbf{B}_n\cdot\mathbf{B}_s^* - \epsilon\mathbf{E}_n\cdot\mathbf{E}_s^*\right) - \mathbf{E}_n\cdot(\boldsymbol{\nabla}\Theta\times\mathbf{E}_s^*),\\ \boldsymbol{\nabla}\cdot\left(\mathbf{E}_n\times\frac{1}{\mu}\mathbf{B}_s^*\right) &= ik_0\left(\frac{1}{\mu}\mathbf{B}_n\cdot\mathbf{B}_s^* - \epsilon\mathbf{E}_n\cdot\mathbf{E}_s^*\right) + \boldsymbol{\nabla}\Theta\cdot(\mathbf{E}_n\times\mathbf{E}_s^*).\end{aligned} \tag{137}$$

Let us consider the conjugate of Eq. (137) and a change in the names of the modes $n\leftrightarrow s$,

$$\boldsymbol{\nabla}\cdot\left(\mathbf{E}_s^*\times\frac{1}{\mu}\mathbf{B}_n\right) = -ik_0\left(\frac{1}{\mu}\mathbf{B}_n\cdot\mathbf{B}_s^* - \epsilon\mathbf{E}_n\cdot\mathbf{E}_s^*\right) - \boldsymbol{\nabla}\Theta\cdot(\mathbf{E}_n\times\mathbf{E}_s^*). \tag{138}$$

and we sum the equations,

$$\boldsymbol{\nabla}\cdot\left(\mathbf{E}_n\times\frac{1}{\mu}\mathbf{B}_s^* + \mathbf{E}_s^*\times\frac{1}{\mu}\mathbf{B}_n\right) = 0. \tag{139}$$

Now expand into transverse and longitudinal components, where $\mathbf{E}_n\propto e^{ik_{zn}z}$,

$$\boldsymbol{\nabla}_\perp\cdot\left(\mathbf{E}_n\times\frac{1}{\mu}\mathbf{B}_s^* + \mathbf{E}_s^*\times\frac{1}{\mu}\mathbf{B}_n\right)_\perp + i(k_{zn}-k_{zs})\left(\mathbf{E}_n\times\frac{1}{\mu}\mathbf{B}_s^* + \mathbf{E}_s^*\times\frac{1}{\mu}\mathbf{B}_n\right)\cdot\hat{\mathbf{z}} = 0. \tag{140}$$

Integrate into the transverse surface over all $\mathbb{R}^2$. The first term vanishes because the fields tend to zero at infinity,

$$i(k_{zn}-k_{zs})\int_{\mathbb{R}^2}\left(\mathbf{E}_n\times\frac{1}{\mu}\mathbf{B}_s^* + \mathbf{E}_s^*\times\frac{1}{\mu}\mathbf{B}_n\right)\cdot\hat{\mathbf{z}}da_\perp = 0, \tag{141}$$

as $(k_{zn}-k_{zs})\neq 0$ the integral must be zero. Furthermore, if we consider

$$P_{zn} = \frac{1}{2}\int_{\mathbb{R}^2}\mathrm{Re}\left(\mathbf{E}_n\times\frac{1}{\mu}\mathbf{B}_n^*\right)\cdot\hat{\mathbf{z}}da_\perp = \frac{1}{4}\int_{\mathbb{R}^2}\left(\mathbf{E}_n\times\frac{1}{\mu}\mathbf{B}_n^* + \mathbf{E}_n^*\times\frac{1}{\mu}\mathbf{B}_n\right)\cdot\hat{\mathbf{z}}da_\perp. \tag{142}$$

Therefore, we have the orthogonality relation of the modes,

$$\frac{1}{4}\int_{\mathbb{R}^2}\left(\mathbf{E}_n\times\frac{1}{\mu}\mathbf{B}_s^* + \mathbf{E}_s^*\times\frac{1}{\mu}\mathbf{B}_n\right)\cdot\hat{\mathbf{z}}da_\perp = \delta_{ns}P_{zn}, \tag{143}$$

where $\mathbf{E}_n = \mathbf{E}_n(\mathbf{r}_\perp)$ is the EM profile without $e^{i(k_z z-\omega t)}$.

**C. Explicit forms of the Θ-ED modes**

For completeness, we show the expressions of the EM field profiles in all their components, for the case of the first two modes in the first labeling $p=1,2$, corresponding to the $0+$ and $0-$ labeling or QTE-even-1 and QTM-even-1 respectively.

*1. QTE-even-1 Mode*

The total EM field for the QTE-even mode labeled as $p=1$ is,

$$E_{z1}(x) = \frac{B_{e1}}{\mathcal{F}_+}\begin{cases} -e^{\alpha_1(x+L)}\sin\gamma_1 L & \text{for} \quad x\leq -L,\\ \sin\gamma_1 x & \text{for} \quad -L<x<L,\\ e^{-\alpha_1(x-L)}\sin\gamma_1 L & \text{for} \quad x\geq L,\end{cases} \qquad B_{z1}(x) = B_{e1}\begin{cases} -\xi_+ e^{\alpha_1(x+L)}\sin\gamma_1 L & \text{for} \quad x\leq -L,\\ \sin\gamma_1 x & \text{for} \quad -L<x<L,\\ \xi_+ e^{-\alpha_1(x-L)}\sin\gamma_1 L & \text{for} \quad x\geq L,\end{cases}$$

$$E_{x1}(x) = B_{e1}\frac{ik_{z1}}{\mathcal{F}_+}\begin{cases} \frac{1}{\alpha_1}e^{\alpha_1(x+L)}\sin\gamma_1 L & \text{for} \quad x\leq -L, \\ \frac{1}{\gamma_1}\cos\gamma_1 x & \text{for} \quad -L<x<L, \\ \frac{1}{\alpha_1}e^{-\alpha_1(x-L)}\sin\gamma_1 L & \text{for} \quad x\geq L, \end{cases} \quad B_{x1}(x) = B_{e1}ik_{z1}\begin{cases} \frac{\xi_+}{\alpha_1}e^{\alpha_1(x+L)}\sin\gamma_1 L & \text{for} \quad x\leq -L, \\ \frac{1}{\gamma_1}\cos\gamma_1 x & \text{for} \quad -L<x<L, \\ \frac{\xi_+}{\alpha_1}e^{-\alpha_1(x-L)}\sin\gamma_1 L & \text{for} \quad x\geq L, \end{cases}$$

$$E_{y1}(x) = -B_{e1}ik_0\begin{cases} \frac{\xi_+}{\alpha_1}e^{\alpha_1(x+L)}\sin\gamma_1 L & \text{for} \quad x\leq -L, \\ \frac{1}{\gamma_1}\cos\gamma_1 x & \text{for} \quad -L<x<L, \\ \frac{\xi_+}{\alpha_1}e^{-\alpha_1(x-L)}\sin\gamma_1 L & \text{for} \quad x\geq L, \end{cases} \quad B_{y1}(x) = B_{e1}\frac{ik_0}{\mathcal{F}_+}\begin{cases} \frac{\epsilon_1\mu_1}{\alpha_1}e^{\alpha_1(x+L)}\sin\gamma_1 L & \text{for} \quad x\leq -L, \\ \frac{\epsilon_2\mu_2}{\gamma_1}\cos\gamma_1 x & \text{for} \quad -L<x<L, \\ \frac{\epsilon_1\mu_1}{\alpha_1}e^{-\alpha_1(x-L)}\sin\gamma_1 L & \text{for} \quad x\geq L, \end{cases}$$

where the sub-index 1 is the mode and $\alpha_1 = \xi_+\gamma_1\tan\gamma_1 L$. A relevant physical quantity is the EM power along the guide direction,

$$P_z = \frac{1}{2}\int\frac{1}{\mu}\text{Re}\left(\mathbf{E}_\perp\times\mathbf{B}_\perp^*\right)\cdot\hat{\mathbf{z}}\,ds. \tag{144}$$

In the $x$-Slab system, we define the linear power density $p_z \equiv P_z/l_y$. For the $p=1$ mode,

$$p_{z1} = \frac{1}{2}\int_{-\infty}^{\infty}\frac{1}{\mu}\text{Re}\left(\mathbf{E}_{\perp 1}\times\mathbf{B}_{\perp 1}^*\right)\cdot\hat{\mathbf{z}}\,dx. \tag{145}$$

Replacing the EM field we have,

$$p_{z1} = B_{e1}^2\frac{k_0k_{z1}}{2}\left(\frac{\sin^2(\gamma_1 L)}{\mu_1\alpha_1^3}\left(\frac{\mu_1\epsilon_1}{\mathcal{F}_+^2}+\xi_+^2\right)+\frac{(\gamma_1 L+\sin(\gamma_1 L)\cos(\gamma_1 L))}{\mu_2\gamma_1^3}\left(\frac{\mu_2\epsilon_2}{\mathcal{F}_+^2}+1\right)\right). \tag{146}$$

We can normalize the modes by considering that $B_{e1}$ is such that the power density is $p_{z1} = 1$[W/m].

### 2. *QTM-even-1 Mode*

The total EM field for the QTM-even mode labeled as $p=2$ is,

$$E_{z2}(x) = E_{e2}\begin{cases} -e^{\alpha_2(x+L)}\sin\gamma_2 L & \text{for} \quad x\leq -L, \\ \sin\gamma_2 x & \text{for} \quad -L<x<L, \\ e^{-\alpha_2(x-L)}\sin\gamma_2 L & \text{for} \quad x\geq L, \end{cases} \quad B_{z2}(x) = E_{e2}\mathcal{F}_-\begin{cases} -\xi_- e^{\alpha_2(x+L)}\sin\gamma_2 L & \text{for} \quad x\leq -L, \\ \sin\gamma_2 x & \text{for} \quad -L<x<L, \\ \xi_- e^{-\alpha_2(x-L)}\sin\gamma_2 L & \text{for} \quad x\geq L, \end{cases}$$

$$E_{x2}(x) = E_{e2}ik_{z2}\begin{cases} \frac{1}{\alpha_2}e^{\alpha_2(x+L)}\sin\gamma_2 L & \text{for} \quad x\leq -L, \\ \frac{1}{\gamma_2}\cos\gamma_2 x & \text{for} \quad -L<x<L, \\ \frac{1}{\alpha_2}e^{-\alpha_2(x-L)}\sin\gamma_2 L & \text{for} \quad x\geq L, \end{cases} \quad B_{x2}(x) = E_{e2}ik_{z2}\mathcal{F}_-\begin{cases} \frac{\xi_-}{\alpha_2}e^{\alpha_2(x+L)}\sin\gamma_2 L & \text{for} \quad x\leq -L, \\ \frac{1}{\gamma_2}\cos\gamma_2 x & \text{for} \quad -L<x<L, \\ \frac{\xi_-}{\alpha_2}e^{-\alpha_2(x-L)}\sin\gamma_2 L & \text{for} \quad x\geq L, \end{cases}$$

$$E_{y2}(x) = -E_{e2}ik_0\mathcal{F}_-\begin{cases} \frac{\xi_-}{\alpha_2}e^{\alpha_2(x+L)}\sin\gamma_2 L & \text{for} \quad x\leq -L, \\ \frac{1}{\gamma_2}\cos\gamma_2 x & \text{for} \quad -L<x<L, \\ \frac{\xi_-}{\alpha_2}e^{-\alpha_2(x-L)}\sin\gamma_2 L & \text{for} \quad x\geq L, \end{cases} \quad B_{y2}(x) = E_{e2}ik_0\begin{cases} \frac{\epsilon_1\mu_1}{\alpha_2}e^{\alpha_2(x+L)}\sin\gamma_2 L & \text{for} \quad x\leq -L, \\ \frac{\epsilon_2\mu_2}{\gamma_2}\cos\gamma_2 x & \text{for} \quad -L<x<L, \\ \frac{\epsilon_1\mu_1}{\alpha_2}e^{-\alpha_2(x-L)}\sin\gamma_2 L & \text{for} \quad x\geq L. \end{cases}$$

The density power is,

$$p_{z2} = E_{e2}^2\frac{k_0k_{z2}}{2}\left(\frac{\sin^2(\gamma_2 L)}{\mu_1\alpha_2^3}\left(\mu_1\epsilon_1+\mathcal{F}_-^2\xi_-^2\right)+\frac{(\gamma_2 L+\sin(\gamma_2 L)\cos(\gamma_2 L))}{\mu_2\gamma_2^3}\left(\mu_2\epsilon_2+\mathcal{F}_-^2\right)\right). \tag{147}$$

We can also normalize by considering that $E_{e2}$ is such that the power density is $p_{z2} = 1$ [W/m].

---